\documentclass[superscriptaddress,notitlepage]{revtex4-1}
\usepackage{graphicx}

\newcommand{\half}{\frac{1}{2}}

\begin{document}
\title{Topological Quantum Chemistry}
\author{Barry Bradlyn}
\thanks{These authors contributed equally to the preparation of this work.}
\affiliation{Princeton Center for Theoretical Science, Princeton University, Princeton, New Jersey 08544, USA}
\author{L. Elcoro}
\thanks{These authors contributed equally to the preparation of this work.}
\affiliation{Department of Condensed Matter Physics, University of the Basque Country UPV/EHU, Apartado 644, 48080 Bilbao, Spain}
\author{Jennifer Cano}
\thanks{These authors contributed equally to the preparation of this work.}
\affiliation{Princeton Center for Theoretical Science, Princeton University, Princeton, New Jersey 08544, USA}
\author{M.~G. Vergniory}
\thanks{These authors contributed equally to the preparation of this work.}
\affiliation{Donostia International Physics Center, P. Manuel de Lardizabal 4, 20018 Donostia-San Sebasti\'{a}n, Spain}
\affiliation{Department of Applied Physics II, University of the Basque Country UPV/EHU, Apartado 644, 48080 Bilbao, Spain}
\affiliation{Max Planck Institute for Solid State Research, Heisenbergstr. 1,
70569 Stuttgart, Germany.}
\author{Zhijun Wang}
\thanks{These authors contributed equally to the preparation of this work.}
\affiliation{Department of Physics, Princeton University, Princeton, New Jersey 08544, USA}
\author{C. Felser}
\affiliation{Max Planck Institute for Chemical Physics of Solids, 01187 Dresden, Germany}
\author{M.~I.~Aroyo}
\affiliation{Department of Condensed Matter Physics, University of the Basque Country UPV/EHU, Apartado 644, 48080 Bilbao, Spain}
\author{B. Andrei Bernevig}
\thanks{To Whom correspondence should be addressed}
\thanks{Permanent Address: Department of Physics, Princeton University, Princeton, New Jersey 08544, USA }
\affiliation{Department of Physics, Princeton University, Princeton, New Jersey 08544, USA}
\affiliation{Donostia International Physics Center, P. Manuel de Lardizabal 4, 20018 Donostia-San Sebasti\'{a}n, Spain}
\affiliation{Laboratoire Pierre Aigrain, Ecole Normale Sup\'{e}rieure-PSL Research University, CNRS, Universit\'{e} Pierre et Marie Curie-Sorbonne Universit\'{e}s, Universit\'{e} Paris Diderot-Sorbonne Paris Cit\'{e}, 24 rue Lhomond, 75231 Paris Cedex 05, France}
\affiliation{Sorbonne Universit\'{e}s, UPMC Univ Paris 06, UMR 7589, LPTHE, F-75005, Paris, France}
\date{\today}
\begin{abstract}
The past decade's apparent success in predicting and experimentally discovering distinct classes of topological insulators (TIs) and semimetals masks a fundamental shortcoming: out of 200,000 stoichiometric compounds extant in material databases, only several hundred of them are topologically nontrivial. Are TIs \emph{that} esoteric, or does this reflect a fundamental problem with the current piecemeal approach to finding them? To address this, we propose a new and complete electronic band theory that highlights the link between topology and local chemical bonding, and combines this with the conventional band theory of electrons. Topological Quantum Chemistry is a description of the universal global properties of \emph{all} possible band structures and materials, comprised of a \emph{graph} theoretical description of momentum space and a dual \emph{group} theoretical description in real space. We classify the possible band structures for \emph{all} 230 crystal symmetry groups that arise from local atomic orbitals, and show which are topologically nontrivial. We show how our topological band theory sheds new light on known TIs, and demonstrate the power of our method to predict a plethora of new TIs.
\end{abstract}
\maketitle
\section{Introduction}\label{sec:intro}

For the past century, chemists and physicists have advocated fundamentally different perspectives on materials: while chemists have adopted an {intuitive} ``local'' viewpoint of hybridization, ionic chemical bonding and finite-range interactions, physicists have
{described} materials through band-structures or Fermi surfaces in a nonlocal, momentum-space picture. These two descriptions seem disjoint, especially with the advent of {TIs}, {which are} exclusively understood in terms of the nontrivial topology of 
Bloch Hamiltonians throughout the Brillouin zone (BZ) in momentum space. Despite the apparent success that the field has had in  predicting \emph{some} [mostly time-reversal {(TR) invariant]} {TIs}, {conventional band theory} is ill-suited to a \emph{natural} treatment of {TIs}. {Given the paucity of known TIs (less than 400 materials out of 200,000 existent in crystal structure databases!), one may ask whether topological materials are truly so rare, or if this reflects a failing of the conventional theory.}

By their very nature, the topological properties of energy bands are properties \emph{global} in momentum space. The duality between real and momentum (direct and reciprocal) space suggests that properties of bands which are nonlocal in momentum space will manifest locally in real space.  In this paper, we unify the real and momentum space descriptions of solids, and in doing so provide a new, powerful, complete and predictive band theory. 
Our procedure provides a complete understanding of the structure of bands in a material \emph{and} links its topological to the chemical orbitals at the Fermi level. It is therefore a theory of Topological Quantum Chemistry.

Developing a complete theory of topological bands requires an extremely large body of work made up of several ingredients. First, we {compile} \emph{all} the possible ways energy bands in a solid can be connected throughout the BZ to obtain \emph{all} {realizable} band structures in \emph{all} non-magnetic space-groups. Crystal symmetries place strong constraints on the allowed connections of bands. At high symmetry points, $\mathbf{k}_i$, in the BZ, Bloch functions are classified by irreducible representations (irreps) of
the symmetry group of {$\mathbf{k}_i$}, 
which also determine the degeneracy.
Away from these high symmetry points, fewer symmetry constraints exist, and some degeneracies are lowered. This is the heart of the $\mathbf{k}\cdot\mathbf{p}$ approach\cite{Kittel87} to band structure, which gives a good description nearby high symmetry $\mathbf{k}-$points. However, {the global band structure requires patching} together the different $\mathbf{k}\cdot\mathbf{p}$ {theories} at various high symmetry points. Group theory places constraints -- ``compatibility relations''  -- on how this can be done. Each solution to these compatibility relations gives groups of bands with different connectivities, corresponding to different physically-realizable phases of matter (trivial or topological). We solve all compatibility relations {for all} 230 space groups {(SGs)} by mapping connectivity in band theory to the graph-theoretic problem of constructing multipartite graphs. Classifying the allowed connectivities of energy bands becomes a combinatorial problem of graph enumeration: we present a fully tractable, algorithmic solution.

Second, we develop the tools to compute how the real-space orbitals in a material determine the symmetry character of the electronic bands. Given only the Wyckoff positions and the orbital symmetry ($s,p,d$) of the elements/orbitals in a material, we derive the symmetry character of all energy bands at \emph{all} points in the BZ. We do this by extending the notion of a band representation {(BR)}, first introduced in Refs.~\onlinecite{Zak1982,Bacry1988}, to the physically relevant case of materials with spin-orbit coupling (SOC) and/or TR symmetry.  {A BR consists of all} bands linked to localized orbitals respecting the crystal symmetry (and possibly TR). The set of BRs is strictly smaller than the set of groups of bands obtained from our graph theory\cite{Bacry1993}. {We identify a special subset of ``elementary'' BRs (EBRs)\cite{Zak1982,Evarestov1997}, {elaborated upon in the Supplementary Material (SM)}, which are the smallest sets of bands
{derived from} local atomic-like Wannier functions.}{\cite{Marzari2012}}
We work out all the {(10300)} different EBRs for all the SGs, Wyckoff positions, and orbitals, which we will present in a separate data paper.{\cite{GroupTheoryPaper}}

If the number of electrons is a fraction of the number of connected bands (connectivity) forming an EBR, then the system is a symmetry-enforced semimetal. The EBR method allows us to easily identify candidate semimetallic materials. As an amusing fact, we find that the largest possible number of connected bands in an EBR is {24} and hence the smallest possible fraction of filled bands in a semimetal is 1/{24}.  If, however, our graph analysis {reveals an instance where the number of connected bands} is \emph{smaller} than the total number of bands in the EBR, we conclude that a $\mathbf{k}$-space description exists but a Wannier one does not\cite{Soluyanov2011,Soluyanov2012,Read2016}, i.e. the {disconnected} bands are topological. TIs are then those materials with bands that are \emph{not} in our list of elementary components but are in our graph enumeration. We thus reformulate the momentum-space approach to topological indices as a real-space obstruction to the existence of atomic-like Wannier functions. In tandem with our graph-theoretic analysis of band connectivity, we enumerate \emph{all} the ways the transition to a topological phase can occur. Hence, we are able to classify \emph{all} topological crystalline insulators. This leads to previously unrecognized large classes of {TIs}. We show the power of our approach by predicting hundreds of new {TIs} and semimetals.

\section{Graph Theory and Band Structure}\label{sec:graphs}

To construct a band theory which accounts for the global momentum space structure of energy bands, we piece together groups of bands from distinct points in the BZ. Consider a $D$-dimensional crystalline material invariant under a SG $G$, containing elements of the form $\{R|\mathbf{d}\}$, where $R$ is a rotation or rotoinversion, and $\mathbf{d}$ is a translation. A Bravais lattice of translations is generated by a set of linearly independent translations $\{E|\mathbf{t}_i\}$, $i=1 \ldots D$, where $E$ represents the identity. 
Each $\mathbf{k}$ vector in the reciprocal lattice is left invariant by its little group, $G_\mathbf{k}\subset G$, and the Bloch wavefunctions $|u_n(\mathbf{k})\rangle$ transform under a sum of irreps of $G_\mathbf{k}$; bands at high-symmetry $\mathbf{k}$-vectors will have (non-accidental) degeneracies equal to the dimension of these representations {(reps)}. For spinless or spin-orbit free/coupled systems, these are ordinary linear/double-valued reps. Away from high symmetry points, degeneracies are reduced and bands disperse according to conventional $\mathbf{k}\cdot\mathbf{p}$ theory.

Consider two different high-symmetry $\mathbf{k}$ vectors, $\mathbf{k}_1$ and $\mathbf{k}_2$, and a line $\mathbf{k}_t=\mathbf{k}_1+t(\mathbf{k}_2-\mathbf{k}_1)$, $t\in [0,1]$  connecting them. 
{To determine} how the irreps at $\mathbf{k}_1$ connect to the irreps at $\mathbf{k}_2$ to form bands along this line, first note that the little group, $G_{\mathbf{k}_t}$, at any $\mathbf{k}_t$ on the line is a subgroup of both $G_{\mathbf{k}_1}$ and $G_{\mathbf{k}_2}$. Thus, an irrep, $\rho$, of $\mathbf{G}_{\mathbf{k}_1}$ at $\mathbf{k}_1$  will
split (subduce, or restrict) {along the line} to a (direct) sum of irreps $\bigoplus_i \tau_i$ of $G_{\mathbf{k}_t}$; symbolically (here and throughout we use ``$\approx$'' to denote equivalence of representations)
\begin{equation}
\rho\downarrow G_{\mathbf{k}_t}\approx\bigoplus_i\tau_i.\label{eq:comprelexample}
\end{equation}
\noindent These restrictions are referred to as \emph{compatibility relations}. Heuristically, they are found by taking the representation $\rho$ and ``forgetting'' about the symmetry elements of $G_{\mathbf{k}_1}$ which are not in $G_{\mathbf{k}_t}$. As energy bands in a crystal do not discontinuously end, the representations, $\sigma$, of $G_{\mathbf{k}_2}$ at $\mathbf{k}_2$ must also satisfy
\begin{equation}
\sigma\downarrow G_{\mathbf{k}_t}\approx\bigoplus_i\tau_i.
\end{equation}
Each representation, $\tau_i$, corresponds to a (group of) band(s) along the line $\mathbf{k}_t$;  bands coming from $\mathbf{k}_1$ in each irrep $\tau_i$ must join with a group of bands transforming in the same irrep coming from $\mathbf{k}_2$. We refer to each set of such pairings as a solution to the compatibility relations.

{Compatibility relations apply to} \emph{each and every} connection line/plane between each pair of $\mathbf{k}$ points in the Brillouin zone, leading to strong but factorially redundant restrictions on how bands may connect in a crystal. To construct the nonredundant solutions to the compatibility relations, we map the question to a problem in graph theory. 
Each {irrep} at the different symmetry distinct $\mathbf{k}$ vectors labels a node in a graph. In our previous example, {the nodes would be} labelled by $\rho,\sigma, \{\tau_1,\tau_2,\dots \}$ for the $\mathbf{k}_1$, $\mathbf{k}_2$, and $\mathbf{k}_t$ high symmetry points and line, respectively. 
{We draw the edges of the graph by the following} rules: \textbf{1.} Irreps at the same $\mathbf{k}$ vector can never be connected by edges -- our graph is multi-partite. \textbf{2.} Nodes corresponding to irreps at $\mathbf{k}_a$ and $\mathbf{k}_b$ can be connected only if $G_{\mathbf{k}_a}\subseteq G_{\mathbf{k}_b}$ or $G_{\mathbf{k}_b}\subseteq G_{\mathbf{k}_a}$ (i.e. $\mathbf{k}$-vectors are compatible). \textbf{3.} Edges must be consistent with the compatibility relations. For instance, Eq.~(\ref{eq:comprelexample}) {corresponds to} an edge from the node labelled by $\rho$ to each node labelled by $\tau_i$. We refer to such a graph as a \emph{connectivity graph}.

\begin{figure*}[ht]
\includegraphics[width=0.5\textwidth]{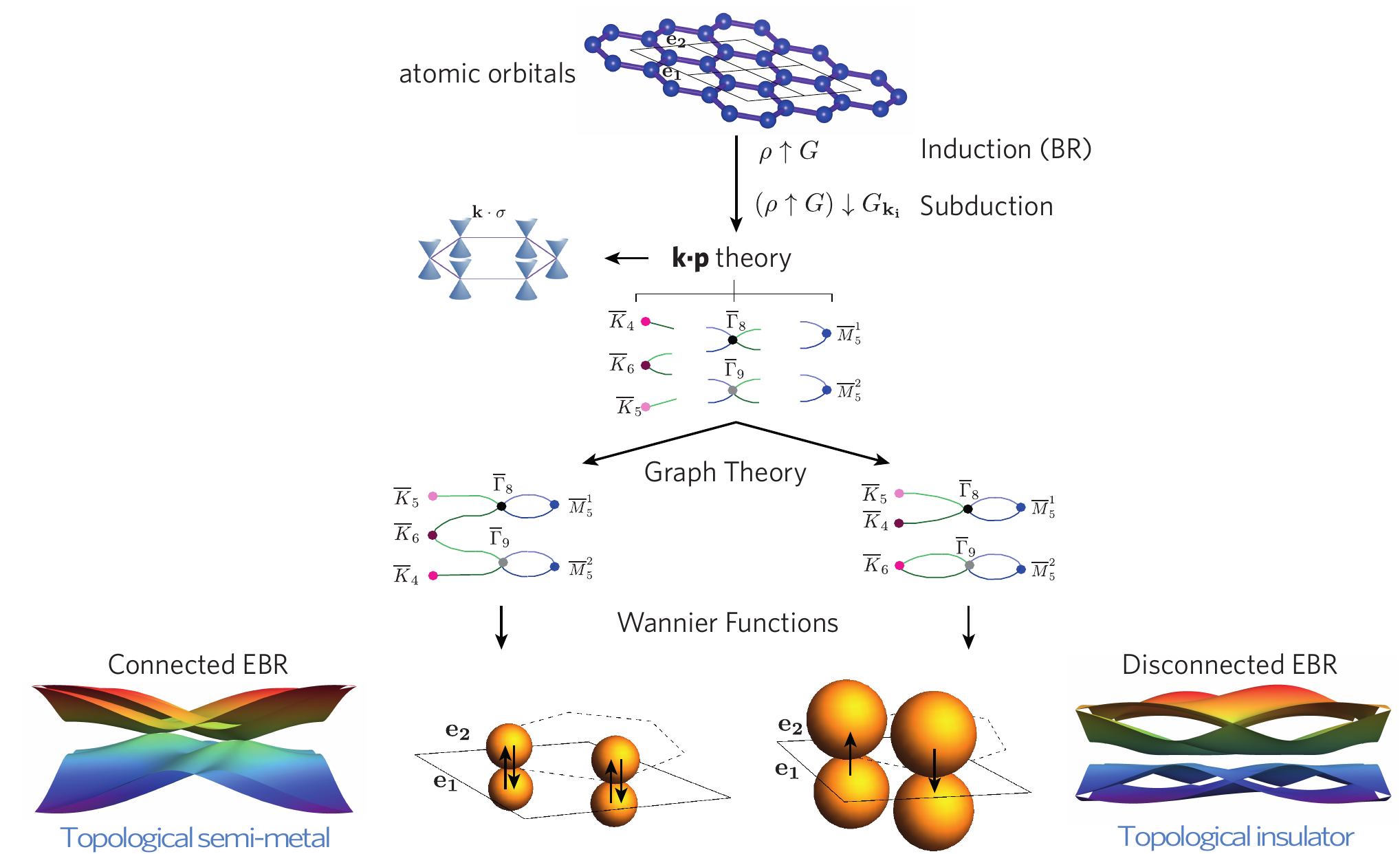}
\caption{\textbf{Schematic: how our theory applies to graphene with SOC}. We begin by inputting the orbitals ($|p_z\uparrow\rangle$, $|p_z\downarrow\rangle$) and lattice positions relevant near the Fermi level. Following the first arrow, we then induce an EBR from these orbitals, which subduces to little group representations at the high symmetry $\Gamma,M$ and $K$ points, shown here as nodes in a graph. Standard $\mathbf{k}\cdot\mathbf{p}$ theory allows us to deduce the symmetry and degeneracy of energy bands in a small neighborhood near these points - the different colored edges emanating from these nodes. The graph theory mapping allows us to solve the compatibility relations along these lines in two topologically distinct ways. On the left, we obtain a graph with one connected component, indicating that in this phase graphene is a symmetry-protected semimetal; the Wannier functions for the four connected bands coincide with the atomic orbial Wannier functions. In contrast, the graph on the right has two disconnected components, corresponding to the topological phase of graphene by Def.~1. {The spin up and spin down localized Wannier functions for the valence band are localized on distinct sites of hexagonal lattice, and so break TR-symmetry in real space\cite{Soluyanov2011}.}}\label{fig:graphenechart}
\end{figure*}

We developed an algorithm (described in the SM) that outputs \emph{all} distinct connectivity graphs for \emph{all} SGs-- a gargantuan task. 
{The factorial complexity is handled by} several subroutines, which ensure that the minimal set of paths in momentum space is considered. Additional filters remove redundant or isomorphic solutions to the compatibility relations.  The tools of graph theory then allow us to partition the nodes of the graph (the little-group irreps) into distinct connected components (subgraphs). Each component corresponds to a connected, isolated group of bands that can describe a set of valence bands in \emph{some} insulating system or protected semimetal, depending on the filling. In particular, such a list consists of \emph{all} (both topologically trivial and nontrivial) valence band groups. The familiar example of graphene with SOC is given in Fig.~\ref{fig:graphenechart} and the SM. We now define and classify topologically nontrival bands in terms of \emph{localized} Wannier functions.

\section{Topologically (non)trivial bands}\label{sec:ebrs}

Consider a group of connected bands in the spectrum of a crystal Hamiltonian separated by a gap from all others. Using existing machinery, {to determine whether} this group is topologically nontrivial requires discovering topological invariants (indices or Wilson loops) from the analytic structure of the Bloch eigenfunctions. We now prove that the \emph{algebraic} global structure of the energy spectrum itself (including connectivities) contains a complete classification of topological materials.  We define:

\begin{itemize}
\item\textbf{Definition 1.} An insulator (filled group of bands) is {\bf topologically nontrivial} if it cannot be continued to {\bf any} atomic limit without either closing a gap or breaking a symmetry.
\end{itemize}

To every isolated group of energy bands, we associate a set of Wannier functions -- orbitals obtained by Fourier transforming linear combinations of the Bloch wavefunctions. 
{In an atomic limit, the Wannier functions are exponentially localized, respect the symmetries of the crystal (and possibly TR) and coincide in most cases (however, see Section~\ref{sec:chemistry}) with the exponentially localized atomic orbitals at infinite (atomic limit) separation.}
Under the action of the crystal symmetries, different atomic sites are distributed into orbits, belonging to Wyckoff Positions (WPs); we denote the points in a Wyckoff orbit in a unit cell as $\{\mathbf{q}_i\}$. In analogy with the 
symmetry group of a $\mathbf{k}$-vector, to each site $\mathbf{q}_i$ there is a finite subgroup, $G_{\mathbf{q}_i}$, of the full SG, $G$, which leaves $\mathbf{q}_i$ invariant, called the \emph{site-symmetry group.} For example, the A, B sites in graphene {belong to WP} $2b$ (the multiplicity $2$ refers to the number of symmetry-related sites in the unit cell); its site-symmetry group is isomorphic to $C_{3v}$. Wannier functions at each site $\mathbf{q}_i$ transform under some rep, $\rho_i$, of $G_{\mathbf{q}_i}$. Crucially, through the mathematical procedure of induction, the real-space transformation properties of these localized Wannier functions determine the little group reps of the bands at every point in the BZ: the action of the SG on the full lattice (rather than just the unit cell)  of Wannier functions gives an infinite-dimensional rep. Its Fourier transform gives the $\mathbf{k}$ dependent matrix rep of all symmetry elements. This restricts to reps of the little group of each $\mathbf{k}$-vector. Following Zak \cite{Zak1982}, we refer to this as a band representation (BR), $\rho_{iG}$, induced\cite{Fulton2004} in the space-group $G$ by the rep $\rho_i$ of $G_{\mathbf{q}_i}$:
\begin{equation}
\rho_{iG}=\rho_i\uparrow G.
\end{equation}
The above is true with or without TR symmetry. {BRs} which also respect time-reversal symmetry in real space are \emph{physical} band representations {(PBRs)}.

By inducing {BRs}, we enumerate all groups of bands with exponentially localized and symmetric Wannier functions. Each such group forms a BR, and every band representation is  a sum of EBRs. {We have identified the 10300 EBRs: }

\begin{itemize}
\item\textbf{Proposition 1.} A band representation $\rho_G$ is elementary if and only if it can be induced from an irreducible representation $\rho$ of a {maximal (as a subgroup of the space group. See the SM)} site-symmetry group $G_\mathbf{q}$ which does not appear in a list of exceptions{\cite{Bacry1993,Michel2001} given in the SM}.
\end{itemize}

We prove a similar statement for physically elementary band representations (PEBRs) and work out their extensive list of exceptions.\cite{GroupTheoryPaper}  In analogy with irreps, an (P)EBR cannot be written as a direct sum of other (P)BRs. A BR which is not elementary is {a ``composite'' BR (CBR)}. Crucially, \emph{any} topologically trivial group of bands is equivalent to some {(P)BR}, a conclusion of Definition~1. Conversely, any topologically nontrivial group of bands cannot be equivalent to any of the enumerated {(P)BRs} (a caveat is discussed in Section~\ref{sec:chemistry}).

We thus conclude that the Wannier functions of a topologically trivial group of bands are smoothly continuable into {an} atomic limit, exponentially localized, and transform under a BR. In a topological material, the Wannier functions for the valence (group of) bands either \textbf{1.} fail to be exponentially localizable, or \textbf{2.} break the crystal symmetry. An example of \textbf{1} is the Chern insulator: a nonvanishing Chern number is an obstruction to exponentially localized Wannier functions.{\cite{Brouder2007}} The Kane-Mele model of graphene\cite{Kane04} (Section~\ref{sec:classify}), is an example of \textbf{2}: in the ${Z}_2$ nontrivial phase, exponentially localized Wannier functions for the valence bands necessarily break TR symmetry\cite{Soluyanov2011} (when the valence and conduction bands are taken together, atomic-like Wannier functions can be formed). Hence, in order to transition to the atomic limit a gap must close.

{We have tabulated all the (P)EBR's induced from every maximal {WP (i.~e.~a WP with maximal site-symmetry group)} for all $230$ SGs, with and without SOC and/or TR symmetry. This data allows us to enumerate all topologically trivial band structures. 
In an accompanying publication\cite{GroupTheoryPaper} we describe the myriad group-theoretic data (subduction tables, etc) \emph{for each of the $5646$ EBRs and $4757$ PEBRs that we find}. {To generate this data, we generalized the well-known induction algorithm based on Frobenius reciprocity and presented in Refs.~\onlinecite{Evarestov1997} and \onlinecite{Bilbao3} to the case of double-valued representations. We present the full details of the computational methods in Ref.~\onlinecite{GroupTheoryPaper}.} Additionally, the data can be accessed through programs hosted on the Bilbao Crystallographic Server\cite{progbandrep}. We also give a summary table of all EBRs and PEBRs in Sec.~VII of the Supplementary Material}

{This allows us to give, for the first time, a classification of TIs that is both \emph{descriptive} and  \emph{predictive}. {Rather than providing a classification (with a topological index, for instance) of the topological phases in each space group, divorced from predictive power, we instead formulate a procedure to determine whether a band structure is topologically trivial or topologically nontrivial, and enumerate the possible non-trivial band structures for each space group.}
 Given the band structure of a material, we can compare isolated energy bands to our tabulated list of (P)EBRs. Any group of bands that transforms as a (P)EBR is topologically trivial. Those groups of bands that remain are guaranteed to be topologically nontrivial.
 Conversely, knowing just the valence orbitals and crystal structure of a material, we can immediately determine  under which -- if any -- EBRs the bands near the Fermi level transform. If graph theory reveals that these EBRs can be disconnected, we deduce that this phase is topological.} {While a disconnected EBR itself serves as a topological index, standard techniques can be used to diagnose which (if any) of the more standard K-theoretic topological indices\cite{Freed2013} are nontrivial.} In the subsequent sections, we outline this recipe, and present hundreds of new predicted topological materials. 

{Our identification of topological crystalline phases also goes beyond recently proposed classifications based on symmetry eigenvalues of occupied bands in momentum space, first proposed in Ref.~\onlinecite{kanegraphs}, and applied in a modified form to three dimensional systems in Ref.~\onlinecite{ashvincomplete}. A shortcoming common to both methods is that while they produce a list of topological indices for each space group, they provide no insight into how to find or engineer materials in any nontrivial class. Second, in constrast to the claims of Ref.~\onlinecite{ashvincomplete},
by focusing only on eigenvalues in momentum space rather than the real-space structure of Wannier functions (or equivalently, the analytic structure of Bloch functions in momentum space), essential information about the topological properties of certain band structures is lost. For instance, the topological phases of SG $P6mm$ ($183$) presented in the Supplementary Material fall outside the scope of Ref.~\onlinecite{ashvincomplete}. Viewed in this light, our classification based on band representations generalizes the notion of a topological eigenvalue invariant in such a way as to capture these missing cases.
}

\section{Classification of {TIs}}\label{sec:classify}

Combining the notion of a BR with the connectivity graphs, we identify several broad classes of TI's, {distinguished by the number of relevant EBRs at the Fermi level when transitioning from a topologically trivial to a nontrivial phase.}
{If, in a trivial phase, the Fermi level sits in a single (P)EBR,} the material is necessarily a semimetal. If tuning an external control parameter (strain, SOC, etc), opens a gap at the Fermi level, the material \emph{necessarily} becomes topological. {This is because} if an {(P)}EBR splits into a disconnected valence and conduction band, then neither the valence nor the conduction band can form BRs: only both together form a (P)EBR. 
{This situation occurs exactly when an EBR}
can be consistently realized in a disconnected way in the BZ. 
This is precisely the sort of task suited for the graph-theoretic machinery of Section~\ref{sec:graphs}! 
The archetypal example for this behavior is the quantum spin Hall transition in the Kane-Mele model of graphene with both next nearest neighbor ``Haldane''\cite{haldanemodel} and Rashba SOC, {which we illustrate schematically in Fig.~\ref{fig:graphenechart}}. 
This model hosts two {spinful} $p$-orbitals per hexagonal unit cell, for a total of four bands forming an EBR. In the Rashba SOC regime, all four bands are connected and the material is a gapless semimetal. 
Turning up Haldane SOC opens a band gap.
By the preceding analysis, this gap \emph{must} be topological. {In Ref.~\onlinecite{GroupTheoryPaper}, we give all (hundreds of) cases {-- each specified by an orbital, Wyckoff position, and SG -- } where this can occur. We give the full details in the SM} 

{The second class of TI's is defined by the presence of more than one relevant EBR at the Fermi level.}
The trivial phase of such a material {can be} an insulator, with EBRs above and below the Fermi level. 
Without loss of generality, we consider one EBR in the conduction band and one in the valence band; generically, any transition involving more than two EBRs can be resolved into a sequence of pairwise transitions. 
A topological phase transition occurs when a gap closes and reopens after a band inversion, such that neither the valence bands nor the conduction bands form a BR.  
In the trivial phase, the little group irreps of the filled bands at each $\mathbf{k}$ point are those of the valence band EBR. After the topological phase transition, the little group irreps at each $\mathbf{k}$ point are \emph{not} consistent with an EBR. This mechanism {describes} the zeitgeist $3D$ {TI} Bi$_2$Se$_3$.{\cite{Zhang09,Xia09}} Without, or with very small, SOC, Bi$_2$Se$_3$ is a trivial insulator; its valence and conduction bands transform in two distinct EBRs. {Increasing} SOC pushes the bands together; at a critical strength, the gap between the valence and conduction bands closes at the $\Gamma$ {point of the BZ}. Above this critical value, a gap reopens, with certain states (labelled by irreps of $G_\Gamma$) exchanged between the valence and conduction bands. {Ultimately}, neither the valence bands nor the conduction bands transform as EBRs and the insulator is topological as per Def.~1\cite{Winkler2016}. If, on the other hand, a full gap does not reopen after band inversion, then we are left with symmetry-protected semimetal a la Cd$_3$As$_2$\cite{LiuJiangZhouEtAl2014} or Na$_3$Bi\cite{LiuZhouZhangEtAl2014}.

{When the phase transition is driven by SOC, we label the classes by $(n,m)$, where $n$ is the number of EBRs at the Fermi level in the trivial phase (without SOC) and $m$ is the number in the topological phases (after SOC is turned on). These phases are indicated in Table~\ref{table:titable}, along with material examples.}

To summarize the theoretical results: in Section~\ref{sec:graphs} we showed that the constraints placed by group theory in \emph{momentum space} on the allowed connectivity of bands can be solved via a mapping to graphs. We constructed all possible allowed isolated band groupings for all $230$ SGs. We then showed that our group-theoretic analysis in \emph{real space} determines -- through the notion of BRs -- which of these isolated band groups are described by localized symmetric Wannier functions (topologically trivial insulators.) {It follows that any} other group of valence bands in an insulator {necessarily constistute a TI}. Importantly, our theory also shows that there exist different classes of topologically trivial insulators, which cannot be adiabatically continued to one another. {We now link this important fact to orbital hybridization. }

\section{Chemical Bonding, Hybridization, and  Non-Equivalent Atomic Limits}\label{sec:chemistry}

{Given a} topologically trivial crystal, it is tempting -- but wrong -- to assume that the electronic Wannier functions, like the constituent atomic orbitals, are localized at the atomic positions.  Basic chemistry {informs us} that orbitals from different atomic sites can hybridize to form bonding and antibonding ``molecular'' orbitals, centered away from any individual atom\cite{hoffmann1987chemistry}. 
{In a crystal formed of these tightly bound molecular (rather than atomic) units, the valence and conduction BRs are induced from the (generically maximal) WPs of the molecular orbitals, rather than from the atomic orbitals; consequently, the valence and conduction band Wannier functions are localized at the molecular orbital WPs, away from the atoms.}

Thus, orbital hybridization, when viewed in the solid-state, represents the {required} transition between two symmetry distinct atomic limit phases. In both phases localized, symmetric Wannier functions exist; the distinction lies in where the orbitals sit in the atomic limit\cite{Zakphase,ksv}. In the first atomic limit, the orbitals lie on the atomic sites. In the second, however, the orbitals do not coincide with the atoms. This phase has been called topological\cite{sshistop,bernevigbook}, but we {refer to} it as an ``obstructed'' atomic limit, {since} it is also described by localized Wannier states\cite{Kivelson1982, Read2016}. The prototypical example is the $1D$ Su-Schrieffer-Heeger (Rice-Mele) chain, whose two phases are distinguished by their hybridization pattern, an electric dipole moment that is $0$ ($\half$) in the trivial (nontrivial) phase. Pumping between two different atomic limits always leads to a nontrivial cycle with observable transport quantities.\cite{RiceMele,Atala13,Nakajima16}
A similar phenomenon is observed in the newly discovered quadrupole insulators.\cite{hughesbernevig2016} {Signatures of the obstructed atomic limit also appear in the real-space entanglement structure of the insulating ground state in finite systems\cite{Fang2013}, and these signatures persist to systems containing only a single molecule.\cite{Tubman2014}}

{We now describe orbital hybridization with EBRs.}
{The valence and conduction bands} are each described by an EBR. {Taken together, the two EBR's comprise a} {CBR}. In both the first and the second atomic limit, as well as at the critical point between them, the {CBR} does not change, although the individual EBR's of the valence and conduction bands do.  We denote the {CBR} in the first atomic limit as $\sigma_v\uparrow G\oplus \sigma_c\uparrow G$, where $\sigma_v$ and $\sigma_c$ are irreps of the site symmetry group, $G_a$, of the atomic sites. In the second atomic limit, the {CBR} is $\rho_v\uparrow G \oplus \rho_c \uparrow G$, where $\rho_v$ and $\rho_c$ are irreps of the site symmetry group, $G_m$, of the 
molecular sites.
{As we show in the SM, a symmetry-preserving transition} can only happen when the two site symmetry groups, $G_{a}$ and $G_m$, have a common subgroup $G_0$ with a representation $\eta$ such that
\begin{equation}
\eta\uparrow G_a\approx\sigma_v\oplus\sigma_c,\; \eta\uparrow G_m\approx\rho_v\oplus\rho_m.
\end{equation}
{This equation indicates} that there is a line joining the atomic sites, and that Wannier functions localized along the line give the same set of bands as those localized at either endpoint. Because of this, the Wannier functions {for the valence band} can move from the atomic to the molecular sites while preserving all symmetries {upon passing through a critical point}.
{In the SM, we illustrate this using the example of $sp$ orbital hybridization.} {Note, furthermore, that we can define an analogous notion of an obstructed atomic limit for systems lacking translational symmetry, using properties of the point group only. In this way, the preceding discussion generalizes straightforwardly to finite-sized crystals, molecules, and even quasicrystals. In the latter case, obstructions to the naive atomic limit are precisely covalent bonds.} We conclude that hybridization, and thus chemical bonding, can be treated as a phase transition.

\section{Algorithmic materials search}\label{sec:matsearch}

{We demonstrate the power of our theory by proposing two algorithms that use databases -- such as the Inorganic Crystal Structure Database (ICSD)\cite{ICSD} -- {which tabulate the occupied} {WPs} for each element in a chemical compound, along with simple energy estimates, in order to identify many new classes of {TIs} and semimetals. 
We distinguish between the number of EBRs at the Fermi level, i.e., the two classes defined in Section~\ref{sec:classify} and summarized in Table~\ref{table:titable}.
We further distinguish between cases with and without SOC, which can be treated separately by existing ab-initio methods. The SOC strength can be viewed as a control parameter driving the topological phase transition.{For EBRs without SOC, we count bands on a per-spin basis: all states are doubly degenerate.} }
Further new topological semimetals, such as a series of ones at filling -1/8, can be obtained by our method. 

\subsection{Single PEBRs at the Fermi Level}
{As described in Section~\ref{sec:classify} and Table~\ref{table:titable}, a $(1,1)$ type TI occurs when a single PEBR is realized as a sum of two band groups disconnected in momentum space.}  {Here, we utilize the fact that 
$(1,1)$ TIs can be realized by} band representations induced from $1D$ site symmetry group irreps that are EBRs but not PEBRs. In this case, the Wannier functions for the valence band involve a single orbital per site, and hence do not respect the twofold time-reversal symmetry degeneracy  \emph{in real space}: since their Wannier states break time-reversal symmetry the material is necessarily a TI. 

\begin{figure*}[ht]
\includegraphics[width=0.5\textwidth]{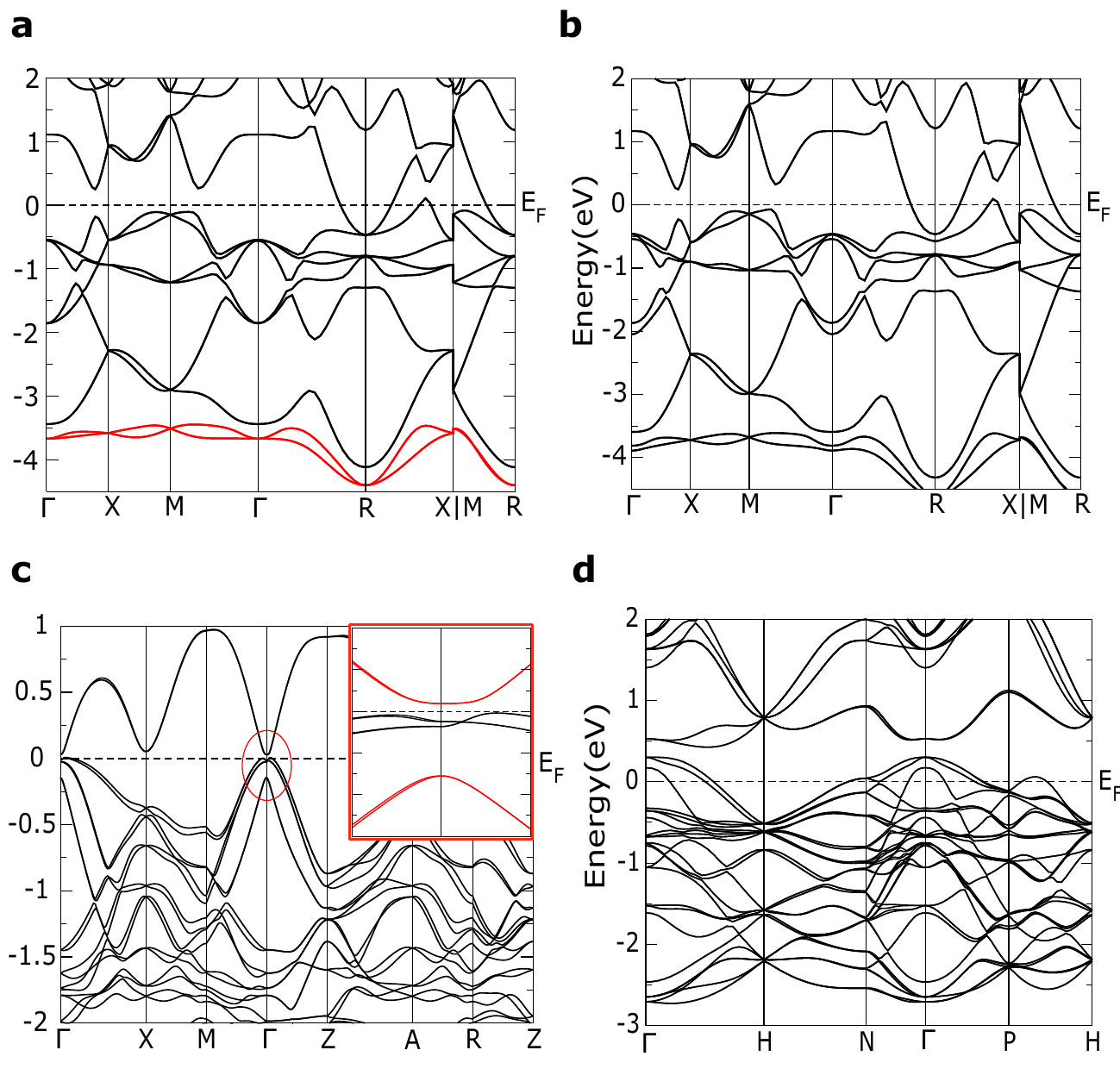}
\caption{{\textbf{Representative band structures for new material predictions}. \textbf{a} Band structure of Pb$_2$O in SG $Pn\bar{3}m$ ($224$). The red group of bands $~3eV$ below the Fermi level originate from a topologically disconnected PEBR. \textbf{b} Band structure for Pb$_2$O under uniaxial strain along the $z$-axis. This distortion opens a topological gap near the Fermi level. \textbf{c}} Band structure for the topologically nontrivial compound Cu$_3$SbS$_4$. {The conduction band, induced from a $1D$ site-symmetry irrep, is not a PEBR.} The inset shows a zoomed in view of the gap at $\Gamma$. \textbf{d} Band structure for the twenty-fourfold connected (semi-)metal Cu$_3$TeO$_6$ in SG $Ia\bar{3}$ (206). The bands at the Fermi level form a 24-band PEBR which is half-filled.}\label{fig:2}
\end{figure*}

{An example is furnished by lead suboxide Pb$_2$O\cite{pb20ref}, a material whose topological properties were until now unexplored. As discussed in the Supplementary Material, this is a non-symmorphic cubic crystal in the space group $Pn\bar{3}m$ ($224)$. Although metallic, this material features a topologically disconnected PEBR far below the Fermi level, shown in Fig.~\ref{fig:2}\textbf{a}. However, we can consider the application of $z-$axis uniaxial strain, under which the crystal symmetry is lowered to the tetragonal subgroup $P4_2/nnm$ (134) of the original space group. There are fewer symmetry constraints imposed on the band structure in this space group, and in particular degeneracies protected by threefold rotational symmetry will be broken. This allows for a gap to open at the Fermi level, leading us to predict that strained Pb$_2$O will be a topological insulator with a small Fermi pocket, as shown in Fig.~\ref{fig:2}\textbf{b}. As an aside, we note that this analysis shows that by using group-subgroup relations, we can make predictions about the topological character of strained systems when the unstrained band structure is known.}

Cu$_3$SbS$_4$, {a candidate TI\cite{cu3sbs4ref,hasancuti}, is also an example of a $(1,1)$ type material}. In Fig.~\ref{fig:2}\textbf{c} we show the calculated band structure. The states near the Fermi level form a single EBR coming from the Cu $d$-orbital electrons. {With SOC, the EBR is gapped,} leading to topologically nontrivial valence and conduction bands. A trivial spurious EBR also lies energetically within the topological gap (shown in black in the inset to Fig.~\ref{fig:2}\textbf{c}), causing the effective transport gap to be smaller than the topological gap (similar to HgTe).  {There are $35$ additional materials in this Cu$_2$ABX$_4$ class of materials.}

We use these considerations to {design a systematic method to} search for $(1,1)$ type TI's. First, we identify all disconnected PEBRs {from} our data paper Ref.~\onlinecite{GraphDataPaper}, which {is organized by WP and site-symmetry irrep;} in the SM we indicate which orbital types give rise to these representations. Next, we cross-reference with the ICSD\cite{ICSD}, which yields a list of candidate materials. This list can be further reduced {by restricting to semi-metals (since an insulator would not yield a topological gap near the Fermi level after turning on SOC).}
Finally, an electron counting analysis, using only the atomic-limit orbital energies, will determine whether or not the topologically relevant {BRs} will lie near the Fermi level. 
{Carrying out this procedure led us to identify the Cu$_2$ABX$_4$ material class introduced in the previous paragraph.} 
We also provide a list\cite{suppmatt} of materials guaranteed to be semi-metals by the minimal-insulating-filling criteria\cite{Watanabe16} (a sufficient, but far from necessary condition for a semi-metal).

\subsection{Multiple PEBRs}

\begin{figure}[ht]
\centering
\includegraphics[width=0.5\textwidth]{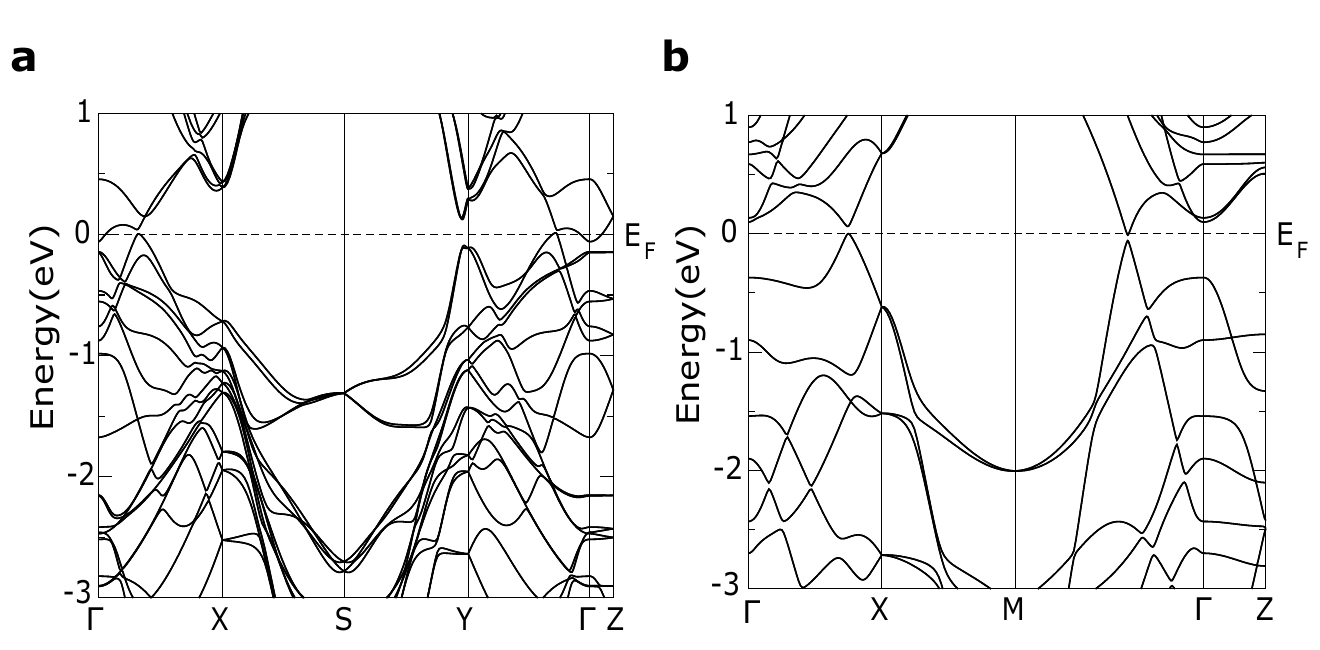}
\caption{\textbf{Representative band structures for topologically nontrivial insulators in the Bi- square net structure type.} \textbf{a} shows the band structure of the $3D$ weak {TI} SrZnSb$_2$ in SG \textit{Pnma} (62), where there is a small in-plane distortion in the Sb square net. \textbf{b} shows the band structure of the $3D$ weak {TI} ZrSnTe in SG \textit{P4/nmm} (129).}\label{fig:3}
\end{figure}

{As described in Section~\ref{sec:classify}, a CBR can be disconnected}
in such a way that neither the valence nor the conduction bands form PEBRs. When these are double-valued (spinful) BRs, we classify them by whether the BR is composite or elementary without SOC. In the 
first case, a single, connected, {SOC-free} EBR decomposes after turning on SOC into a sum of two PEBRs that are disconnected in a topologically nontrivial way. These constitute $(1,2)$ type TIs. This class includes materials composed of layered Bi$^{-1}$ square nets and related structures, which we discuss further in the SM. The relevant states at the Fermi level are the Bi $p_x$ and $p_y$ orbitals, which induce a single EBR when {SOC} is neglected. These materials are filling-enforced semimetals with a symmetry-protected nodal (degenerate) line at the Fermi level. 
{SOC fully gaps the nodal line, causing the EBR to disconnect. Consequently, this is a topological gap}.
Fig.~\ref{fig:3} depicts band structures for the representative {TIs} SrZnSb$_2$ and ZrSnTe\cite{129tis,zrsnte}.

A similar materials search program to that described in the previous subsection can be implemented for the $(1,2)$ type of Table~\ref{table:titable}. 
{SOC-}free EBRs which decompose into {CBRs} with SOC can be quickly identified {from Ref.~\onlinecite{GroupTheoryPaper}.} With these in hand, a list of materials with the proper orbitals and occupied {WPs} can be compiled from the ICSD\cite{ICSD}. Finally, a{n SOC-}free electron counting analysis will reveal which of these materials are semi-metallic without SOC. These are prime candidates to become {TIs} when SOC is included. 
{This systematic search has led us} to identify {over 300} $(1,2)$ materials in {SG $P4/nmm$ ($129$)\cite{129tis,zrsnte,schoop2015}, as well as 58 new candidate TIs in SG \textit{Pnma} ($62$)}. {SG \textit{Pnma} $(62)$ arises from an in-plane distortion of SG $P4/nmm$ ($129$), and again demonstrates that our group-theoretic approach allows us to predict the topological character of materials upon structural distortion.}
{We present these materials in the SM.}

Finally, {we discuss the known TIs} Bi$_2$Se$_3$ and KHgSb. In these materials, without SOC the band representation at the Fermi level is gapped and composite. 
With infinitesimal SOC, the band representation remains gapped and composite.
{A topological phase transition occurs only when SOC is strong enough to drive a band inversion that exchanges states with distinct little group reps at $\Gamma$ between the conduction and valence bands.}
{Because such a transition depends on the strength of SOC -- unlike in the $(1,1)$ and $(1,2)$ cases, where the transition is at infinitesimal SOC -- Bi$_2$Se$_3$ and KHgSb are described by our method but could not be unequivocally predicted.}
 \subsection{Partially filled bands and semimetals}
 Although not the main focus of this work, we also note that our method allows for the prediction of metals and semimetals. Going beyond standard methods of counting multiplicities of occupied WPs\cite{hoffmann1987chemistry}, we can predict symmetry-protected semimetals by looking for partially-filled, connected EBRs induced from high-dimensional site-symmetry representations. In this way we have found the A$_{15}$B$_4$ family\cite{batteryref} of sixteen-fold connected metals in SG $I\bar{4}3d$ (220) with A$=$Cu,Li,Na and B$=$Si,Ge,Sn,Pb. Through charge-transfer, the sixteen bands in this PEBR are $7/8$ filled; this exotic filling holds promise for realizing novel physics when interactions are included. The band structure for Li$_{15}$Ge$_4$ is shown in Fig.~\ref{fig:2}\textbf{d}. There is a symmetry-protected threefold degeneracy near the Fermi level at the $P$ point\cite{Bradlyn2016} Finally, we calculate that Cu$_3$TeO$_6$ in SG $Ia\bar{3}$ (206)\cite{24foldref} has a half filled, connected \emph{twenty-four} band PEBR at the Fermi level when interactions are neglected. Our EBR theory reveals that this is the highest symmetry-enforced band connectivity in any SG. 

{\subsection{Systematic materials search: Summary}

 While the materials presented above represent the proof-of-principle for our material search strategy, a full, systematic search of the entire materials database based on our criteria reveals a myriad of new topological insulator and semimetal candidates. While we defer a full discussion of our new materials predictions to a forthcoming work, we shall here list some of our more promising candidate materials, as identified through our systematic search, and verified with ab-intio DFT calculations. In SG P$\bar{3}m1$ ($164$) we predict that CeAl$_2$Ge$_2$, CeISi, BiTe, and Nb$_3$Cl$_8$ will be topological insulators. In SG $P4/mmm$ ($123$) we have LiBiS$_2$ and AgSbTe$_2$. A particularly promising family of topological materials is given by TiAsTe, ZrSbTe, HfSbTe, Hf$_3$Ni$_4$Ge$_4$, Sr$_3$Li$_4$Sb$_4$, Ba$_2$Bi$_3$, and Ba$_3$Al$_2$Ge$_2$ in SG $I/mmm$ ($71$). Additionally, we find NaAu$_2$ in SG $Fd\bar{3}m$ ($227$), LaPd$_2$O$_4$ in SG $I4_1/a$ ($88$), BaGe$_2$Ru$_2$ in SG $Fddd$ ($70$), Ni$_2$Ta$_2$Te$_4$ in SG $Pbam$ ($53$), Ag$_2$Ca$_4$Si$_6$ in SG $Fmmm$ ($69$), Ta$_2$Se$_8$I in SG $I422$ ($97$), SnP in SG $I4mm$ ($107$), and Tl$_4$CuTe$_3$ in SG $I4/mcm$ ($140$) each of which we predict hosts novel topological bands.}

\section{Conclusion}\label{sec:conclusion}

We have combined group theory, chemistry, and graph theory to provide the framework of Topological Quantum Chemistry.  
{We provide a complete description of the Wannier-Bloch duality between real and momentum space, in the process}
linking the extended (physics) versus local (chemistry) approaches to electronic states.  Our theory is descriptive and predictive: we can algorithmically search for and predict new {TIs} and semimetals. In a series of accompanying Data papers, we present the group-theoretic data and graph-theoretic algorithms necessary to deduce the conclusions of Sections~\ref{sec:graphs} and \ref{sec:ebrs}, and to implement the materials search described in Section~\ref{sec:matsearch}. By taking the ideas presented in this paper to their logical conclusion, we arrive at a new paradigm, {which applies not only to TIs, but to semimetals and to band theory in general.} 
 The synthesis of symmetry and topology, of localized orbitals and Bloch wavefunctions, allows for a full understanding of noninteracting solids, which we have only begun to explore in this work. Furthermore, our emphasis on the symmetry of localized orbitals opens a promising avenue to incorporate magnetic groups or interactions into the theory of topological materials. 

{\bf Data Availability:} All data supporting the conclusions of this work is hosted on the Bilbao Crystallographic Server (\url{http://cryst.ehu.es}). All information about EBRs, PEBRs, and their connectivity graphs can be accessed via the BANDREP application\cite{progbandrep}. The algorithms used to generate this data, as well as a guide to the use of all relevant programs, can be found in the accompanying data papers, Refs.~\onlinecite{GroupTheoryPaper,GraphDataPaper}.

{\bf Acknowledgements:} BB would like to thank Ivo Souza, Richard Martin, and Ida Momennejad for fruitful discussions. MGV would like to thank Gonzalo Lopez-Garmendia for help with computational work. BB, JC, ZW, and BAB acknowledge the hospitality of the Donostia International Physics Center, where parts of this work were carried out. JC also acknowledges the hospitality of the Kavli Institute for Theoretical Physics, and BAB also acknowledges the hospitality and support of the \'{E}cole Normale Sup\'{e}rieure and Laboratoire de Physique Th\'{e}orique et Hautes Energies. The work of MVG was supported by FIS2016-75862-P and FIS2013-48286-C2-1-P national projects of the Spanish MINECO. The work of LE and MIA was supported by the Government of the Basque Country (project IT779-13)  and the Spanish Ministry of Economy and Competitiveness and FEDER funds (project MAT2015-66441-P). ZW and BAB, as well as part of the development of the initial theory and further ab-initio work, were supported by the Department of Energy de-sc0016239, Simons Investigator Award, the Packard Foundation, and the Schmidt Fund for Innovative Research. The development of the practical part of the theory, tables, some of the code development, and ab-initio work was funded by NSF EAGER Grant No. DMR-1643312, ONR - N00014-14-1-0330, and NSF-MRSEC DMR-1420541. 

{\bf Author Contributions:} BB, LE, JC, MGV, and ZW contributed equally to this work. BB, JC, ZW and BAB provided the theoretical analysis, with input from CF. JC developed specific models to test the theory. LE and MIA performed the computerized group-theoretic computations. BB, LE and MGV devised and developed the graph algorithms, as well as the EBR connectivities; LE and MGV performed the computerized graph theory computations. ZW discovered the new materials presented in this paper with input from CF, and performed all first-principles calculations. 

{\bf Author Information:} Reprints and permissions information is available at www.nature.com/reprints. The authors declare no competing financial interests. Correspondence and requests for materials should be addressed to bernevig@princeton.edu.

\bibliography{connectivity-supp}

\pagebreak

\begin{table*}[!]
\begin{tabular}{c|c|c}
Number of PEBRs without SOC & Number of PEBRs with SOC & Examples \\
\hline
1 & 1 & Graphene, Bismuthene, Cu$_3$SbS$_4$ \\
1 & 2 & Bi$^{1-}$ square nets, Cu$_2$SnHgSe$_4$ \\
2 & 2 & Bi$_2$Se$_3$, KHgSb
\end{tabular}
\caption{{\textbf{Summary of topological phase transitions between a system without SOC to a TI with SOC, for BRs induced from up to two irreps of the site-symmetry group of occupied {WPs}}. When there is a single EBR in the first column, the system is necessarily a semimetal without SOC -- these materials become {TIs} for \emph{arbitrarily} small SOC.
The first column gives the number of PEBRs directly straddling the Fermi level without SOC {(per spin)}. The second column shows how this number of band representations changes when SOC is turned on. }}

\label{table:titable}
\end{table*}

\end{document}


\title{Supplementary Material for Topological Quantum Chemistry}
\maketitle
\onecolumngrid
\section{Band Representations and Wannier Functions}\label{sec:bandreps}

Here we expand upon the notion of {band representations} as discussed in the main text. First introduced by Zak\cite{Zak1982}, a band representation for a space group $G$ is a set of energy bands {$E_n(\mathbf{k})$} {spanned} by a given collection of (exponentially) localized Wannier orbitals. To be consistent with the crystal symmetries, the localization centers of these Wannier orbitals {(Wannier centers)} must form an orbit under the action of $G$: given a Wannier function centered at a point $\mathbf{q}_1$ {in the unit cell}, there is a Wannier function centered at $g\mathbf{q}_1$ for all $g\in G$, {including} all positions related to these by lattice translations. The set $\{\mathbf{q}_\alpha\}$ of these Wannier centers form an orbit {under the space group action and hence can be labelled by a} {\emph{Wyckoff position}} of the \emph{space} group $G$. Note that every system representable with a tight-binding model has such a real-space [direct space] description. 

Before we show how to construct a band representation from a set of localization centers {of the Wannier orbitals}, we will carefully define our terminology. Note that we use the {conventional origin choice (origin choice $2$)} for all space groups as given by the Bilbao Crystallographic Server\cite{Bilbao1,Bilbao2,Bilbao3}. {Our terminology follows that of Refs.~\onlinecite{Bacry1988,Evarestov1997,Cracknell,Bilbao1,Bilbao2,Bilbao3}. For basic facts about the theory of finite groups, we refer the reader to Refs.~\onlinecite{Serre,Fulton2004}}.
\begin{defn}
A {\bf symmetry site} $\mathbf{q}$ is any point in the unit cell of a crystal. The set of symmetry operations $g\in G$ that leave $\mathbf{q}$ fixed  (absolutely, not up to lattice translations) is called the {\bf stabilizer group}, or {\bf site-symmetry group} $G_\mathbf{q}\subset G$. By definition, a site-symmetry group is isomorphic to a crystallographic point group. A site-symmetry group is called {\bf non-maximal}
if there exists a finite group $H$, such that $G_\mathbf{q}\subset H\subset G$; a site-symmetry group that is not non-maximal is {\bf maximal}. 
\end{defn}

Note that the translation part of $g\in G_\mathbf{q}$ may include lattice translations, {so long as it keeps {the point} $\mathbf{q}$ fixed. Nonetheless, $G_\mathbf{q}$ must be isomorphic to a \emph{point} group.} 

\begin{defn}
The orbit $\{\mathbf{q}_\alpha=g_\alpha \mathbf{q} | {g_\alpha \notin G_\mathbf{q}} \}, \alpha=1,\dots,n$ of a symmetry site $\mathbf{q}$ modulo lattice translations {are classified by} a {\bf Wyckoff position} of multiplicity $n$. Note that we define the multiplicity with respect to the primitive, rather than the conventional cell. The stabilizer groups $G_{\mathbf{q}_\alpha}$ are all isomorphic and conjugate to the stabilizer group $G_\mathbf{q}\equiv G_{\mathbf{q}_1}$. We say that a Wyckoff position is maximal if the stabilizer group $G_\mathbf{q}$ is maximal.
\label{def:Wyckoff}  
\end{defn}
As an example,  if $\mathbf{q}$ is {a general} point in the unit cell {with trivial stabilizer group, $G_\mathbf{q}=\{E|000\}$}, then $\mathbf{q}$ belongs to the ``general'' Wyckoff position with multiplicity equal to the order of the point group of the space group. This is not a maximal position in general, but it {is} \emph{a} position.

Let us now return to the problem of constructing a band representation. Without loss of generality, consider the case of Wannier functions localized on symmetry sites $\{\mathbf{q}_\alpha| \alpha = 1,... , n\}$  classified by a single Wyckoff position {of multiplicity $n$}. Then the $n_q$ functions localized on the site $\mathbf{q}\equiv\mathbf{q}_1$ transform under some representation $\rho$  of the site-symmetry group $G_\mathbf{q}$, with dimension $n_q$. For the time being we do not specify whether or not $\rho$ is irreducible; we will show later that we need only concern ourselves with the irreducible representations (irreps). Crystal symmetry dictates that there are $n_q$ orbitals localized on the other equivalent symmetry sites $\mathbf{q}_\alpha$ {in the orbit}, and that these transform under the {conjugate} representation defined by
\begin{equation}
{\rho_\alpha(h)=\rho(g_\alpha^{-1} h g_\alpha)}
\end{equation}
for $h\in G_{\mathbf{q}_\alpha}$. 
One can see that $g_\alpha^{-1} h g_\alpha \in G_q $ because $h\in G_{\mathbf{q}_\alpha} \Rightarrow  h q_\alpha = q_\alpha  \Rightarrow h g_\alpha q = g_\alpha q$.
We can thus index our Wannier functions as {$W_{i\alpha}(\mathbf{r}-\mathbf{t}_\mu)$}, where $i=1\dots n_q$ indexes the functions localized on symmetry site $\mathbf{q}_\alpha+\mathbf{t}_\mu$ and $\mathbf{t}_\mu$ is a lattice vector. 

Finally, the elements $g_\alpha,\;\alpha\neq 1$ act by permuting the different symmetry sites {$\mathbf{q}_\alpha$}. Taking all of these facts together allows us to define the {\bf induced representation}\cite{Fulton2004} $\rho_G\equiv\rho\uparrow G\equiv \mathrm{Ind}_{G_\mathbf{q}}^{G}\rho$ of the space group $G$ induced from the representation $\rho$ of $G_\mathbf{q}$.  This representation is {$n_q\times n\times N$ dimensional (assuming periodic boundary conditions), where $N\rightarrow\infty$ is the number of (primitive) unit cells in the system}, and the
representation matrices have a block structure, with {$n^2\times N^2$} blocks of $n_q\times n_q$ dimensional submatrices; a group element $g$ whose matrix representative has a nonvanishing $(\alpha\beta)$ block maps $\mathbf{q}_\beta$ to $\mathbf{q}_\alpha$.

For our purposes, it is most convenient to work with the Fourier transforms
\begin{equation}
a_{i\alpha}(\mathbf{k},\mathbf{r})=\sum_{\mu}e^{i\mathbf{k}\cdot \mathbf{t}_\mu}W_{i\alpha}(\mathbf{r}-\mathbf{t}_\mu),
\end{equation}
{where the sum is over the lattice vectors and $\alpha = 1,..., n$. In this way we can exchange our infinite $N^2\times n^2\times n_q^2$ matrices for finite-dimensional $n^2\times n_q^2$ matrix-valued functions of $\mathbf{k}$, {which takes $N^2$ values in the first Brillouin zone (BZ)}}. {Any translationally-invariant, quadratic Hamiltonian acting in the Hilbert space of these Wannier functions commutes with these matrices.}
The concrete formula for the induced representation matrices $\rho_G(g)$ can then be defined as follows:
\begin{defn}\label{def:br}
The {\bf band representation} $\rho_G$ induced from the {$n_q-$dimensional} representation $\rho$ of the site-symmetry group $G_\mathbf{q}$ of a particular point $\mathbf{q}$, {whose orbit belongs to the Wyckoff position $\{\mathbf{q}_\alpha \equiv g_\alpha \mathbf{q} | g_\alpha \notin G_\mathbf{q}\, {\mathrm{for}\, \alpha\neq 1} \}$ of multiplicity $n$,} is defined for all $h\in G$ by the action
\begin{equation}
(\rho_G(h)a)_{i\alpha}(\mathbf{k},\mathbf{r})=e^{-i(h\mathbf{k})\cdot\mathbf{t}_{\beta\alpha}}{\sum_{i'=1}^{n_\mathbf{q}}}\rho_{i'i}(g_{\beta}^{-1}\{E|-\mathbf{t}_{\beta\alpha}\}hg_\alpha)a_{i'\beta}(h\mathbf{k},\mathbf{r}),
\label{eq:inducedrep}
\end{equation}
{here $\alpha,\beta,i,j$ are matrix indices, where for each choice of $\alpha$ the index $\beta$ is determined by the {unique}} coset of $G$ that contains $hg_\alpha$:
\begin{equation}
hg_\alpha =\{E|\mathbf{t}_{\beta\alpha}\}g_{\beta}g
\label{eq:defbeta} 
\end{equation}
for some $g\in G_\mathbf{q}$
 and Bravais lattice vector $\mathbf{t}_{\beta\alpha}$.
By moving $g_\alpha$ to the right-hand-side of Eq~(\ref{eq:defbeta}), it is evident that 
$h\mathbf{q_\alpha} = \{E|\mathbf{t}_{\beta\alpha}\}g_{\beta}g g_\alpha^{-1} \mathbf{q}_\alpha = \{E|\mathbf{t}_{\beta\alpha}\}g_{\beta}g\mathbf{q} = \{E|\mathbf{t}_{\beta\alpha}\}g_{\beta}\mathbf{q} = \{E|\mathbf{t}_{\beta\alpha}\}\mathbf{q}_\beta$. The second and fourth equalities follow from the definition of $\mathbf{q}_{\alpha,\beta}$ and the third equality follows from $g\in G_\mathbf{q}$. Thus,
\begin{equation}
\mathbf{t}_{\beta\alpha}=h\mathbf{q}_\alpha-\mathbf{q}_{\beta}. \label{eq:Rab}
\end{equation}
If $\rho$ is an $n_q$-dimensional representation of $G_\mathbf{q}$, and if the Wyckoff multiplicity of the position $\{\mathbf{q}_\alpha\}$ is $n$, then there are $n\times n_q$ {energy bands} in the band representation $\rho_G$.
\end{defn}
(This is a special case of the general induction procedure; a similar formula can be used for inducing the representation of any group from one of its subgroups.) {Notice that we did not need to specify a particular Hamiltonian. Our discussion applies to any Hamiltonian that respects the crystal symmetry and acts on a local Hilbert space of Wannier functions.}
All of the above holds for either spinless or spin-orbit coupled systems. When spin-orbit coupling is negligible, we consider the single valued (or spinless) linear representations of the site symmetry group{; we double everything to account for the trivial spin degeneracy}. For spin-orbit coupled systems, we must use the double-valued (spinor) representations.

Note that a band representation is formally infinite dimensional since it depends on the momentum $\mathbf{k}$ ({it has as many dimensions as the number of unit cells in the crystal}), while the irreducible representations of the space groups are indexed by discrete sets of $\mathbf{k}$ vectors. As such, every band representation is formally reducible, {as it decomposes as} an infinite direct sum (over $\mathbf{k}$ points) of space group irreps at each $\mathbf{k}$ point. However, it is the band representations, rather than the space group irreps, that tell us about the global band structure topology in a crystal. 
Hence, we will be interested in the decomposition of band representations into sums of other band representations. 

{First, we must specify how to tell if two band representations are equivalent. Given two band representations $\rho_G$ and $\sigma_G$, a necessary condition for {their} equivalence up to now\cite{Bacry1988,RegRep} is that at all points in the BZ they restrict to the same little group representations. However, for the study of topological phases, we need a stronger form of equivalence. We define a form of \emph{homotopy equivalence} that makes explicit the smoothness properties needed for discussing topological phase transitions. Namely, we say
\begin{defn}\label{def:equiv}
Two band representations $\rho_G^\mathbf{k}$ and $\sigma_G^\mathbf{k}$ are {\bf equivalent} iff there exists a {unitary} matrix-valued function $S(\mathbf{k},t,g)$ smooth in $\mathbf{k}$ and continuous in $t$ such that for all $g\in G$
\begin{enumerate}
\item $S(\mathbf{k},t,g)$ is a band representation for all $t\in[0,1]$,
\item $S(\mathbf{k},0,g)=\rho_G^\mathbf{k}(g)$, and
\item $S(\mathbf{k},1,g)=\sigma_G^\mathbf{k}(g)$ 
\end{enumerate}
\end{defn}
{Note that since $S$ is continuous in $t$, any property of a band representation evolves continuously under the equivalence $S$. In particular, the Wilson loop {(i.~e.~Berry phase)} matrices\cite{Zakphase,Aris2014,Aris2016} computed from the bands in the representation $\rho_G^{\mathbf{k}}$ are homotopic to the Wilson loop matrices computed in the representation $\sigma_G^{\mathbf{k}}$. As such, two equivalent band representations cannot be distinguished by any quantized Wilson loop invariant.} {\blue This is a constructive formulation of the type of equivalence noted in Refs.~\onlinecite{Michel1992} and \onlinecite{RegRep}}

We can also give a more constructive view of equivalence. {Consider} two symmetry sites, $\mathbf{q}, \mathbf{q}'$, which have {distinct} site symmetry groups, $G_{\mathbf{q}}$ and $G_{\mathbf{q}'}$, respectively, with nonempty intersection, $G_0=G_{\mathbf{q}}\cap G_{\mathbf{q}'}$. 
Since $G_0$ is the intersection of two distinct stabilizer groups, it is a stabilizer group of some {(lower)} symmetry site $\mathbf{q}_0$.
This symmetry site will have a variable parameter that interpolates between the symmetry sites $\mathbf{q}$ and $\mathbf{q}'$; this allows us to easily identify $\mathbf{q}_0$ from a table of Wyckoff positions{, which we have done for all the Wyckoff positions of all the $230$ space groups\cite{}.} If $G_0$ is an index $m_q$ subgroup in $G_\mathbf{q}$, then the associated Wyckoff position {with stabilizer group $G_0$} has multiplicity $m_q$ times that of $\mathbf{q}$. Furthermore, $G_\mathbf{q}$ has a coset decomposition in terms of $m_q$ cosets of $G_0$; analogous statements hold when we view $G_0$ as an index $m_q'$ subgroup of $G_{\mathbf{q}'}$. We can use this coset decomposition to induce representations of $G_\mathbf{q}$ and $G_{\mathbf{q}'}$ from representations of $G_0$, in much the same way as outlined above. Given a representation $\rho$ of $G_0$, the band representations $(\rho\uparrow G_{\mathbf{q}})\uparrow G$ and $(\rho\uparrow G_{\mathbf{q}'})\uparrow G$ are equivalent. The existence of a homotopy $S$ implementing this equivalence is guaranteed by the fact that the symmetry site $\mathbf{q}_0$ can be continuously moved from $\mathbf{q}$ to $\mathbf{q}'$ without violating the crystal symmetries.

Using Definition \ref{def:equiv} of equivalence, we define
\begin{defn}\label{def:composite}
A band representation is called {\bf composite} if it {is equivalent to} the direct sum of other band representations. A band representation that is not composite is called {\bf elementary}.
\end{defn}

Using the fact that induction {of representations}, $\uparrow$, commutes with direct sums, and that induction factors through subgroup inclusion\cite{Bacry1993}, we deduce that elementary band representations are induced from irreducible representations of maximal site symmetry groups. These conditions are necessary, however they are not sufficient. It may still be the case that a band representation induced from a maximal site symmetry irrep is equivalent [in the sense of Def.~(\ref{def:equiv})] to a composite band representation. We catalogue all such exceptions in Table~\ref{table:sbr} for the space groups {(first found in Ref.~\onlinecite{Bacry1988})}, and in Table~\ref{table:dbr} for the double space groups. {The full decription of how this data was obtained will be presented in the accompanying Data paper\cite{GroupTheoryPaper}, and the data itself is accessible through the BANDREP program on the Bilbao Crystallographic Server\cite{progbandrep}. We have
\begin{prop}
A band representation $\rho_G$ is elementary if and only if it can be induced from an irreducible representation $\rho$ of a {maximal} site-symmetry group $G_\mathbf{q}$ which does not appear in the first column of Table~\ref{table:sbr} or \ref{table:dbr}.
\end{prop} 
}

Up to this point, we have not commented on time-reversal symmetry. We may include antiunitary time-reversal symmetry as an element in any site-symmetry group, {as} it acts locally in real [direct] space, i.~e.~ it commutes with all space group elements. For spinless systems, time-reversal squares to $+1$, while for spinful systems it squares to $-1$. We call site-symmetry representations which are compatible with the action of time-reversal \emph{physical} representations. Note that all physically irreducible site-symmetry representations are even-dimensional for spinful systems by Kramers's theorem. The entire discussion thusfar holds mutatis mutandis for physical band representations, physical equivalence, and physically elementary band representations, by generalizing Defs.~\ref{def:br},\ref{def:equiv}, and \ref{def:composite} to the TR-symmetric case. Taking time-reversal symmetry into account, we find that only the band representations below the double line in Table~\ref{table:sbr} fail to be physically elementary. The rest of the entries in Table~\ref{table:sbr}, as well as all the entries in Table~\ref{table:dbr} are composite without TR symmetry, but physically elementary. Moreover, {we find} there are additional physical exceptions for spinless systems with time-reversal symmetry, catalogued in Table~\ref{table:zakwaswrong}. {The machinery we used to obtain these tables will be explained in detail in Ref.~\onlinecite{GroupTheoryPaper}; they represent an exhaustive search of all induced representations for all space groups. Summarizing, we have with TR that
\begin{prop}
A spinless (i.e. single-valued) band representation $\rho_G$ is physically elementary if and only if it can be induced from a physically irreducible representation $\rho$ of a {maximal} site-symmetry group $G_\mathbf{q}$ which does not appear in {the first column of} Table~\ref{table:sbr} {below the double line, and if it does not appear in Table~\ref{table:zakwaswrong}}.

A spinful (i.e. double-valued) band representation $\rho_G$ is physically elementary if and only if it can be induced from a physically irreducible representation $\rho$ of a {maximal} site-symmetry group $G_\mathbf{q}$.
\label{prop:physelementary}
\end{prop}
}

\section{Connectivity Graphs}
In this Appendix, we review the necessary background for constructing the connectivity graphs associated with elementary band representations. After reviewing compatibility relations in more detail than presented in the main text, we outline our algorithm for computing the allowed connectivities of elementary band representations using the notion of spectral graph partitioning. This allows us to develop an approach to the classification and search for TIs significantly more general than those found in recent proposals\cite{kanegraphs,globaltop}. {A more complete account of the machinery and related data -- which takes more than 100 pages -- will be given in Ref.~\onlinecite{graphdatapaper}, but the following represents a good introduction to our method and its results.}

\subsection{Compatibility Relations}\label{ex:sg216comprel}

Recall that in the textbook theory of energy bands{\cite{Cracknell}}, global band topology {-- the various interconnections between different bands throughout the BZ --} is inferred from the representations of the little groups $G_\mathbf{k}$ through the use of so-called ``compatibility relations''. Specifically, irreducible little group representations at high-symmetry $\mathbf{k}$-points, lines, and planes in the BZ are reducible along high-symmetry lines, planes, and volumes (the general $\mathbf{k}$-point) respectively; the compatibility relations determine which representations can be consistently connected along these subspaces. Since the little groups along high-symmetry surfaces are subgroups of the little groups of their boundaries, the compatibility relations can be determined by starting with the little group representation on a high-symmetry surface, and restricting (subducing) to the representations of the higher dimensional surfaces that it bounds.

As an example, let us examine space group $P\bar{4}3m$ (215). This is a symmorphic space group with primitive cubic Bravais lattice, and point group $T_d$.  The group $T_d$ is generated by a threefold rotation, {$C_{3,111}$, a fourfold roto-inversion, $IC_{4z}\equiv S_4^-$ , and a mirror reflection, $m_{1\bar{1}0}$}. Consider the high-symmetry point $\Gamma=(0,0,0)$ in the BZ, and the line $\Lambda=(k,k,k)$ emanating from it. The point group of the little group $G_\Gamma$ of $\Gamma$, known as the {\bf little co-group} $\bar{G}_\Gamma$, is isomorphic to the point group of the space group, while the little co-group $\bar{G}_\Lambda$ of $\Lambda$ is generated by $C_{3,111}$ and $m_{1\bar{1}0}$, and thus isomorphic to the group $C_{3v}$. As such, irreps of $G_\Gamma$ restrict (or subduce) to representations of the little group $G_\Lambda$ of the line $\Lambda$. The compatibility relations enumerate all such restrictions. For example, let us consider first the little group $G_\Gamma$. 
We note that since the space group $P\bar{4}3m$ (215) is symmorphic, the representations of the little groups $G_\mathbf{k}$ are trivially determined by the representations of the little co-groups $\bar{G}_\mathbf{k}$. 
Here and throughout, we will simplify notation by giving representation matrices and character tables for the little co-groups where appropriate. 
The four dimensional double-valued $\bar{\Gamma}_8$ representation of $\bar{G}_\Gamma\approx T_d$ (this is the spin-$3/2$ representation), specified by
\begin{equation*}
\bar{\Gamma}_8(C_{3, 111})= 
\frac{\sqrt{2}}{4}e^{-i\pi/4}
\left(\begin{array}{cccc}
-i & -\sqrt{3} & i\sqrt{3} & 1\\
-i\sqrt{3} & -1 & -i & -\sqrt{3}\\
-i\sqrt{3} & 1 & -i & \sqrt{3}\\
-i & \sqrt{3} & i\sqrt{3} & -1
\end{array}\right),\;\; 
\bar{\Gamma}_8(IC_{4z}) = \left(
\begin{array}{cccc}
 -\sqrt[4]{-1} & 0 & 0 & 0 \\
 0 & -(-1)^{3/4} & 0 & 0 \\
 0 & 0 & \sqrt[4]{-1} & 0 \\
 0 & 0 & 0 & (-1)^{3/4} \\
\end{array}
\right)
\end{equation*}
\begin{equation}\label{eq:td4drep}
\bar{\Gamma}_8(m_{1\bar{1}0})= \left(
\begin{array}{cccc}
 0 & 0 & 0 & (-1)^{1/4} \\
 0 & 0 & -(-1)^{3/4}& 0 \\
 0 & -(-1)^{1/4} & 0 & 0 \\
 (-1)^{3/4} & 0 & 0 & 0 \\
\end{array}
\right)
\end{equation}
 The matrix for $C_{3,111}$ has eigenvalues $-1,-1,e^{i\pi/3},e^{-i\pi/3}$, while the matrix for $m_{1\bar{1}0}$ has eigenvalues $-i,-i,i,i$.

 Next, we note that there are three double-valued representations of $G_\Lambda$, conventionally labelled $\bar{\Lambda}_4$, $\bar{\Lambda}_5$, and $\bar{\Lambda}_6$. The matrix representative of $\{C_{3,111}|000\}$ in each of these representations is given by
 \begin{equation}
 \bar{\Lambda}_4(\{C_{3,111}|000\})=\bar{\Lambda}_5(\{C_{3,111}|000\})=-1,\; \bar{\Lambda}_6(\{C_{3,111}|000\})=\left(\begin{array}{cc} e^{-i\pi/3} &0 \\ 0 & e^{i\pi/3}\end{array}\right),
 \end{equation}
 and the matrices for $\{m_{1\bar{1}0}|000\}$ are
 \begin{equation}
\bar{\Lambda}_4(\{m_{1\bar{1}0}|000\})=-i,\;\bar{\Lambda}_5(\{m_{1\bar{1}0}|000\})=i,\; \bar{\Lambda}_6(\{m_{1\bar{1}0}|000\})=\left(\begin{array}{cc} 0 & -1 \\ 1 & 0\end{array}\right)
 \end{equation}

 By comparing eigenvalues, we deduce that the $\bar{\Gamma}_8$ representation of $G_\Gamma$ must restrict to the $\bar{\Lambda}_4\oplus\bar{\Lambda}_5\oplus\bar{\Lambda}_6$ representation of $G_\Lambda$. The compatibility relation for the $\bar{\Gamma}_8$ representation at $\Gamma\rightarrow\Lambda$ is thus,
\begin{equation}
\bar{\Gamma}_8\downarrow G_\Lambda\approx\bar{\Lambda}_4\oplus\bar{\Lambda}_5\oplus\bar{\Lambda}_6\label{eq:sg216comprel}
\end{equation}

\subsection{Graph Theory Review}\label{sec:graphreview}

{Compatibility relations like these must be satisfied at each and every high-symmetry point, line, and plane throughout the BZ. In particular (as discussed in the main text), in order to connect the little group representations of pairs of high-symmetry $\mathbf{k}$-points, and so form global energy bands, we must ensure that the compatibility relations are satisfied along the lines and planes joining the two points. In general, there will be many ways to form global energy bands consistent with the compatibility relations, each yielding a physically distinct realizable band structure. Our goal is to systematically classify all these valid band structures.}
Since the compatibility relations are a purely group-theoretic device with meaning independent of any choice of Hamiltonian, {we can accomplish this task by introducing} more refined graph-theoretic picture of band connectivity, as presented in the main text.

To begin, we introduce some graph-theoretic terminology.
\begin{defn}\label{def:partition}
A {\bf partition} of a graph is a subset, $V_0$, of nodes such that no two nodes in $V_0$ are connected by an edge. 
\end{defn}
In our construction, each partition will correspond to a high-symmetry $\mathbf{k}$-point, and irreps of the little group of each $\mathbf{k}$-point will be represented as nodes, as shown in Fig.~\ref{fig:graphschematic}
\begin{defn}\label{def:degree}
The {\bf degree of a node} $v_0$ in a graph is the number of edges that end on $v_0$.
\end{defn}
These definitions allow us to formalize the notion of a connectiviy graph as introduced in the main text, in particular, 

\begin{defn}\label{def:compgraph}
Given a collection of little group representations, $\mathcal{M}$, (i.e. bands) forming a (physical) band representation for a space group $G$, we construct the {\bf connectivity graph} $C_\mathcal{M}$ as follows: 
{we associate a node, $p^a_{\mathbf{k}_i}\in C_\mathcal{M}$, in the graph to each representation $\rho_{\mathbf{k}_i}^a\in\mathcal{M}$ of the little group $G_{\mathbf{k}_i}$ of every high-symmetry manifold (point, line, plane, and volume), $\mathbf{k}_i$.}
If an irrep occurs multiple times in $\mathcal{M}$, there is a separate node for each occurence.

{The degree of each node, $p^a_{\mathbf{k}_i}$, is $P_{\mathbf{k}_i}\cdot \mathrm{dim}(\rho^a_{\mathbf{k}_i})$, where $P_{\mathbf{k}_i}$ is the number of high-symmetry manifolds connected to the point $\mathbf{k}_i$}:  $\mathrm{dim}(\rho^a_{\mathbf{k}_i})$ edges lead to each of these other $\mathbf{k}-manifolds$ in the graph, one for each energy band. 
When the manifold corresponding to $\mathbf{k}_i$ is contained within the manifold corresponding to $\mathbf{k}_j$, as in a high-symmetry point that lies on a high-symmetry line, their little groups satisfy $G_{\mathbf{k}_j} \subset G_{\mathbf{k}_i}$. {For each node $p_{\mathbf{k}_i}^a$, we compute
\begin{equation}
\rho_{\mathbf{k}_i}^a\downarrow G_{\mathbf{k}_j}\approx\bigoplus_{b}\rho_{\mathbf{k}_j}^b.
\end{equation}
We then connect each node $p_{\mathbf{k}_j}^b$ to the node $p_{\mathbf{k}_i}^a$ with $\mathrm{dim}(\rho^b_{\mathbf{k}_j})$ edges.
}
\end{defn}
{We give an illustration of these concepts in Fig.~\ref{fig:graphschematic}.}

\begin{figure}[t]
\includegraphics[height=3in]{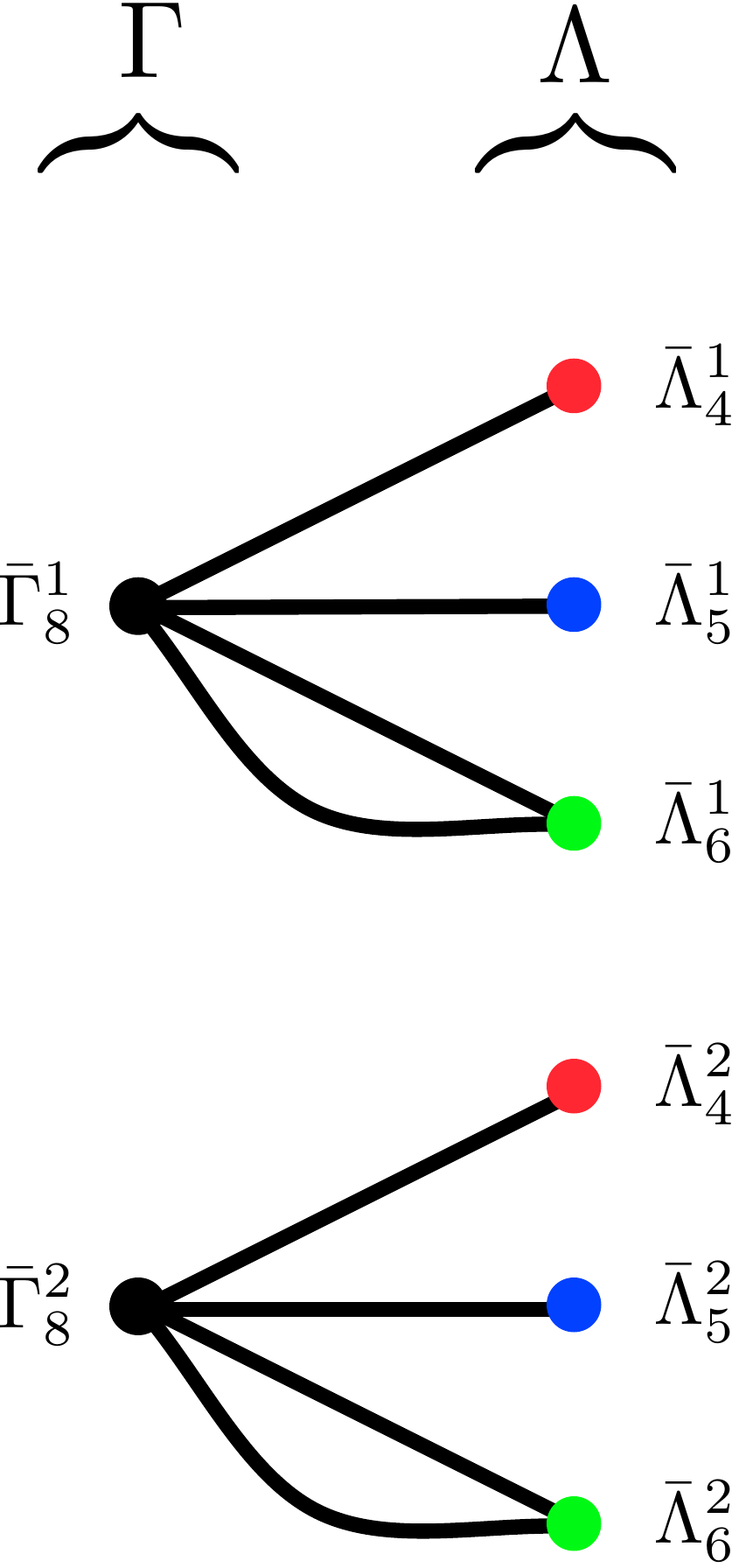}
\caption{{Subgraph of a connectivity graph corresponding to the compatibility relations along $\Gamma$ and $\Lambda$ for  $P\bar{4}3m$ (215) as discussed in Sec.~\ref{ex:sg216comprel}. There are two partitions in the graph labelled by $\Gamma$ and $\Lambda$. In the $\Gamma$ partition there are two nodes indicated by black circles, labelled $\bar{\Gamma}_8^1$ and $\bar{\Gamma}_8^2$, each corresponding to a copy of the $\bar{\Gamma}_8$ little group representation. Similarly, in the $\Lambda$ partition, there are two nodes corresponding to copies of the $\bar{\Lambda}_4$ little group representation and indicated by red circles; two nodes corresponding to the $\bar{\Lambda}_5$ representation and indicated by blue circles; and two nodes corresponding to the $\bar{\Lambda}_6$ representation and indicated by green circles. The nodes are connected by edges (represented by black lines) consistent with the compatibility relation Eq.~(\ref{eq:sg216comprel}). Because there are only two partitions in this subgraph, $P_\Gamma=P_\Lambda=1$ (c.~f.~Def.~\ref{def:compgraph}) for all nodes. The degree of each node in the $\Gamma$ partition is $4=P\cdot\mathrm{dim}(\bar{\Gamma}_8)$. Similarly, since $\mathrm{dim}(\bar{\Lambda}_6$)=2, the degree of the nodes $\bar{\Lambda}_6^1$ and $\bar{\Lambda}_6^2$ is $2$.   
The remaining nodes in the $\Lambda$ partition have degree $1$, since they correspond to $1D$ representations. Note, for example, that if the $\Lambda$ line was also connected to another high symmetry $\mathbf{k}$-point (labelled $L$, for instance), then $P_\Lambda=2$, and the degree of each node in the $\Lambda$ partition would double.}}\label{fig:graphschematic}
\end{figure}

The advantage of this graph-theoretic approach to topological phase transitions is that it is algorithmically tractable. Using the {460} tables of compatibility relations {which we have generated and will publish in the accompanying Ref.~\onlinecite{GroupTheoryPaper}}, we have algorithmically constructed all compatiblity graphs consistent with the {$5646$} allowed elementary band representations, {as well as for the $4757$ independent physically elementary band representations}. {While this may naively seem to be a hopeless task, we have developed several algorithms, {outlined in Section~\ref{sec:algorithms}, and in more detail in the accompanying Ref.~\onlinecite{graphdatapaper}}, which reduce the problem to analyzing a computationally tractable $\sim 10000$ graphs per band representation.} To decompose these into connected groups of valence and conduction bands, we use standard results of spectral graph theory\cite{GraphThy}. In particular, recall that 

\begin{defn}
The adjacency matrix, $A$, of a graph with $m$ nodes is an $m\times m$ matrix, where the $(ij)$'th entry is the number of edges connecting node $i$ to node $j$.
\end{defn}
In addition, 
\begin{defn}\label{def:degreemat}
The degree matrix, $D$, of a graph is a diagonal matrix whose $(ii)$'th entry is the degree of the node $i$.
\end{defn}
We can then form the Laplacian matrix
\begin{equation}
L\equiv D-A
\end{equation}
We make use of the following fact about the spectrum of $L$:

\begin{prop}
For each connected component of a graph, there is a zero eigenvector of the Laplacian. Furthermore, the components of this vector are $1$ on all nodes in the connected component, and $0$ on all others.
\end{prop}
The proof of this statement follows directly from the observation that the sum of entries in any row of the Laplacian matrix is by definition zero, coupled with the observation that if $L_{ij}\neq 0$, then nodes $i$ and $j$ lie in the same connected component\cite{GraphThy}. {We give an example of this method applied to graphene below in Section~\ref{subsec:graphenegraph}}.

\subsection{Connectivity Graphs}\label{sec:algorithms}

We apply this graph-theoretic machinery to the connectivity graphs (defined in Sec.~II of the main text) associated to elementary band representations. {We start with all the little group representations at high-symmetry points and lines contained in a given EBR}. Because the representation is elementary, we know that the connectivity graphs will have either one connected component, or will decompose into a set of topological band groups, {as explained in the main text}. All connectivity graphs with more than one connected component, if they exist, will then correspond to {topological phases. }

In order to construct the Laplacian matrix, we separate the task into two steps. We first construct all possible adjacency matrices, and then we subtract the degree matrix from each of them. Since the adjacency matrices have a block structure, with nonzero blocks determined by the compatibility relations, we first build each block submatrix separately.
We start by identifying the maximal $\mathbf{k}$-vectors in the BZ. In analogy to maximal Wyckoff positions, these are the $\mathbf{k}$ vectors whose little co-groups are maximal subgroups of the point group of the space group. A valid submatrix will then be created based on {our derived} compatibility  and site-symmetry tables\cite{GroupTheoryPaper}.The rows represent the maximal $\mathbf{k}$-vectors and the columns represent the connecting (non-maximal) lines and/or planes. The entries in the submatrix fulfill the following rules: we can only allow one nonzero entry per column, and the sum of the entries in each row equals the dimension of the corresponding little-group representation. Given a single valid submatrix, {all others can be obtained by permuting the columns.}

{With these submatrices, we} build up the full adjacency matrix row by row. In doing so, we must ensure that we account for all possible connections along non-maximal lines and planes. Additionally, we would like to avoid overcounting configurations that differ only by a relabelling of representations along non-maximal $\mathbf{k}$-vectors. {We have developed two main tools to do this. First, although Def.~\ref{def:compgraph} for the connectivity graphs makes use of all high-symmetry manifolds in the BZ, many of them provide redundant information. We thus consider for each space group only the minimal set of paths in $\mathbf{k}$-space necessary. We derived these for each space group by searching first for the paths in the BZ connecting all maximal $\mathbf{k}$-vectors along the highest symmetry surfaces possible, and then pruning connections which add no additional symmetry constraints. For non-symmorphic space groups, it is also necessary to consider paths connecting maximal $\mathbf{k}$-vectors in different unit cells of the reciprocal lattice, to account for the monodromy of representations\cite{Herring1942}. Second, we select from the set of valid submatrices in each block of the adjacency matrix only those that yield non-isomorphic connectivity graphs. A detailed discussion of the algorithm we used is given in Ref.~\onlinecite{graphdatapaper}. The topologically distinct connectivity graphs for each elementary band representation can be accessed through the BANDREP program on the Bilbao Crystallographic Server\cite{progbandrep}.}

\section{Example: Graphene}

In this Appendix, we illustrate the application of our representation- and graph-theoretic methods through the example of graphene with spin-orbit coupling -- the primordial topological insulator. We begin in Subsection~\ref{subsec:graphenegrps} by reviewing the crystal structure and symmetries of the honeycomb lattice. Next, in Subsection~\ref{subsec:graphenebr} we construct explicitly the elementary band representation realized by spin-orbit coupled $p_z$ orbitals in a hexagonal lattice, and hence deduce the full symmetry content of graphene \emph{irrespective of any microscopic model}. In Subsection~\ref{subsec:graphenegraph} we apply our connectivity graph machinery to this band representation, allowing us to catalogue the different allowed topological phases of graphene. Finally, in Subsection~\ref{subsec:grapheneham}, we show how these cases can be physically realized, and comment on the relationship of our approach to older work. {This gives a much needed practical example of how to use our formalism.}. Because we are interested in spin-orbit coupled systems, for the remainder of this section we will employ primarily double-valued point and space group representations unless otherwise specified.

\subsection{Space group symmetries}\label{subsec:graphenegrps}
The $2D$ honeycomb lattice of graphene has as its symmetry group the wallpaper group $p6mm$ {(No. $17$, the most symmetric triangular wallpaper group)}. This is a symmorphic group with primitive lattice basis vectors,
\begin{align}
\mathbf{e}_1&=\frac{\sqrt{3}}{2}\hat{\mathbf{x}}+\frac{1}{2}\hat{\mathbf{y}} \\
\mathbf{e}_2&=\frac{\sqrt{3}}{2}\hat{\mathbf{x}}-\frac{1}{2}\hat{\mathbf{y}},
\end{align}
which are pictured in Fig~\ref{fig:basisvectors}. Note that the Bilbao Crystallographic Server\cite{Bilbao1,Bilbao2,Bilbao3} uses 
\begin{equation}
\mathbf{e}_1'=\mathbf{e}_2,\; \mathbf{e}_2'=\mathbf{e}_1-\mathbf{e}_2
\end{equation}
as an alternative choice of primitive lattice vectors.

\begin{figure}
\centering
\subfloat[]{
	\includegraphics[width=1.5in]{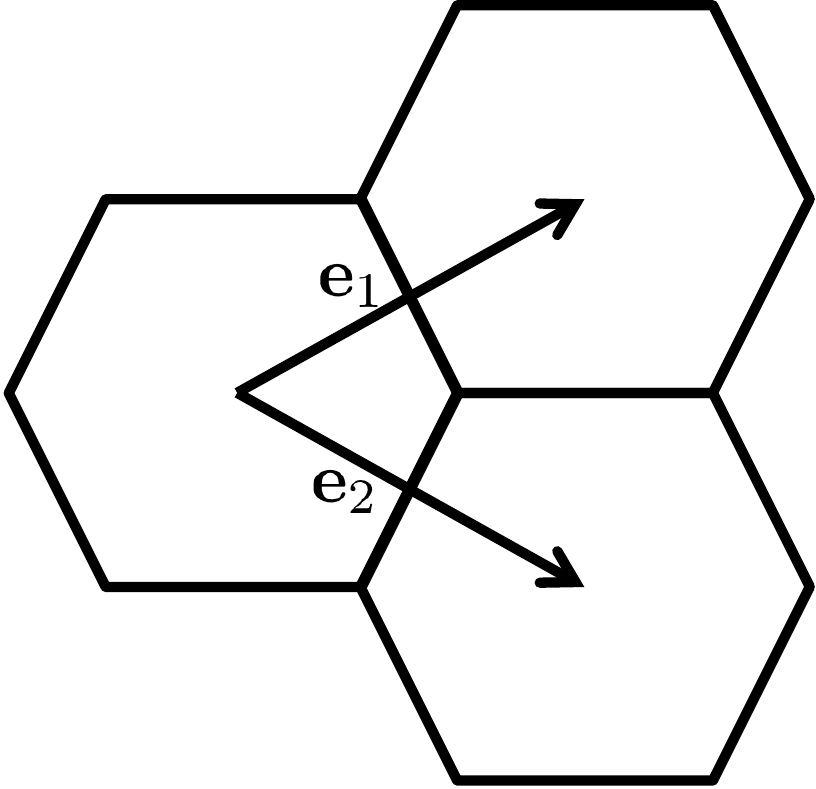}
	\label{fig:basisvectors}
}
\hspace{.5in}
\subfloat[]{
	\includegraphics[width=1.5in]{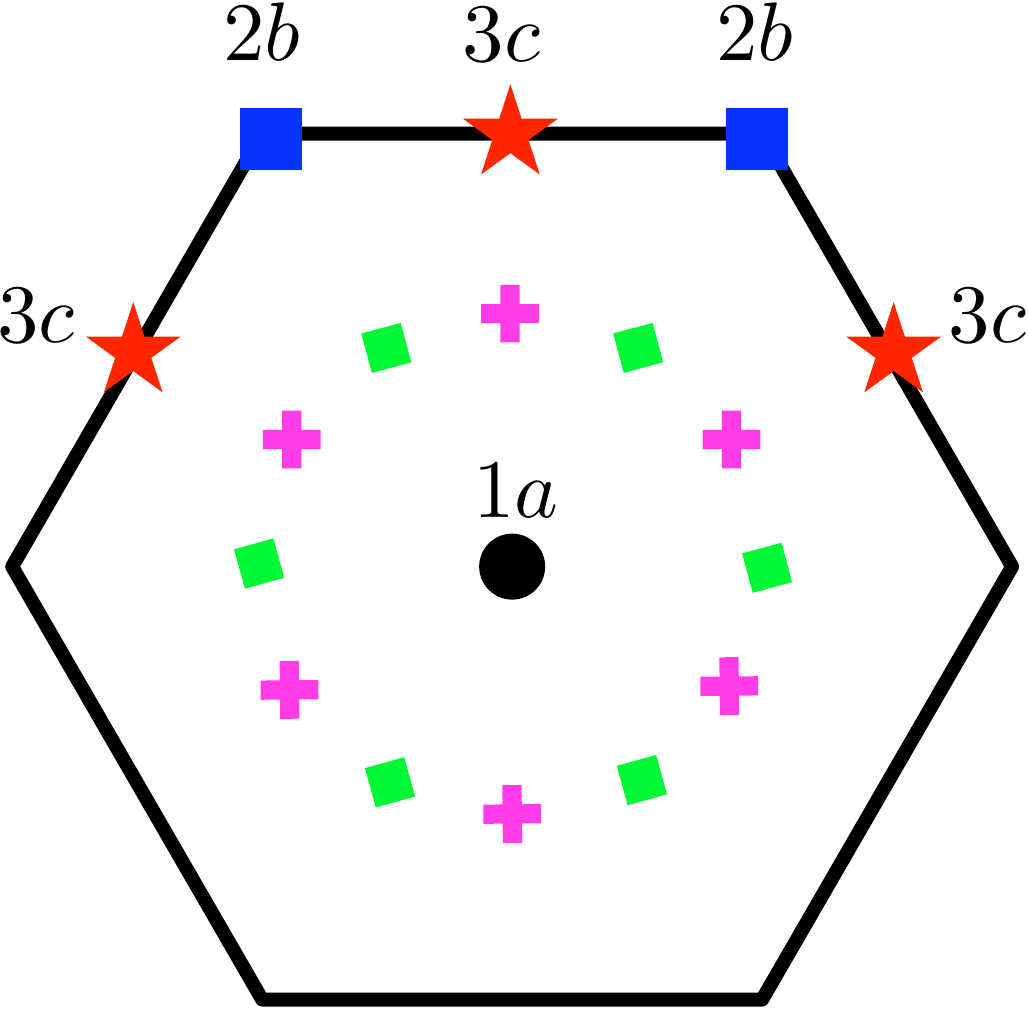}
	\label{fig:Wyckoff}
}
\caption{{Lattice basis vectors (a) and Wyckoff positions (b) of the hexagonal lattice. The (maximal) $1a$, $2b$ and $3c$ Wyckoff positions are indicated by {a black dot, blue squares, and red stars, respectively}. The {non-maximal} $6d$ and $6e$ positions are indicated by {purple} crosses and {green} diamonds, respectively. The multiplicity is determined by the index of the stabilizer group with respect to the point group $C_{6v}$ ($6mm$).}}
\end{figure}

The point group is $C_{6v}$, and is generated by
\begin{align}
C_{3z}:& (\mathbf{e}_1,\mathbf{e}_2)\rightarrow(-\mathbf{e}_2,\mathbf{e}_1-\mathbf{e}_2)\label{eq:grpaction1}\\
C_{2z}:& (\mathbf{e}_1,\mathbf{e}_2)\rightarrow(-\mathbf{e}_1,-\mathbf{e}_2)\\
m_{1\bar{1}}:&(\mathbf{e}_1,\mathbf{e}_2)\rightarrow(\mathbf{e}_2,\mathbf{e}_1),\label{eq:grpaction2}
\end{align}
where the subscript $1\bar{1}$ denotes that the mirror line has normal vector $\mathbf{e}_1-\mathbf{e}_2$. Although this set of generators is overcomplete ({a minimal set of generators is $\{C_{6z},m_{1\bar{1}}\}$}), it is convenient for our purposes. The three-dimensional space group with the symmetries catalogued above is space group $P6mm$ (183), which differs only in the addition of a third translation vector; we recover the $2D$ symmetry group by taking the length of this lattice vector to infinity. We note, however, that when we consider the $2D$ wallpaper group as embedded in three-dimensional space, we have some freedom when it comes to imposing extra symmetries such as inversion $I$. For this particular wallpaper group, we see that the combination $m_z=IC_{2z}$ fixes every point in the $2D$ lattice, but acts on the spin degree of freedom as a rotation by $\pi$ about the $z$ axis. As such, imposing inversion symmetry on graphene is tantamount to imposing the conservation of $S_z$. This will become important when we consider spin-orbit coupling. In general, however, we view spin conservation as non-essential, and restrict ourselves in most cases to the symmetries of  $P6mm$ (183). 

The honeycomb lattice has three maximal Wyckoff positions, {as shown in Fig~\ref{fig:Wyckoff}}. In graphene, the carbon atoms sit at the $2b$ position, with symmetry sites {$\{\mathbf{q}^b_1,\mathbf{q}^b_2\}=\{(\frac{1}{3}\frac{1}{3}),(\bar{\frac{1}{3}}\bar{\frac{1}{3}})\}$}. {Here} and throughout $\bar{x}=-x$. The stabilizer group $G_{\mathbf{q}^b_1}$ is isomorphic to the group $C_{3v}$; it is generated by the elements $\{m_{1\bar{1}}|00\}$ and $\{C_{3z}|01\}$. It is an index two subgroup of the point group $C_{6v}$, and the quotient group $C_{6v}/C_{3v}$ is generated by the coset {that contains $C_{2z}$} (regardless of whether we are using point groups or double point groups, this quotient group is isomorphic to the abelian group with two elements, since $C_{2z}^2=\bar{E}\in C_{3v}$).

In the BZ, we take for our primitive reciprocal-lattice basis vectors
\begin{align}
\mathbf{g}_1&=2\pi\left(\frac{\sqrt{3}}{3}\hat{\mathbf{x}}+\hat{\mathbf{y}}\right) \\
\mathbf{g}_2&=2\pi\left(\frac{\sqrt{3}}{3}\hat{\mathbf{x}}-\hat{\mathbf{y}}\right),
\label{eq:reciprocal}
\end{align}
which are shown in Fig~\ref{fig:reciprocal}.
We will be primarily interested in the little group representations at three high symmetry points in the BZ. 
The first is the $\Gamma$ point, with coordinates $(00)$. The little co-group $\bar{G}_{\Gamma}$ is, as always, the point group {$C_{6v}$.} Next, there are the three time-reversal invariant $M$ points (that is, points $\mathbf{k}$ such that $-\mathbf{k}\equiv\mathbf{k}$ modulo a reciprocal lattice vector), which we denote $M$, $M'$ and $M''$. These have coordinates $(\frac{1}{2}0)$, $(\frac{1}{2}\half)$ and $(0\half)$ respectively. For the remainder of this appendix we need only concern ourselves with the first of these, and so we will refer to it unambiguously as ``the'' $M$ point; the others are related to it by $C_{3z}$ symmetry. It has little co-group {$\bar{G}_M$, which is isomorphic to $C_{2v}$ and generated by $C_{2z}$ and $C_{3z}m_{1\bar{1}}$.} Finally, there are the $K$ and $K'$ points -- the focus of most topological investigations in graphene. 
We will focus here primarily on the $K$ point which has coordinates $(\frac{1}{3}\frac{2}{3})$; the $K'$ point can be obtained by a $\pi/3$ rotation.
The little co-group {$\bar{G}_{K}$ is isomorphic to $C_{3v}$ and} is generated by $C_{3z}$ and $C_{2z}m_{1\bar{1}}$. 
The high symmetry points are shown in Fig~\ref{fig:reciprocal}. {In t}ables \ref{table:c6v}, \ref{table:c3v}, and \ref{table:c2v} {we} give the character tables for the irreducible representations of the little co-groups $\bar{G}_\Gamma,\bar{G}_K,$ and $\bar{G}_M$ respectively. We indicate double-valued (spinor) representations with a bar over the representation label. As mentioned previously, these character tables fully determine the representations of the corresponding little groups, since  $P6mm$ (183) is symmorphic.

\begin{table}[h]
\begin{tabular}{c|ccccccc}
Rep & $E$ & $C_{3z}$ & $C_{2z}$ & $C_{6z}$ & $m_{1\bar{1}}$ & $C_{6z}m_{1\bar{1}}$ & $\bar{E}$ \\
\hline
$\Gamma_1$ &1 &1 &1 &1 &1 &1 &1\Tstrut \\
$\Gamma_2$ &1 &1 &1 &1 &-1 &-1 &1 \\
$\Gamma_3$ &1 &1 &-1 &-1 &-1 &1 &1 \\
$\Gamma_4$ &1 &1 &-1 &-1 &1 &-1 &1 \\
$\Gamma_5$ &2 &-1 &2 &-1 &0 &0 &2 \\
$\Gamma_6$ &2 &-1 &-2 &1 &0 &0 &2 \\
$\bar{\Gamma}_7$ &2 &-2 &0 &0 &0 &0 &-2 \\
$\bar{\Gamma}_8$ &2 &1 &0 &-$\sqrt{3}$ &0 &0 &-2 \\
$\bar{\Gamma}_9$ &2 &1 &0 &$\sqrt{3}$ &0 &0 &-2 \\
\end{tabular}
\caption{The character table for the little co-group $\bar{G}_\Gamma\approx C_{6v}$ of the $\Gamma$ point. The irreps $\Gamma_1$-$\Gamma_6$ are all single valued, while $\bar{\Gamma}_7,\bar{\Gamma}_8,$ and $\bar{\Gamma}_9$ are double valued. $\bar{\Gamma}_9$ is the spin-$\half$ representation, $\bar{\Gamma}_7$ is the $|S=3/2,m_z=\pm 3/2\rangle$ representation, and $\bar{\Gamma}_8$ is the $|S=5/2,m_z=\pm 5/2\rangle$ representation, {all distinguishable by the action of $C_{6z}$}.}
\label{table:c6v}
\end{table}
\begin{table}[h]
\begin{tabular}{c|cccc}
Rep & $E$ & $C_{3z}$ & $C_{2z}m_{1\bar{1}}$ & $\bar{E}$ \\
\hline
$K_1$ & 1 & 1 & 1 & 1 \Tstrut\\
$K_2$ & 1 & 1 & -1 & 1\\
$K_3$ & 2 & -1 & 0 & 2 \\
$\bar{K}_4$ & 1 & -1 & -i & -1 \\
$\bar{K}_5$ & 1 & -1 & i & -1 \\
$\bar{K}_6$ & 2 & 1 & 0 & -2
\end{tabular}
\caption{Character table for the little co-group $\bar{G}_K\approx C_{3v}$ of the $K$ point. There are three single-valued representations $K_1$--$K_3$, and three double valued representations $\bar{K}_4$--$\bar{K}_6$. The one-dimensional representations $\bar{K}_4$ and $\bar{K}_5$ are complex conjugates of each other. The two dimensional $\bar{K}_{6}$ representation is the spin-$\half$ representation, while the one-dimensional $\bar{K}_4$ and $\bar{K}_5$ representations act in the space spanned by $|S=3/2, m_z=3/2\rangle\pm i|S=3/2,m_z=-3/2\rangle$ respectively.}\label{table:c3v}
\end{table}
\begin{table}[h]
\begin{tabular}{c|ccccc}
 Rep & $E$ & $C_{2z}$ & $m$ & $C_{2z}m$ &$\bar{E}$ \\
 \hline
$M_1$ & 1 & 1 & 1 & 1 & 1\Tstrut \\
$M_2$ & 1 & 1 & -1 & -1 & 1 \\
$M_3$ & 1 & -1 & -1 & 1 & 1 \\
$M_4$ & 1 & -1 & 1 & -1 & 1 \\
$\bar{M}_5$ & 2 & 0 & 0 & 0 & -2
\end{tabular}
\caption{Character table for the little co-group $\bar{G}_M\approx C_{2v}$ of the $M$ point, for both single and double-valued representations. The single-valued representations $M_1$--$M_4$ are all one dimensional. The unique double-valued representation, $\bar{M}_5$, is the two-dimensional spin-$\half$ representation.  In terms of the Pauli matrices, it is given concretely as $\bar{M}_5(C_{2z})=i\sigma_z,\bar{M}_5(m)=i\sigma_y$.}
\label{table:c2v}
\end{table}

\subsection{$p_z$ Orbitals and the Elementary band representation}\label{subsec:graphenebr}

In graphene, the relevant orbitals near the Fermi level are the two spin species of the $p_z$ orbitals at the $2b$ Wyckoff position. Let us focus on the orbitals $\{|p_z\uparrow\rangle_1,|p_z\downarrow\rangle_1\}$ at the $\mathbf{q}^b_1$ site. These transform according to an irreducible double-valued (spinor) representation $\rho$ of the site symmetry group $G_{\mathbf{q}^b_1}=C_{3v}$: in the space of these orbitals, $\{C_{3z}|01\}$ acts as a rotation about the $z$-axis in spin space, and $m_{1\bar{1}}$ acts as a spin-flip. Furthermore, time-reversal symmetry $T$ acts as a spin flip times complex conjugation. Symbolically,
\begin{equation}
\rho(\{C_{3z}|01\})=e^{i\pi/3 s_z},\; \rho(m_{1\bar{1}})=is_x,\; \rho(T)=is_y\mathcal{K},
\end{equation}
where $\{s_0,s_x,s_y,s_z\}$ are Pauli matrices that act in the space of spin $\uparrow\downarrow$ ($s_0$ is the identity matrix), and $\mathcal{K}$ is complex conjugation. Similarly, the $p_z$ orbitals at the $\mathbf{q}^b_2$ site transform in an equivalent representation obtained by conjugation by $C_{2z}$.

Because $\rho$ is a physically irreducible representation of the site-symmetry group of a maximal Wyckoff position (and does not appear in Table~\ref{table:dbr}), it induces a physically elementary band representation, $\rho^\mathbf{k}_G=\rho\uparrow G$. It has four bands, coming from the four orbitals per unit cell. To construct the matrices $\rho^\mathbf{k}_G$ we directly examine the action of the point group elements on the spin and location of orbitals. Let's focus first on $\{C_{3z}|00\}$. Its action on orbitals at the $\mathbf{q}^b_1$ site can be deduced from above; for orbitals at the $\mathbf{q}^b_2$ site, it acts as a rotation in spin space, and takes $\{C_{3z}|00\}\mathbf{q}^b_2=\mathbf{q}^b_2+\mathbf{e}_2$. Introducing a set of Pauli matrices $\{\sigma_0,\sigma_x,\sigma_y,\sigma_z\}$ which act in the sublattice basis ($\sigma_0$ is the identity matrix), this allows us to write
\begin{equation}
\rho^\mathbf{k}_G(\{C_{3z}|00\})=e^{i\pi/3 s_z}\otimes e^{i(\mathbf{k}\cdot\mathbf{e}_2)\sigma_z},\label{eq:brC3}
\end{equation}
where $\otimes$ is the usual tensor product. Similarly, $m_{1\bar{1}}$ acts as a spin flip at the $\mathbf{q}^b_2$ site, and also leaves this point invariant. Hence\footnote{Note that this choice for the representative of $m_{1\bar{1}}$ differs by a unitary transformation from the more conventional basis where $m_{1\bar{1}}$ is represented by $i\sigma_y$, a $\pi$ spin rotation about the $\mathbf{e}_1-\mathbf{e}_2=\hat{\mathbf{y}}$ axis. This basis choice is necessary for the Hamiltonian Eq.~(\ref{eq:graphenerashbaham}) to match the well known expression of Kane and Mele\cite{Kane04}.}
\begin{equation}
\rho^\mathbf{k}_G(\{m_{1\bar{1}}|00\})=is_x\otimes\sigma_0.\label{eq:brM}
\end{equation}
Next, since time-reversal acts independent of position, we know immediately that
\begin{equation}
\rho^\mathbf{k}_G(T)=is_y\otimes\sigma_0\mathcal{K}.\label{eq:brT}
\end{equation}
Lastly, we need to examine $\{C_{2z}|00\}$. This interchanges the two sublattices (orbitals), and so acts as $\sigma_x$ in sublattice space. In spin space, it acts as a rotation by $\pi${, and commutes with T, the time-reversal operator}. Thus we deduce
\begin{equation}
\rho^\mathbf{k}_G(\{C_{2z}|00\})=is_z\otimes\sigma_x.\label{eq:brC2}
\end{equation}
\begin{figure}
\centering
	\includegraphics[width=1.2in]{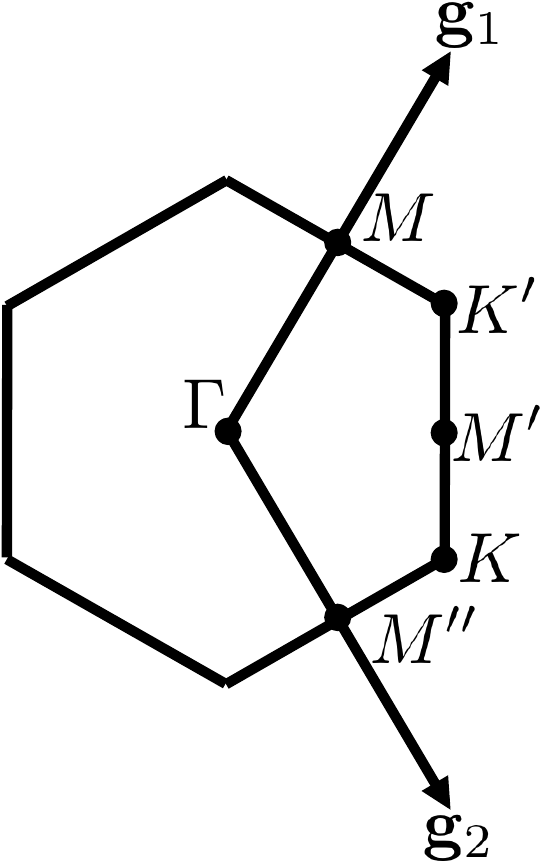}
	\caption{Reciprocal lattice basis vectors and high symmetry points of the hexagonal lattice.}
	\label{fig:reciprocal}
\end{figure}
Representation matrices in hand, it is now a simple matter of comparison with Tables \ref{table:c6v}, \ref{table:c3v}, \ref{table:c2v} to determine the little group representations at each high symmetry point. First, at the $\Gamma$ point all point group elements are in the little co-group, and we see that the matrices $\rho^\mathbf{k}_G$ restrict to
\begin{equation}
\rho_G^\Gamma(\{C_{3z}|00\})=e^{i\pi/3s_z}\otimes\sigma_0,\; \rho^\Gamma_G(\{m_{1\bar{1}}|00\})=is_x\otimes\sigma_0,\; \rho_G^\Gamma(\{C_{2z}|00\})=is_z\otimes\sigma_x,\;\rho_G^\Gamma(T)=is_y\otimes\sigma_0\mathcal{K}.
\end{equation}
Comparing the trace of each of these unitary matrices to the characters in Table~\ref{table:c6v}, we see that
\begin{equation}
\left(\rho\uparrow G\right)\downarrow G_\Gamma\approx\bar{\Gamma}_8\oplus\bar{\Gamma}_9.
\end{equation}
[Although TR is not mentioned in Table~\ref{table:c6v} or \ref{table:c2v}, the $\bg_8$ and $\bg_9$ representations of $G_\Gamma$ and the $\bar{M}_5$ representation of $G_M$ satisfy Kramers's theorem consistent with our choice of time-reversal matrix Eq.~(\ref{eq:brT}).] Next, the little co-group at $K$ is generated by $C_{3z}$ and $C_{2z}m_{1\bar{1}}$ which in this band representation are given by
\begin{equation}
\rho_G^K(\{C_{3z}|00\})=e^{i\pi/3s_z}\otimes e^{2\pi i /3\sigma_z},\; \rho^K_G(C_{2z}m_{1\bar{1}})=-is_y\otimes\sigma_x. \label{eq:Kptinducedreps}
\end{equation}
Upon taking traces and comparing with Table~\ref{table:c3v} we deduce
\begin{equation}
\left(\rho\uparrow G\right)\downarrow G_K\approx\bar{K}_4\oplus\bar{K}_5\oplus\bar{K}_6.
\end{equation}
Finally, using the fact that there is only a unique double-valued representation $\bar{M}_5$ allowed at the $M$ point, we deduce by simple dimension counting that
\begin{equation}
\left(\rho\uparrow G\right)\downarrow G_M\approx\bar{M}_5\oplus\bar{M}_5.
\end{equation}

Thus, we have deduced, \emph{from symmetry alone}, the little group representations of the energy bands induced by the $p_z$ orbitals in graphene. For future convenience, we summarize this in Table~\ref{table:graphenebrreps}. In the spirit of our programme described in the main text, the next step is to analyze how these energy bands are permitted to connect throughout the BZ. {We will accomplish this with the aid of the graph theory method outlined in Section~\ref{sec:graphreview}.}

\begin{table}[t]
\begin{tabular}{c|c|c|c}
BR & $\Gamma$ & $K$ & $M$ \\
\hline
$\rho\uparrow G$ & $\bar{\Gamma}_8\oplus\bar{\Gamma_9}$ & $\bar{K}_4\oplus\bar{K}_5\oplus\bar{K}_6$ & $\bar{M}_5\oplus\bar{M}_5$\Tstrut
\end{tabular}
\caption{Little group representations for the energy bands induced from $p_z$ orbitals in graphene.}\label{table:graphenebrreps}
\end{table}

\subsection{Graph Analysis}\label{subsec:graphenegraph}

To construct the connectivity graphs -- and hence determine the allowed topological phases of graphene, we shall follow the procedure outlined in the main text and in Section~\ref{sec:graphreview}. To do this, we need to first examine the compatibility relations along the lines joining $\Gamma,K,$ and $M$ in the BZ. Once we have determined the compatibility relations for the little group representations occurring in the band representation $\rho\uparrow G$ of $p_z$ orbitals, we will explicitly construct the distinct connectivity graphs for the system. In particular, we show that there is a fully connected {protected (at half-filling)} semi-metallic phase, as well as a disconnected topological insulating phase.

Let us begin with the line $\Sigma=k\mathbf{g}_1,\;{k\in[0,\half]}$ which {links} $\Gamma$ and $M$. The little co-group $\bar{G}_\Sigma$ of this line is the abelian group $C_s$ generated by $C_{3z}m_{1\bar{1}}${\cite{PointGroupTables}}. It has two double-valued representations denoted by $\bar{\Sigma}_3$ and $\bar{\Sigma}_4$, distinguished by whether {the group generator} is represented by  $\pm i$, respectively. Consider the little group representations appearing in the $\rho\uparrow G$ band representation in Table~\ref{table:graphenebrreps}. {From Table~\ref{table:c6v}, we see that in both the $\bar{\Gamma}_8$ and $\bar{\Gamma}_9$ representations at $\Gamma$, the character $\chi(C_{3z}m_{1\bar{1}})=0$ (it is in the same conjugacy class as $m_{1\bar{1}}$ in the table). From this we deduce that each of these representations restricts to the direct sum $\bar{\Sigma}_3\oplus\bar{\Sigma}_4$ on the line $\Sigma$. A similar analysis shows that the representation $\bar{M}_5$ at the $M$ point subduces also to $\bar{\Sigma}_3\oplus\bar{\Sigma}_4$}. We summarize this in the compatibility relations (\ref{eq:sigmacomprels}):
\begin{align}
\bar{\Gamma}_8\downarrow G_\Sigma&=\bar{\Sigma}_3\oplus\bar{\Sigma}_4 \nonumber \\
\bar{\Gamma}_9\downarrow G_\Sigma&=\bar{\Sigma}_3\oplus\bar{\Sigma}_4 \nonumber \\
\bar{M}_5\downarrow G_\Sigma&=\bar{\Sigma}_3\oplus\bar{\Sigma}_4\label{eq:sigmacomprels}.
\end{align}

Next, we look at the line $T=(\half+k)\mathbf{g}_1+2k\mathbf{g}_2,{k\in[-\frac{1}{6},0]}$, which connects the $K$ and $M$ points. The little co-group $\bar{G}_T$ of this line is also isomorphic to $C_s$, this time generated by the mirror $C_{6z}m_{1\bar1}$. As above, we denote its two irreps by $\bar{T}_3$ and $\bar{T}_4$. {By looking at the characters of the little group representations $\bar{K}_4,\bar{K}_5$, and $\bar{K}_6$, from Table~\ref{table:c3v}, we deduce that}
\begin{align}
\bar{K}_4\downarrow G_T&=\bar{T}_3 \nonumber\\
\bar{K}_5\downarrow G_T&=\bar{T}_4 \nonumber\\
\bar{K}_6\downarrow G_T &=\bar{T}_3\oplus\bar{T}_4.\label{eq:KTcomprels}
\end{align}
The restriction of the representation $\bar{M}_5$ into representations of $C_s$ was computed {in Eq.~(\ref{eq:sigmacomprels})}, and so
\begin{equation}
\bar{M}_5\downarrow G_T=\bar{T}_3\oplus\bar{T}_4\label{eq:MTcomprels}.
\end{equation}

Finally, we examine the line $\Lambda=k\mathbf{g}_1+2k\mathbf{g}_2, {k\in[0,\frac{1}{3}]}$, which connects the points $\Gamma$ and $K$. Like the previous cases, the little co-group of this line is $C_s$, this time generated by $C_{2z}m_{1\bar{1}}$, {and the compatibility relations are }

\begin{align}
\bar{K}_4\downarrow G_\Lambda&=\bar{\Lambda}_3 \nonumber\\
\bar{K}_5\downarrow G_\Lambda&=\bar{\Lambda}_4 \nonumber\\
\bar{K}_6\downarrow G_\Lambda &=\bar{\Lambda}_3\oplus\bar{\Lambda}_4 \nonumber \\
\bar{\Gamma}_8\downarrow G_\Lambda&=\bar{\Lambda}_3\oplus\bar{\Lambda}_4 \nonumber \\
\bar{\Gamma}_9\downarrow G_\Lambda&=\bar{\Lambda}_3\oplus\bar{\Lambda}_4.\label{eq:LDcomprels}
\end{align}

The only remaining $\mathbf{k}$-surface in the BZ is the general position, denoted $GP=k_1\mathbf{g}_1+k_2\mathbf{g}_2$. However, it has a trivial little group with only one {1$D$} double-valued irreducible representation $\bar{GP}_2$. {Because of this, the compatibility relations are trivial -- all representations restrict to copies of $\bar{GP}_2$, and group theory provides no restrictions on the connectivity. Thus this surface} does not add anything new to the connectivity analysis and we omit it here. 

We can now construct the degree, adjacency, and Laplacian matrices, {defined in Section~\ref{sec:graphreview},} consistent with these compatibility relations. In each of these matrices, the rows and columns (i.e. the nodes in the connectivity graph) are labelled by the different irreps occurring in the band representation. In this example, we have the following nodes: $\bar{\Gamma}_8,\bar{\Gamma}_9,\bar{\Sigma}_3^1,\bar{\Sigma}_3^2,\bar{\Sigma}_4^1,\bar{\Sigma}_4^2,\bar{\Lambda}_3^1,\bar{\Lambda}_3^2,\bar{\Lambda}_4^1,\bar{\Lambda}_4^2,\bar{K}_4,\bar{K}_5,\bar{K}_6,\bar{T}_3^1,\bar{T}_3^2,\bar{T}_4^1,\bar{T}_4^2,\bar{M}_5^1,\bar{M}_5^2$. {We reiterate here that, as per Def.~\ref{def:compgraph}, representations at high-symmetry points \emph{and} lines correspond to nodes in our graph.} Note that if a representation occurs more than once {(as is the case along the lines $\Sigma,T$, and $\Lambda$, as well as at the point $M$)}, there is a distinct node for each copy that appears, which we label here with a superscript.

We begin first with the degree matrix $D$, {as defined in Def.~\ref{def:degreemat}}. Let us denote by $d(\sigma)$ the degree of the node labelled by representation $\sigma$ in the connectivity graph. We know from Def.~\ref{def:compgraph} that $d(\sigma)$ is given by $\mathrm{dim}(\sigma)$ times the number of distinct compatiblity tables in which $\sigma$ appears. {For example, $\mathrm{dim}(\bar{\Gamma}_8)=2$}. Furthermore, it connects to $P=2$ other high-symmetry lines (in the notation of Def.~\ref{def:compgraph}), $\Sigma$ and $\Lambda$. Following this prescription, the entry $d(\bar{\Gamma}_8)=2\times2=4$. Carrying out this procedure for all little group representations, we find the nonzero (diagonal) entries $D$ which we summarize in Table~\ref{table:183degmat}.
\begin{table}[h]
\begin{tabular}{c|ccccccccccccccccccc}
& $\bar{\Gamma}_8$&$\bar{\Gamma}_9$&$\bar{\Sigma}_3^1$&$\bar{\Sigma}_3^2$&$\bar{\Sigma}_4^1$&$\bar{\Sigma}_4^2$&$\bar{\Lambda}_3^1$&$\bar{\Lambda}_3^2$&$\bar{\Lambda}_4^1$&$\bar{\Lambda}_4^2$&$\bar{K}_4$&$\bar{K}_5$&$\bar{K}_6$&$\bar{T}_3^1$&$\bar{T}_3^2$&$\bar{T}_4^1$&$\bar{T}_4^2$&$\bar{M}_5^1$&$\bar{M}_5^2$\\
\hline
$d(\rho)$ & $4$ & $4$ & $2$ & $2$ &$2$ & $2$ & $2$ &$2$ &$2$ &$2$&$2$ &$2$ &$4$&$2$ &$2$ &$2$&$2$&$4$&$4$\Tstrut
\end{tabular}
\caption{Nonzero entries in the degree matrix for the $\bar{\rho}_6^{2b}\uparrow G$ band representation in  $P6mm$} (183)\label{table:183degmat}
\end{table}

Next, we construct the allowed adjacency matrices for this band representation. The adjacency matrices all have a sparse block structure -- blocks connecting different $\mathbf{k}$-points are nonzero only if the points are compatible. We see then that the only nonzero blocks are the $\Gamma-\Sigma$, $\Gamma-\Lambda$, $K-\Lambda$, $K-T$, $M-T$, and $M-\Sigma$ blocks. Furthermore, up to relabellings of identical representations, i.e. $\bar{M}_5^1\leftrightarrow\bar{M}_5^2,\,\bar{\Lambda}_3^1\leftrightarrow\bar{\Lambda}_3^2$, etc., there are only four distinct adjacency matrices {(we elaborate on these details in Ref.~\onlinecite{graphdatapaper})}. {These fall into two groups which differ by the exchange $\bar{\Gamma}_8\leftrightarrow\bar{\Gamma}_9$, since $\bar{\Gamma}_8$ and $\bar{\Gamma}_9$ have identical compatibility relations along both $\Sigma$ and $\Lambda$. For brevity, we write here only the two independent matrices from which the remaining two can be obtained by {the exchange $\bar{\Gamma}_8\leftrightarrow\bar{\Gamma}_9$}.} They are
\begin{equation}
A_1=\begin{blockarray}{cccccccccccccccccccc}
\bar{\Gamma}_8&\bar{\Gamma}_9&\bar{\Sigma}_3^1&\bar{\Sigma}_3^2&\bar{\Sigma}_4^1&\bar{\Sigma}_4^2&\bar{\Lambda}_3^1&\bar{\Lambda}_3^2&\bar{\Lambda}_4^1&\bar{\Lambda}_4^2&\bar{K}_4&\bar{K}_5&\bar{K}_6&\bar{T}_3^1&\bar{T}_3^2&\bar{T}_4^1&\bar{T}_4^2&\bar{M}_5^1&\bar{M}_5^2&\\
\begin{block}{(cc|cccc|cccc|ccc|cccc|cc)c}
0 & 0 & 1 & 0 & 1 & 0 & 1 & 0 & 1 & 0 & 0 & 0 & 0 & 0 & 0 & 0 & 0 & 0 & 0 & \bar{\Gamma}_8 \\
0 & 0 & 0 & 1 & 0 & 1 & 0 & 1 & 0 & 1 & 0 & 0 & 0 & 0 & 0 & 0 & 0 & 0 & 0 & \bar{\Gamma}_9 \\
\cline{1-19}
1 & 0 & 0 & 0 & 0 & 0 & 0 & 0 & 0 & 0 & 0 & 0 & 0 & 0 & 0 & 0 & 0 & 1 & 0 & \bar{\Sigma}_3^1 \\
0 & 1 & 0 & 0 & 0 & 0 & 0 & 0 & 0 & 0 & 0 & 0 & 0 & 0 & 0 & 0 & 0 & 0 & 1 & \bar{\Sigma}_3^2 \\
1 & 0 & 0 & 0 & 0 & 0 & 0 & 0 & 0 & 0 & 0 & 0 & 0 & 0 & 0 & 0 & 0 & 1 & 0 & \bar{\Sigma}_4^1 \\
0 & 1 & 0 & 0 & 0 & 0 & 0 & 0 & 0 & 0 & 0 & 0 & 0 & 0 & 0 & 0 & 0 & 0 & 1 & \bar{\Sigma}_4^2\\
\cline{1-19}
1 & 0 & 0 & 0 & 0 & 0 & 0 & 0 & 0 & 0 & 1 & 0 & 0 & 0 & 0 & 0 & 0 & 0 & 0 & \bar{\Lambda}_3^1 \\
0 & 1 & 0 & 0 & 0 & 0 & 0 & 0 & 0 & 0 & 0 & 0 & 1 & 0 & 0 & 0 & 0 & 0 & 0 & \bar{\Lambda}_3^2 \\
1 & 0 & 0 & 0 & 0 & 0 & 0 & 0 & 0 & 0 & 0 & 0 & 1 & 0 & 0 & 0 & 0 & 0 & 0 & \bar{\Lambda}_4^1 \\
0 & 1 & 0 & 0 & 0 & 0 & 0 & 0 & 0 & 0 & 0 & 1 & 0 & 0 & 0 & 0 & 0 & 0 & 0 & \bar{\Lambda}_4^2 \\
\cline{1-19}
0 & 0 & 0 & 0 & 0 & 0 & 1 & 0 & 0 & 0 & 0 & 0 & 0 & 1 & 0 & 0 & 0 & 0 & 0 & \bar{K}_4 \\
0 & 0 & 0 & 0 & 0 & 0 & 0 & 0 & 0 & 1 & 0 & 0 & 0 & 0 & 0 & 0 & 1 & 0 & 0 & \bar{K}_5 \\
0 & 0 & 0 & 0 & 0 & 0 & 0 & 1 & 1 & 0 & 0 & 0 & 0 & 0 & 1 & 1 & 0 & 0 & 0 & \bar{K}_6 \\
\cline{1-19}
0 & 0 & 0 & 0 & 0 & 0 & 0 & 0 & 0 & 0 & 1 & 0 & 0 & 0 & 0 & 0 & 0 & 1 & 0 & \bar{T}_3^1 \\
0 & 0 & 0 & 0 & 0 & 0 & 0 & 0 & 0 & 0 & 0 & 0 & 1 & 0 & 0 & 0 & 0 & 0 & 1 & \bar{T}_3^2 \\
0 & 0 & 0 & 0 & 0 & 0 & 0 & 0 & 0 & 0 & 0 & 0 & 1 & 0 & 0 & 0 & 0 & 1 & 0 & \bar{T}_4^1 \\
0 & 0 & 0 & 0 & 0 & 0 & 0 & 0 & 0 & 0 & 0 & 1 & 0 & 0 & 0 & 0 & 0 & 0 & 1 & \bar{T}_4^2 \\
\cline{1-19}
0 & 0 & 1 & 0 & 1 & 0 & 0 & 0 & 0 & 0 & 0 & 0 & 0 & 1 & 0 & 1 & 0 & 0 & 0 & \bar{M}_5^1 \\
0 & 0 & 0 & 1 & 0 & 1 & 0 & 0 & 0 & 0 & 0 & 0 & 0 & 0 & 1 & 0 & 1 & 0 & 0 & \bar{M}_5^2 \\
\end{block}
\end{blockarray}
\end{equation}
and
\begin{equation}
A_2=\begin{blockarray}{cccccccccccccccccccc}
\bar{\Gamma}_8&\bar{\Gamma}_9&\bar{\Sigma}_3^1&\bar{\Sigma}_3^2&\bar{\Sigma}_4^1&\bar{\Sigma}_4^2&\bar{\Lambda}_3^1&\bar{\Lambda}_3^2&\bar{\Lambda}_4^1&\bar{\Lambda}_4^2&\bar{K}_4&\bar{K}_5&\bar{K}_6&\bar{T}_3^1&\bar{T}_3^2&\bar{T}_4^1&\bar{T}_4^2&\bar{M}_5^1&\bar{M}_5^2&\\
\begin{block}{(cc|cccc|cccc|ccc|cccc|cc)c}
0 & 0 & 1 & 0 & 1 & 0 & 1 & 0 & 1 & 0 & 0 & 0 & 0 & 0 & 0 & 0 & 0 & 0 & 0 & \bar{\Gamma}_8 \\
0 & 0 & 0 & 1 & 0 & 1 & 0 & 1 & 0 & 1 & 0 & 0 & 0 & 0 & 0 & 0 & 0 & 0 & 0 & \bar{\Gamma}_9 \\
\cline{1-19}
1 & 0 & 0 & 0 & 0 & 0 & 0 & 0 & 0 & 0 & 0 & 0 & 0 & 0 & 0 & 0 & 0 & 1 & 0 & \bar{\Sigma}_3^1 \\
0 & 1 & 0 & 0 & 0 & 0 & 0 & 0 & 0 & 0 & 0 & 0 & 0 & 0 & 0 & 0 & 0 & 0 & 1 & \bar{\Sigma}_3^2 \\
1 & 0 & 0 & 0 & 0 & 0 & 0 & 0 & 0 & 0 & 0 & 0 & 0 & 0 & 0 & 0 & 0 & 1 & 0 & \bar{\Sigma}_4^1 \\
0 & 1 & 0 & 0 & 0 & 0 & 0 & 0 & 0 & 0 & 0 & 0 & 0 & 0 & 0 & 0 & 0 & 0 & 1 & \bar{\Sigma}_4^2\\
\cline{1-19}
1 & 0 & 0 & 0 & 0 & 0 & 0 & 0 & 0 & 0 & 1 & 0 & 0 & 0 & 0 & 0 & 0 & 0 & 0 & \bar{\Lambda}_3^1 \\
0 & 1 & 0 & 0 & 0 & 0 & 0 & 0 & 0 & 0 & 0 & 0 & 1 & 0 & 0 & 0 & 0 & 0 & 0 & \bar{\Lambda}_3^2 \\
1 & 0 & 0 & 0 & 0 & 0 & 0 & 0 & 0 & 0 & 0 & 1 & 0 & 0 & 0 & 0 & 0 & 0 & 0 & \bar{\Lambda}_4^1 \\
0 & 1 & 0 & 0 & 0 & 0 & 0 & 0 & 0 & 0 & 0 & 0 & 1 & 0 & 0 & 0 & 0 & 0 & 0 & \bar{\Lambda}_4^2 \\
\cline{1-19}
0 & 0 & 0 & 0 & 0 & 0 & 1 & 0 & 0 & 0 & 0 & 0 & 0 & 1 & 0 & 0 & 0 & 0 & 0 & \bar{K}_4 \\
0 & 0 & 0 & 0 & 0 & 0 & 0 & 0 & 1 & 0 & 0 & 0 & 0 & 0 & 0 & 1 & 0 & 0 & 0 & \bar{K}_5 \\
0 & 0 & 0 & 0 & 0 & 0 & 0 & 1 & 0 & 1 & 0 & 0 & 0 & 0 & 1 & 0 & 1 & 0 & 0 & \bar{K}_6 \\
\cline{1-19}
0 & 0 & 0 & 0 & 0 & 0 & 0 & 0 & 0 & 0 & 1 & 0 & 0 & 0 & 0 & 0 & 0 & 1 & 0 & \bar{T}_3^1 \\
0 & 0 & 0 & 0 & 0 & 0 & 0 & 0 & 0 & 0 & 0 & 0 & 1 & 0 & 0 & 0 & 0 & 0 & 1 & \bar{T}_3^2 \\
0 & 0 & 0 & 0 & 0 & 0 & 0 & 0 & 0 & 0 & 0 & 1 & 0 & 0 & 0 & 0 & 0 & 1 & 0 & \bar{T}_4^1 \\
0 & 0 & 0 & 0 & 0 & 0 & 0 & 0 & 0 & 0 & 0 & 0 & 1 & 0 & 0 & 0 & 0 & 0 & 1 & \bar{T}_4^2 \\
\cline{1-19}
0 & 0 & 1 & 0 & 1 & 0 & 0 & 0 & 0 & 0 & 0 & 0 & 0 & 1 & 0 & 1 & 0 & 0 & 0 & \bar{M}_5^1 \\
0 & 0 & 0 & 1 & 0 & 1 & 0 & 0 & 0 & 0 & 0 & 0 & 0 & 0 & 1 & 0 & 1 & 0 & 0 & \bar{M}_5^2 \\
\end{block}
\end{blockarray}
\end{equation}
These matrices differ only in their $K-\Lambda$ and $K-T$ blocks. As a consistency check, we verify that the sum of elements in the row or column labelled by $\sigma$ is equal to $d(\sigma)$ from Table~\ref{table:183degmat}; {thus, the degree matrix $D$ satisfies $D_{ij}=\delta_{ij}\sum_\ell A_{i\ell}$}. We show each of these graphs pictorially in Figure~\ref{fig:183graphs}. Although the graph method does not impose any constraints on the energies of the irreducible representations, we are free to interpret {and visualize} the vertical positioning of the nodes of the graph as the energy of the respective energy bands. Doing so gives Fig.~\ref{fig:183graphs} the alternative interpretation as a plot of the band structure!

We can now construct the Laplacian matrices $L_1=D-A_1$ and $L_2=D-A_2$ associated to these two graphs. To save space we will not write these out explicitly. We find that the null space of the matrix $L_1$ is spanned by the unique vector
\begin{equation}
\psi_1=\left(\begin{array}{ccccccccccccccccccc}1&1&1&1&1&1&1&1&1&1&1&1&1&1&1&1&1&1&1\end{array}\right)^T
\label{eq:connectedvector}
\end{equation}
indicating that the graph described by the matrix $A_1$ has a single connected component {consisting of all the nodes in the graph}. On the other hand, we find that the null space of $L_2$ is spanned by
\begin{align}
\psi_2^1&=\left(\begin{array}{ccccccccccccccccccc}1&0&1&0&1&0&1&0&1&0&1&1&0&1&0&1&0&1&0\end{array}\right)^T\label{eq:discvector1}\\
\psi_2^2&=\left(\begin{array}{ccccccccccccccccccc}0&1&0&1&0&1&0&1&0&1&0&0&1&0&1&0&1&0&1\end{array}\right)^T
\label{eq:discvector2}
\end{align}
indicating that the graph described by the matrix $A_2$ has two connected components. Consulting our ordering of representations in Table~\ref{table:183degmat}, we see that the first connected component contains the little group representations $\bar{\Gamma}_8,\bar{\Sigma}_3^1,\bar{\Sigma}_4^1,\bar{\Lambda}_3^1,\bar{\Lambda}_4^1,\bar{K}_4,\bar{K}_5,\bar{T}_3^1,\bar{T}_4^1$ and $\bar{M}_5^2$, while the other connected component contains the remainder $\bar{\Gamma}_9,\bar{\Sigma}_3^2,\bar{\Sigma}_4^2,\bar{\Lambda}_3^2,\bar{\Lambda}_4^2,\bar{K}_6,\bar{T}_3^2,\bar{T}_4^2$ and $\bar{M}_5^1$. (Interchanging $\bar{\Gamma}_8$ and $\bar{\Gamma}_9$ also results in a valid disconnected graph). Since each of these connected components comes from splitting an elementary band representations, they each describe a topological group of bands, and hence a topological insulator. 

This is consistent with the results of Ref.~\onlinecite{Soluyanov2011}, which found that the Wannier functions in the valence bands in the topological phase of graphene were of the form $|p_z \uparrow+\downarrow \rangle$ localized on the $A$ sites, and $|p_z \uparrow-\downarrow \rangle$ localized on the $B$ sites (the two points in the unit cell of the $2b$ Wyckoff orbit); by examining the action of $C_{3z}$ on these orbitals we conclude immediately that they do not, {by themselves}, form a representation space (carrier space) of the site-symmetry group $G_{\mathbf{q}^b_1}$. {In fact, no set of spin-$1/2$ Wannier functions formed from $s$ or $p$ orbitals and with one orbital per site can respect the spatial symmetries, since $C_{3v}$ has only two dimensional double-valued representations with $m_z=\pm1/2$ (even when time-reversal symmetry is neglected) as per Table~\ref{table:c3v}.} The fact that graphene is a strong -- rather than just a crystalline -- topological insulator is revealed in the fact that these Wannier functions are also not time-reversal invariant.

\subsection{Hamiltonian Analysis}\label{subsec:grapheneham}

We justify the preceding analysis concretely by considering a tight-binding model of $p_z$ or ($s$) orbitals centered on $2b$ sites with the most general Rashba and Haldane type SOC interactions. In particular, we will show how different classes of spin-orbit coupling terms can drive a transition between the two phases indicated in Fig~\ref{fig:183connected} and \ref{fig:183disconnected}. We will use the basis of spin and sublattice (orbital) Pauli matrices {(including the identity matrices)} $s_i\otimes\sigma_j{, \; i,j=0,1,2,3}$ introduced previously in Sec.~\ref{subsec:graphenebr}.
In this basis, we can expand any Bloch Hamiltonian in terms of sixteen Hermitian basis elements. We call a term in the Hamiltonian an SOC term if it does not act as the identity in spin space. If it commutes with $s_z$, it is of ``Haldane'' type. All other spin-orbit coupling is of Rashba type (because any term that breaks spin conservation in this basis is also not invariant under $C_{2z}I$, when inversion is taken to act in three dimensions. This is true for any two-dimensional system embedded in three dimensional space, c.~f.~ Ref.~\cite{Kane04}) The most general Haldane-type SOC term is
\begin{equation}
H_{HSOC}(\mathbf{k})=d_0(\mathbf{k})s_z\otimes\sigma_0+d_x(\mathbf{k})s_z\otimes\sigma_x+d_y(\mathbf{k})s_z\otimes\sigma_y+d_z(\mathbf{k})s_z\otimes\sigma_z.
\end{equation}
Looking at the $\Gamma$ point first, $C_{2z}$ symmetry forces $d_y(0)=d_z(0)=0$, while mirror symmetry forces $d_0(0)=d_x(0)=0$. Thus, Haldane spin orbit coupling does not affect the band structure at the $\Gamma$ point. At the $K$ point, however, Eq.~(\ref{eq:Kptinducedreps}) shows that mirror symmetry forces $d_0(K)=d_x(K)=0$, and that $C_{3}$ symmetry forces $d_y(K)=0$. Thus, at the $K$ point, Haldane spin orbit coupling takes the general form
\begin{equation}
H_{HSOC}(K)=\lambda_H s_z\otimes\sigma_z.\label{eq:genHaldane}
\end{equation}

We perform the same analysis for Rashba spin-orbit coupling. We start with the most general spin-non-conserving Hamiltonian,
\begin{equation}
H_{R}(\mathbf{k})=\sum_{i=x,y}\sum_{j=0,x,y,z}d_{ij}(\mathbf{k})s_i\otimes\sigma_j
\label{eq:genRashba}
\end{equation}
At the $\Gamma$ point, $H_R(0)=0$, since $C_{3z}$ fails to commute with every term in Eq~(\ref{eq:genRashba}) {(this is perhaps well-known for the standard Rashba term $(\mathbf{k}\times\vec{\sigma})_z$, however here we have shown it is true for \emph{any} spin-nonconserving term that respects the crystal symmetries)}. At the $K$ point, however, $C_{3z}$ symmetry allows $d_{xy}(K)\neq 0$ and $d_{yx}(K)\neq 0$. Furthermore, mirror symmetry forces {$d_{xy}(K)=-d_{yx}(K)$}. Hence, 
\begin{equation}
H_{R}(K)=\lambda_R(s_x\otimes\sigma_y-s_y\otimes\sigma_x)\label{eq:graphenerashbaham}
\end{equation}
{The terms in Eqs~(\ref{eq:genHaldane}) and (\ref{eq:genRashba}) exhaust the space of possible SOC terms.}

We now analyze the effect of spin orbit coupling $H_H=H_{HSOC}+H_R$ on the band structure.
{We have shown already that symmetry prohibits the spin-orbit coupling from altering the band structure at $\Gamma$.}
At $K$, the eigenvalues of $H_H(K)$ are $\delta E_{\pm}=-\lambda_H\pm2\lambda_R$ and $\delta E_{0}=\lambda_H$; {the latter is two-fold degenerate, corresponding to the $\bar{K}_6$ little group representation, while the former correspond to the $\bar{K}_4$ and $\bar{K}_5$ representations.} The associated eigenvectors are
\begin{equation}
\psi_{\pm}=\frac{1}{\sqrt{2}}\left(|\uparrow \mathbf{q}^b_2\rangle\mp|\downarrow\mathbf{q}^b_1\rangle\right), \psi_{01}=|\uparrow\mathbf{q}^b_1\rangle,\psi_{02}=|\downarrow\mathbf{q}^b_2 \rangle
\label{eq:2bbasis}
\end{equation}
First consider the case {$\lambda_R \pm \lambda_H >0$, }
so that $\delta E_+>\delta E_0>\delta E_-$. 
{(We also could have chosen $\lambda_R\pm \lambda_H <0$, in which case $\delta E_+ < \delta E_0 < \delta E_-$.)}
The Dirac cone at $K$ {that exists} without spin orbit coupling thus splits into a twofold degenerate state sandwiched between two non-degenerate states. 
Since $H_H$ does not change the band structure at $\Gamma$, {we will without loss of generality let the $\bar{\Gamma}_9$ representation sit higher in energy than $\bar{\Gamma}_8$ and the $\bar{K}_5$ representation sit above $\bar{K}_4$. Then the induced band representation is four-fold connected, as shown in Fig~\ref{fig:183connected}. Re-ordering the bands at $\Gamma$ or swapping $\bar{K}_4$ and $\bar{K}_5$ does not change this connectivity (as noted previously in Section~\ref{subsec:graphenegraph}). 
By inspecting Table~\ref{table:graphenebrreps}, we confirm that the bands transform under the four-fold connected elementary band rep $\bar{\rho}^{2b}_6\uparrow G$, induced from $G_{\mathbf{q}^b_1}$, as we asserted at the beginning of this section.
Connecting to the graph theory analysis in Sec~\ref{subsec:graphenegraph}, the four-fold band connectivity corresponds to the case of the single null vector in Eq~(\ref{eq:connectedvector}). This is an enforced semimetal phase.
}

In the opposite regime, 
{where ${\rm sgn}(\lambda_R-\lambda_H) = - {\rm sgn}(\lambda_R + \lambda_H)$}, the situation is more interesting. In this case, at the $K$ point, the twofold degenerate $\psi_{0}$ states sit higher {or lower} in energy than the $\psi_\pm$ states. This ordering implies that bands are now only twofold connected, as shown in Fig~\ref{fig:183disconnected}.
{If we assume the $\bar{\Gamma}_9$ representation of $G_\Gamma$ sits higher in energy than $\bar{\Gamma}_8$ and $\lambda_R -\lambda_H<0$ ($\lambda_R + \lambda_H > 0$), then the conduction bands at the $\Gamma$ point transform under the $\bar{\Gamma}_9$ irrep of $G_\Gamma$ (since $H_{HSOC}(0)=0$), while conduction bands at the $K$ point transform under the $\bar{K}_6$ representation of $G_K$.
Additionally, the valence bands transform under the $\bar{\Gamma}_8$ irrep at the $\Gamma$ point and the $\bar{K}_4\oplus\bar{K}_5$ irrep at the $K$ point.}
{Thus, while it is true that all four bands together still transform under the $\bar{\Gamma}_6\uparrow G$ band representation induced from $G_{\mathbf{q}^b_1}$, for these parameters, the valence bands and the conduction bands alone do not form a band representation}.
This possibility was introduced in the graph theory analysis in Sec~\ref{subsec:graphenegraph} and the two disconnected pieces correspond to the two null vectors in Eqs~(\ref{eq:discvector1}) and (\ref{eq:discvector2}).

\begin{figure}
\centering
\subfloat[]{
        \includegraphics[width=2.3in]{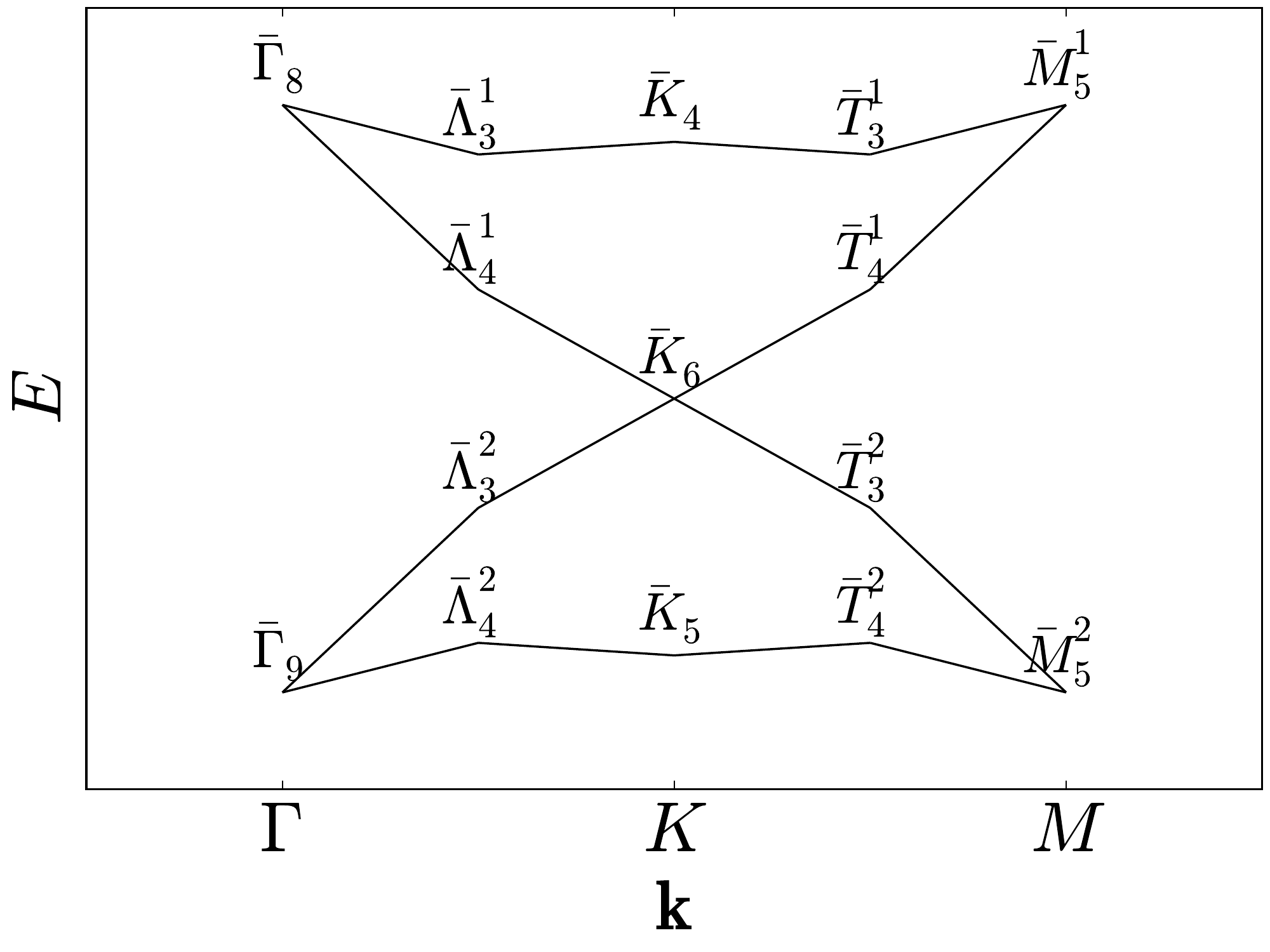}
        \label{fig:183connected}
}
\hspace{.5in}
\subfloat[]{
        \includegraphics[width=2.3in]{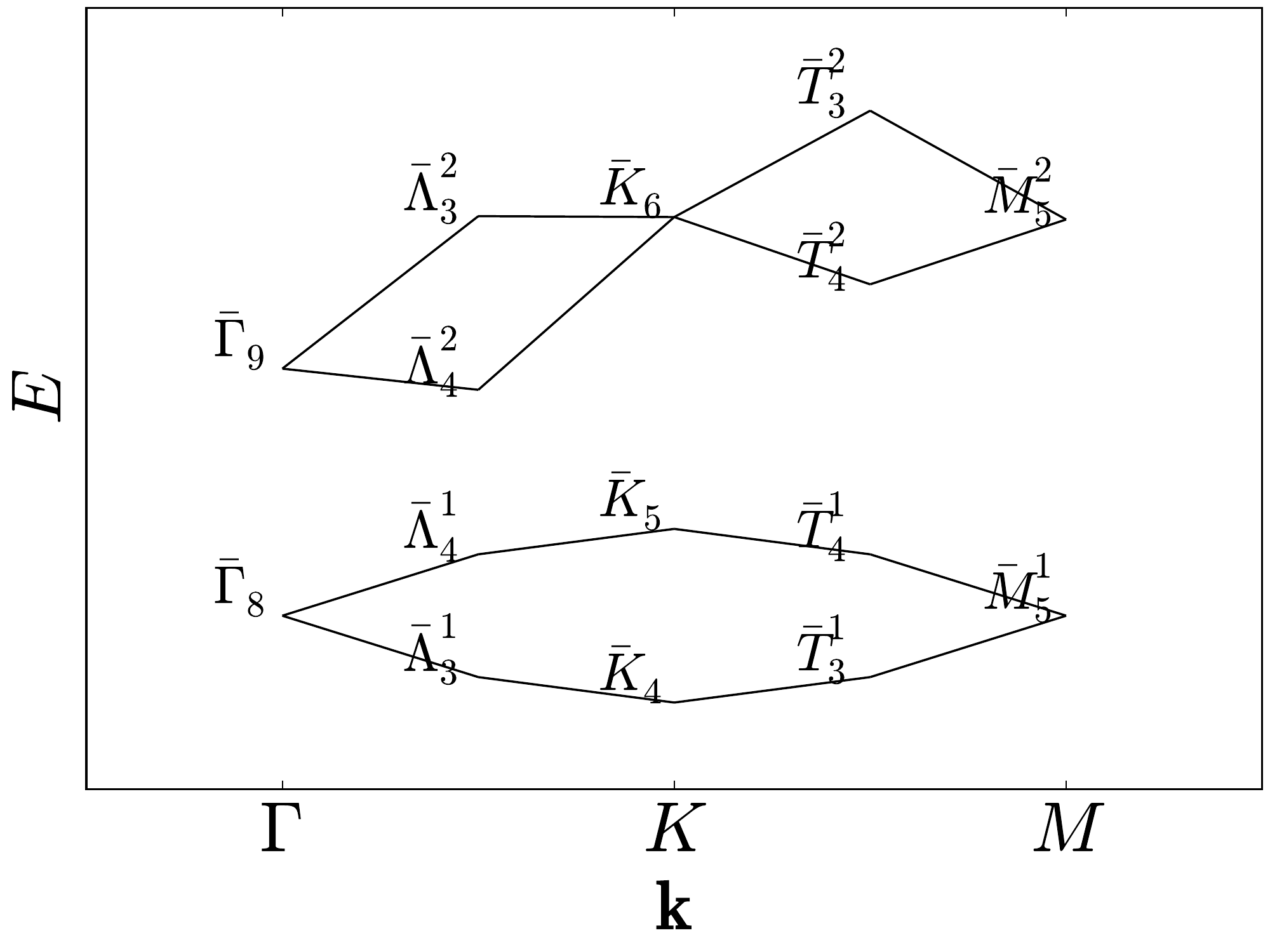}
        \label{fig:183disconnected}
}
     \caption{Band structures corresponding to the connectivity graphs for $P6mm$ (183), with little group representations along points and lines labelled as shown. (a) shows the graph corresponding to the adjacency matrix $A_1$, while (b) shows the graph corresponding to adjacency matrix $A_2$.
}\label{fig:183graphs}
\end{figure}

\section{Hybridization and Topology in the 1D chain}

To elucidate the connection between hybridization, bonding, and topological phases, we will here examine a simple{, well known} model {in a new light}: a 1D inversion symmetric chain with one (spinless) $s$ orbital and one (spinless) $p_x$ orbital per site. While formally equivalent to the Su-Schrieffer-Heeger\cite{ssh1979} and Rice-Mele\cite{RiceMele} models, in this formulation the connection of topology and chemistry is manifested.

We begin by defining our lattice, a schematic diagram of which is shown in Fig.~\ref{fig:1dlattice}.
\begin{figure}[t]
\subfloat[]{
\includegraphics[width=0.7\textwidth]{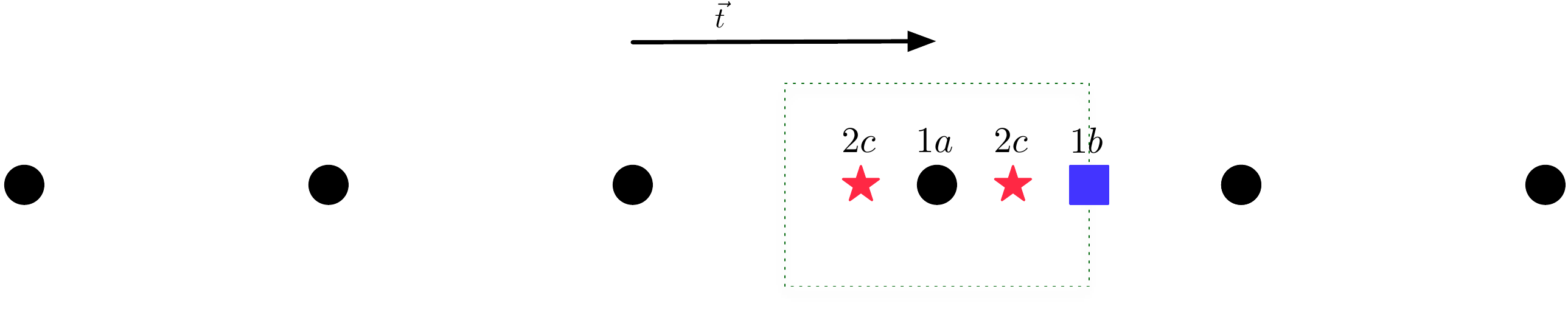}\label{fig:1dlattice}
}\quad
\subfloat[]{
\includegraphics[width=0.3\textwidth]{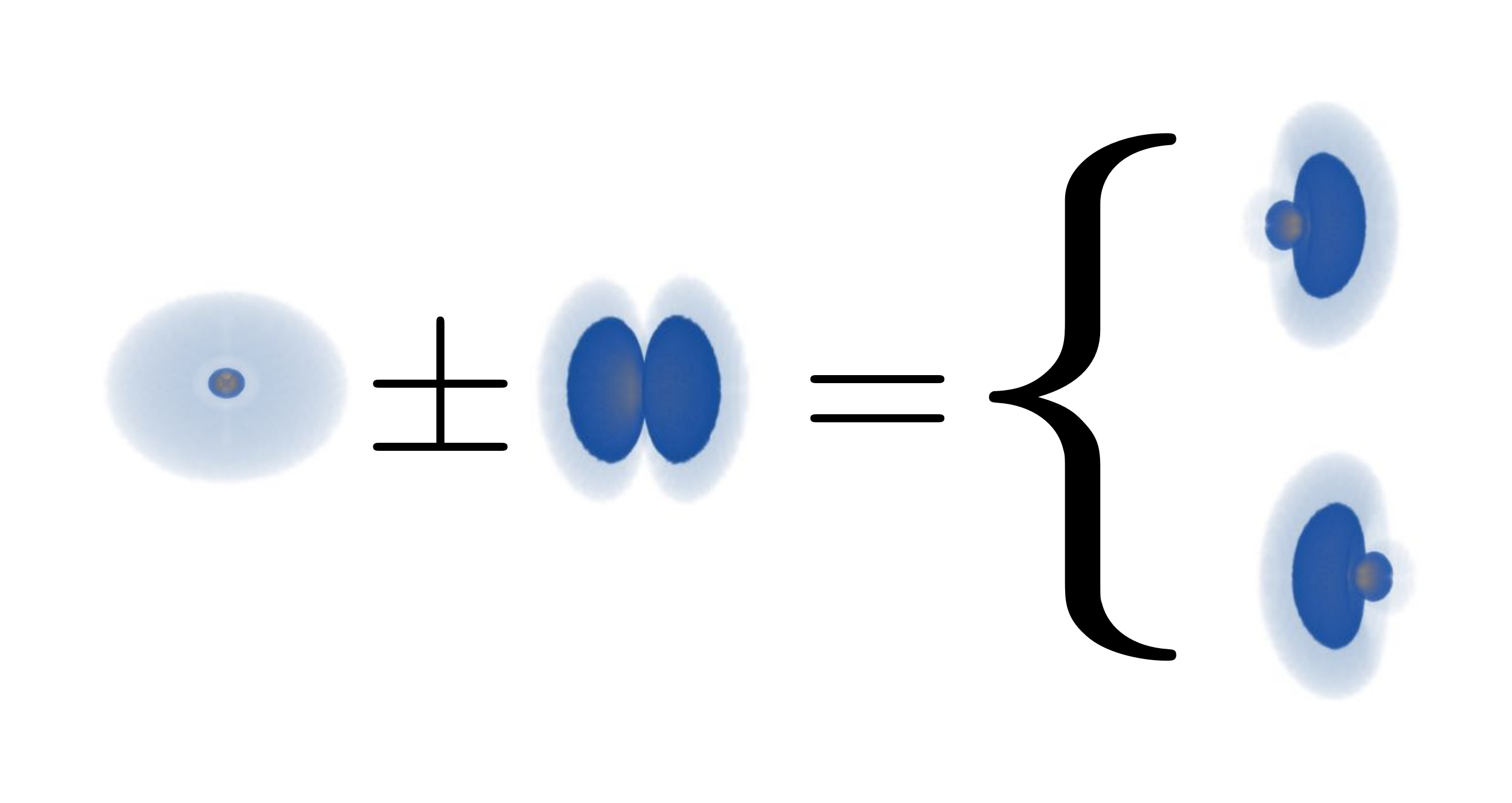}\label{fig:sporbs}	
}
\caption{
The 1D inversion-symmetric chain. (a) shows a schematic diagram of the 1D inversion symmetric lattice, space group $\mathfrak{p} \bar{1}$. The lattice sites are shown in black circles, and the Bravais lattice translation vector is labelled by $\vec{t}$. The green dashed square outlines a single unit cell of the lattice. The lattice site itself serves as our choice of inversion center, and is the $1a$ Wyckoff position. The blue square at the edge of the unit cell is the $1b$ Wyckoff position. The two red stars indicate the points in orbit of the non-maximal $2c$ Wyckoff position. (b) is a schematic representation of $sp$-hybridized orbitals relevant to the transition between the trivial and topological phases. They are obtained as symmetric and antisymmetric linear combination of $s$ and $p$ orbitals.}
\end{figure}
Lattice sites in the figure are represented by black circles. We take as our origin the lattice site within the green dashed {rectangle}, which denotes one unit cell. With this choice of origin, The space group $G$ is generated by
\begin{equation}
G=\langle \{E|\vec{t}\},\{I|0\}, T\rangle,
\end{equation}
where $T$ is spinless time reversal {(although time-reversal is not necessary for the following discussion, we include it here to limit the number of allowed terms in the Hamiltonian)}. There are three distinct Wyckoff positions in the unit cell of this crystal. The first, labelled $1a$, has coordinates $\mathbf{q}^a_1=0$, located at the inversion center. The site symmetry group $G_{\mathbf{q}^a_1}$ is generated by $\{I|0\}$ and $T$. This is a maximal Wyckoff position, since the site symmetry group is isomorphic to the point group of  $\mathfrak{p} \bar{1}$. 
Similarly, the second maximal position is labelled $1b$, and has coordinates $\mathbf{q}^b_1=\half$ in units of the lattice constant. The site-symmetry group $G_{\mathbf{q}^b_1}$ is also isomorphic to the  point group  $\bar{1}$, but now generated by $\{I|1\}$ and $T$.

Finally, there is also the non-maximal Wyckoff position $2c$, with coordinates $\{\mathbf{q}^c_1,\mathbf{q}^c_2\}=\{x,-x\}$. The stabilizer group of either of these sites contains only time-reversal {and the identity element}.

In reciprocal space, the BZ of the crystal is an interval generated by the reciprocal lattice translation $\vec{g}$ satisfying $\vec{g}\cdot\vec{t}=2\pi$. In units of $\vec{g}$, there are two inversion symmetric points in the BZ: $\Gamma=0$ and $X=\half{(\equiv\pi)}$.

Now, we enumerate the elementary band representations for this space group. First, we note that the full point group -- and hence the site-symmetry groups $G_{\mathbf{q}_1^a}$ and $G_{\mathbf{q}_1^b}$ -- have two one-dimensional irreducible representations $\rho_\pm$ distinguished by whether or not the inversion element is represented by $\pm 1$; in both cases time-reversal is represented by complex conjugation. We can carry out the induction procedure for each maximal Wyckoff position. Because the generating element of $G_{\mathbf{q}^a_1}$ contains no translation, the inversion matrix is momentum-independent in band representations induced from this site, and so the inversion eigenvalues at $\Gamma$ and $X$ are identical. Conversely, because the generating element of $G_{\mathbf{q}^b_1}$ contains a lattice translation, the inversion eigenvalues at $\Gamma$ and $X$ differ in parity in band representations induced from this site. We summarize the results in Table~\ref{table:1dbrs}.
\begin{table}[t]
\begin{tabular}{c|c|c|c}
Position & Rep & $\Gamma$ & $X$ \\
\hline
a & $\rho^a_+\uparrow G$ & $+$ & $+$ \Tstrut\\
 & $\rho^a_-\uparrow G$ & $-$ & $-$ \\
 \hline
 \hline
 b & $\rho^b_+\uparrow G$ & $+$ & $-$ \Tstrut\\
  & $\rho^b_-\uparrow G$ & $-$ & $+$
  \end{tabular}
  \caption{Elementary band representations for the 1D space group $\mathfrak{p} \bar{1}$. The first column indicates the Wyckoff position, and the second column the band representation induced from this site. The third column gives the eigenvalue of inversion at the $\Gamma$ point in the BZ. The last column gives the eigenvalue of inversion at the $X$ point.}\label{table:1dbrs}
  \end{table}

 From the table, we note that the composite band representations $(\rho_+^a\oplus\rho_-^a)\uparrow G$ and $(\rho_+^b\oplus\rho_-^b)\uparrow G$ have the same representation content in momentum space. In fact, these composite band representations are equivalent: The intersection $G_{\mathbf{q}^a_1}\cap G_{\mathbf{q}^b_1}=G_{\mathbf{q}^c_1}$ contains only identity and  time-reversal. The unique irrep of this group induces the representations $\rho^a_+\oplus \rho^a_-$ at the $1a$ position, and $\rho^b_+\oplus\rho^b_-$ at the $1b$ position (this follows from the fact that these are the unique two-dimensional site-symmetry representations with zero character of inversion, the so-called \emph{regular representations}\cite{Serre}) Hence this is an equivalence of band representations, which we write as
 \begin{equation}
 (\rho^a_+\oplus\rho^a_-)\uparrow G \approx  (\rho^b_+\oplus\rho^b_-)\uparrow G.\label{eq:1dequiv}
 \end{equation}

 Now let us consider the band structure induced by a single $s$ and single $p_x$ orbital at the $1a$ site of the lattice. Because $s$ orbitals transform in the $\rho_+^a$ representation, while $p$ orbitals transform in the $\rho_-^a$ representation, the full two-band band structure transforms in the $(\rho^a_+\oplus\rho^a_-)\uparrow G$ band representation. Keeping in mind the equivalence Eq.~(\ref{eq:1dequiv}), this means that Wannier functions for the two bands taken together can lie anywhere along the line between the $1a$ and $1b$ position. If the energetics are such that the system is gapped {(which is generically the case in 1D)}, then the exponentially localized valence band {(which is fully occupied at a filling of one electron per unit cell)} Wannier functions will lie \emph{either} on the $1a$ or $1b$ site.

 To see this concretely, let us construct an explicit tight binding model consistent with these symmetries. The most general nearest-neighbor Bloch Hamiltonian induced from an $s$ and a $p$ orbital on the $1a$ site takes the form
 \begin{equation}
 H(k)=-\left[\epsilon+(t_{ss}+t_{pp})\cos (ka)\right]\sigma_z-2t_{sp}\sin (ka)\sigma_y,\label{eq:sshham}
 \end{equation}
where $a$ is the lattice constant, $\epsilon$ is the onsite-energy difference between $s$ and $p$ orbitals, $t_{ss}$ is $s-s$ hopping, $t_{pp}$ is $p-p$ hopping, and $t_{sp}$ is the interorbital hopping {(had we chosen to break time-reversal symmetry, there would be an additional allowed $s-p$ hopping term, which does not affect the fundamental physics)}. This Hamiltonian maps to the Su-Schrieffer-Heeger model\cite{ssh1979} under a rotation $\sigma_z\rightarrow\sigma_x$. In this basis, the inversion matrix is represented as
\begin{equation}
\rho^\mathbf{k}(\{I|0\})=\sigma_z. \label{eq:invmatrix}
\end{equation}

 There are two simple limits of this model which we will analyze. First, consider the case where $t_{ss}=t_{sp}=t_{pp}=0$. In this limit, the spectrum is gapped and given by

\begin{equation}
E_\pm(k)=\pm\epsilon.
\end{equation} 
This limit corresponds to decoupled atoms, and so we expect the valence band Wannier functions to be localized on the $1a$ sites. We can verify this by computing the Wannier center polarization from the Zak phase\cite{Zakphase,ksv}. Indeed, the occupied band eigenfunction 
\begin{equation}
\psi_0(k)=\left(\begin{array}{c}
1 \\
0
\end{array}\right)
\end{equation}
is $k$-independent and periodic, so the Zak phase is 0. From this we deduce that the Wannier functions are localized at the origin of the unit cell, which here is the $1a$ site. We find the same result for the conduction band. In this phase then the Wannier functions are nearly identical to the original atomic orbitals. To see this another way, comparing with Eq.~(\ref{eq:invmatrix}) shows that the valence band transforms in the $\rho_+^a\uparrow G$ band representation, and so the occupied Wannier functions are $s$-like orbitals localized at the $1a$ site.

Now let us consider the opposite limit $\epsilon=0, t_{ss}=t_{pp}=t_{sp}=t$. We find that the spectrum is also flat, with
\begin{equation}
E_{\pm}=\pm2t,
\end{equation}
However now the valence band eigenfunctions are nontrivial. Since the Hamiltonian is a sum of Pauli matrices, we can write immediately that
\begin{equation}
\psi_-(k)=e^{ika/2}\left(
\begin{array}{c}
\cos\frac{ka}{2} \\
i\sin\frac{ka}{2}
\end{array}\right),
\end{equation}
where we have included a prefactor to make the wavefunction periodic. We find for the Zak phase
\begin{equation}
\phi=i\int_{-\pi}^\pi dk \psi_-^\dag \nabla_k\psi_- =\pi,
\end{equation}
from which we deduce that the valence band Wannier functions are localized a half-translation from the origin of the unit cell, which here is at the $1b$ site. We draw the same conclusion by examining the conduction band Wannier functions. Additionally, comparing with Eq.~(\ref{eq:invmatrix}), we see that the occupied band Wannier functions transform in the $\rho_+^{b}\uparrow G$ band representation, and so are $s$-like orbitals centered on the $1b$ site. This is {what is routinely called the ``topological phase''} of the $1D$ chain. {However, based on our Definition 1 in the main text, it represents just another atomic limit; it can be described by symmetric, localized Wannier functions. Between the two atomic limits, there is a phase transition, and hence an ``edge mode,'' as was first pointed out by Shockley\cite{Shockley1939}. The nonzero polarization we have computed is the bulk signature of this edge mode}. We show this for generic parameter values in Figs.~\ref{fig:sshtrivedge}-\ref{fig:sshbulk}. This proves that edge modes can occur at the interface between distinct atomic limits, however they can be pushed into the bulk spectrum in these cases via an appropriate edge potential.

\begin{figure}[t]
\subfloat[]{
\includegraphics[height=2.3in]{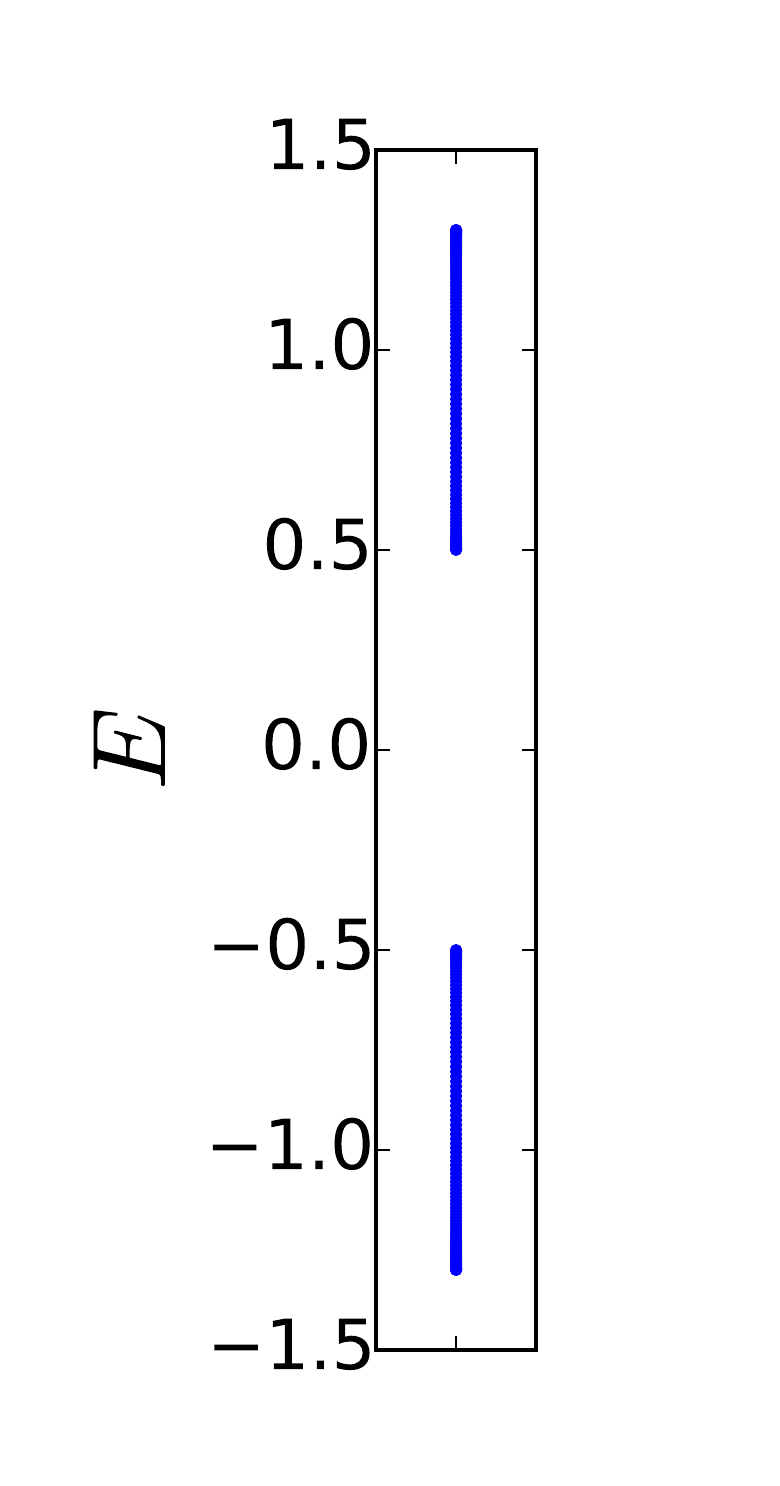}\label{fig:sshtrivedge}
}\quad
\subfloat[]{
\includegraphics[height=2.3in]{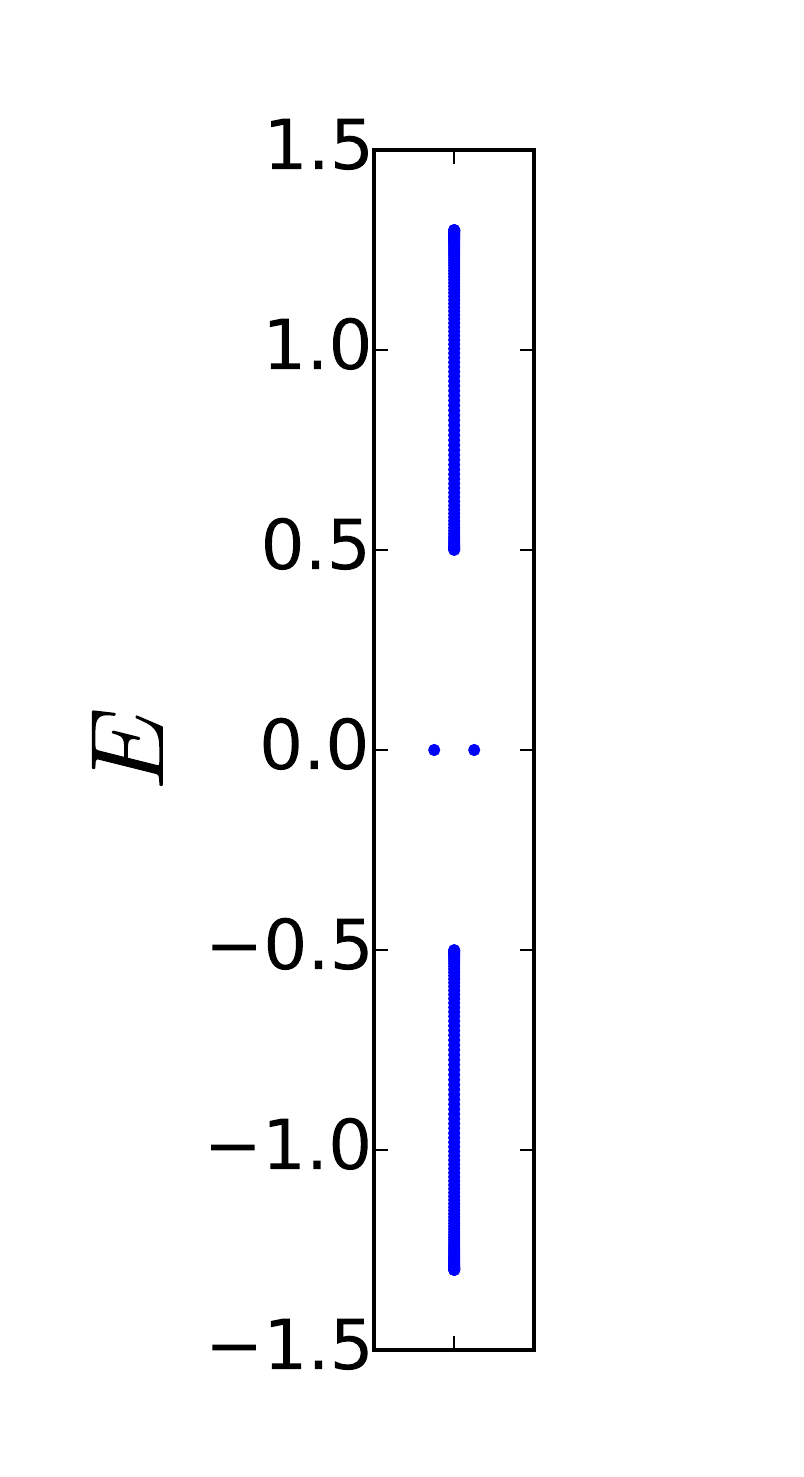}\label{fig:sshtopedge}
}
\subfloat[]{
\includegraphics[height=2in]{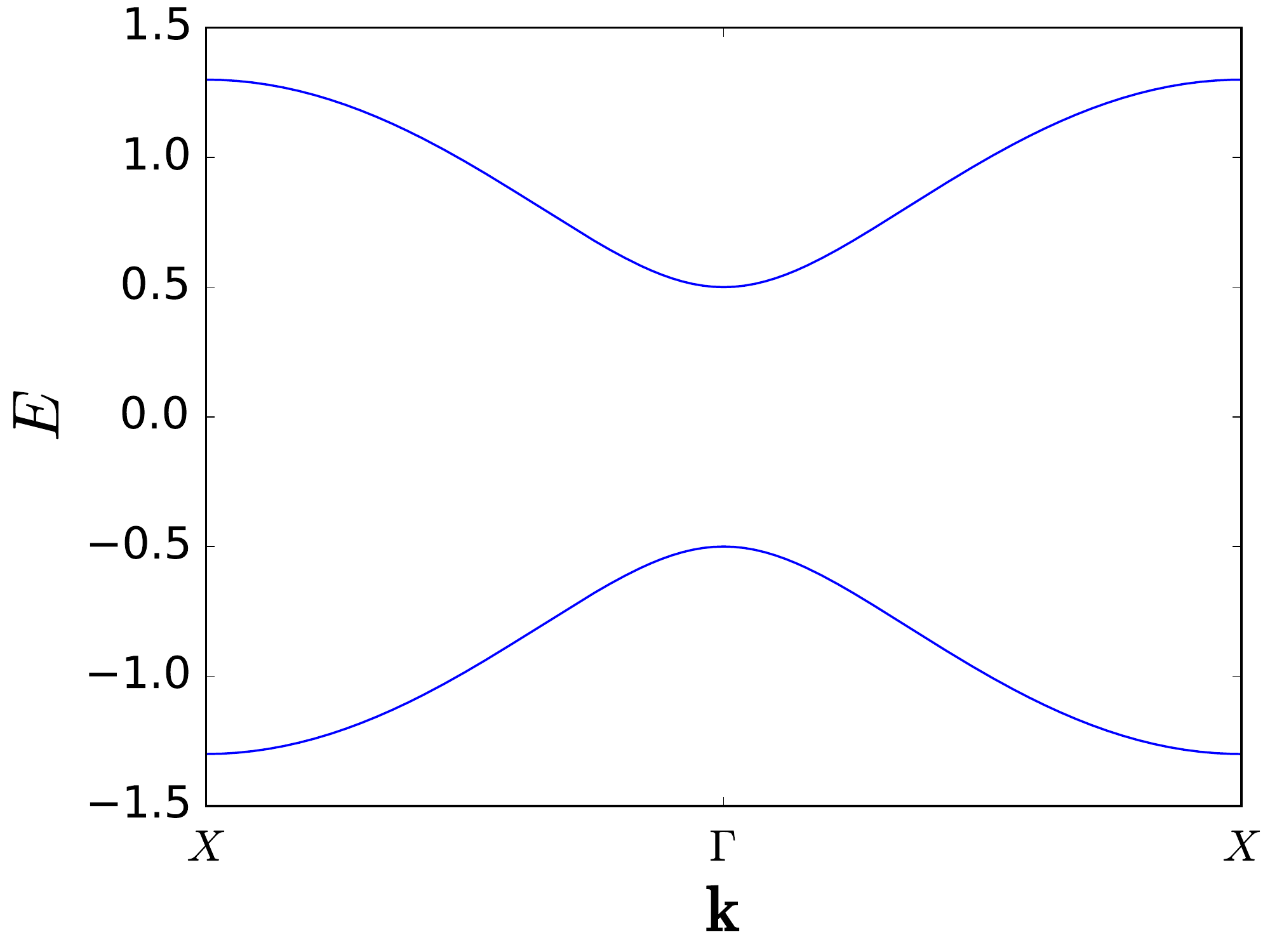}\label{fig:sshbulk}
}\quad
\caption{Spectra for the $1D$ inversion symmetric chain in finite and infinite size. {Because we included only nearest-neighbor hopping and neglected a constant on-site energy in the Hamiltonian~(\ref{eq:sshham}), there is an additional inessential particle-hole symmetry in the spectrum.} (a) shows the spectrum for a chain of $100$ sites in the trivial phase with $\epsilon=0.9$ and $t_{ss}=t_{sp}=t_{pp}=0.2$; note that the spectrum is fully gapped. (b) shows the spectrum for a chain of $100$ sites in the topological phase with $t_{ss}=t_{sp}=t_{pp}=0.45$ and $\epsilon=0.4$. There are a pair of topological edge states, one localized on either edge of the chain. (c) shows the bulk spectrum, which is identical in the two cases.}

\end{figure}

As they are localized at the center of the unit cell, the Wannier functions in the topological phase clearly do not derive from the original atomic orbitals. How then, are we to understand them? The answer comes from chemistry. Given an $s$ and $p$ atomic orbital with nearly the same energy (i.e. $\epsilon\approx 0$), it is convenient to form $sp$ \emph{hybrid orbitals}, {as shown in Fig.~\ref{fig:sporbs}}. When the hopping amplitudes $t$ are also large compared to $\epsilon$, chemical theory tells us we should work in the basis of molecular bonding and antibonding orbitals formed by taking linear combinations of $sp$ orbitals on adjacent atoms. These molecular orbitals are also inversion symmetric, and therefore lie exactly halfway between the atoms, at the $1b$ site\cite{Anderson1984,hoffmann1987chemistry}! We thus see that the topological phase transition between trivial and nontrivial phases of the chain is a \emph{chemical} transition between weak and strong covalent bonding, where the formation of molecular bonding orbitals leads to a quantized charge polarization and edge states.

\section{Survey of material predictions}

{With our theory now developed, we move on to apply our method to find new topological materials. The strategy for this search has already been presented in the main text. Here we summarize our early findings. For topological insulators, we have identified new, broad classes of materials. The first, {Cu$_2$ABX$_4$}, with A$=$Ge,Sn,Sb, B$=$Zn,Cd,Hg,Cu, and X$=$S,Se,Te, was introduced in the main text. There are a total of $36$ materials in this structure type. These compounds are analyzed in detail in Subsection~\ref{subsec:cu2abx4}. 

In Subsection~\ref{subsec:bisquare}, we examine a large class of layered materials with square nets of As, Bi, and Sb. {These fall into two space groups. First, in  $P4/nmm$ (129) there is the class WHM with W$=$Ti,Zr,Hf, or a rare earth metal, H$=$Si,Ge,Sn,Pb, and M$=$O,S,Se,Te as well as the class ACuX$_2$ with A a rare earth metal and X$=$P,As,Sb,Bi. A small number of materials in these families have been shown to be topological before\cite{129tis,zrsnte}, however here we will present a general group-theoretic argument for why they all \emph{must} be topological generically. These arguments additionally allow us to identify 58 new topological insulator candidates in the distorted  $Pnma$ (62): LaSbTe\cite{latesbref}, SrZnSb$_2$, and AAgX$_2$ with A a rare earth metal and X$=$P,As,Sb,Bi.}

In Subsection~\ref{subsec:metal} we present realizations of \emph{sixteen-fold} connected metals, where crystal symmetries force sixteen bands to be connected throughout the BZ. These metals can realize exotic filling fractions ($7/8$ in our example) which may allow for interesting phenomena when interactions are included.

{Using our method, we were also able to identify several other interesting classes of compounds, a detailed analysis of which we defer to future work so as not to overburden the reader. First, we have identified three new Dirac semimetals IrTe$_2$\cite{irte2ref,Pascut2014}, NiTe$_2$, and HfTe$_2$ in  $P\bar{3}m1$ (164), the symmetry group of buckled graphene). While Dirac semimetals similar material families have been analyzed recently by others\cite{ptse2dirac}, here we have used our powerful connectivity theory to find candidate materials with Dirac points at or very near the Fermi level, as shown in Fig.~\ref{fig:IrTe2}. Also in this space group, we identify CNb$_2$\cite{Cnb2ref} as a promising topological insulator candidate. We show its band structure in Fig.~\ref{fig:CNb2}. Additionally, we have identified topological bands below the Fermi level in Pb$_2$O\cite{pb20ref} in  $Pn\bar{3}m$ (224), shown in Fig.~\ref{fig:Pb20}. Furthermore, we predict that under uniaxial strain in the $z$-direction, the strucure distorts to  $P4_2/nnm$ (134), and a topological gap opens near the Fermi level. This is shown in Fig.~\ref{fig:Pb20strain}. Lastly, we find a candidate for a \emph{24-fold connected symmetry protected semimetal}, Cu$_3$TeO$_6$\cite{24foldref}, in  $Ia\bar{3}$ (206). In this material, a twenty-four band EBR is half-filled at the Fermi level, realizing the most interconnected EBR allowed by symmetry. We show the band structure in Fig.~\ref{fig:24fold}.} Additional candidates for exotic metals can be found in Table~\ref{table:semimetals}.
\begin{figure}[t]
\subfloat[]{
	\includegraphics[height=1.6in]{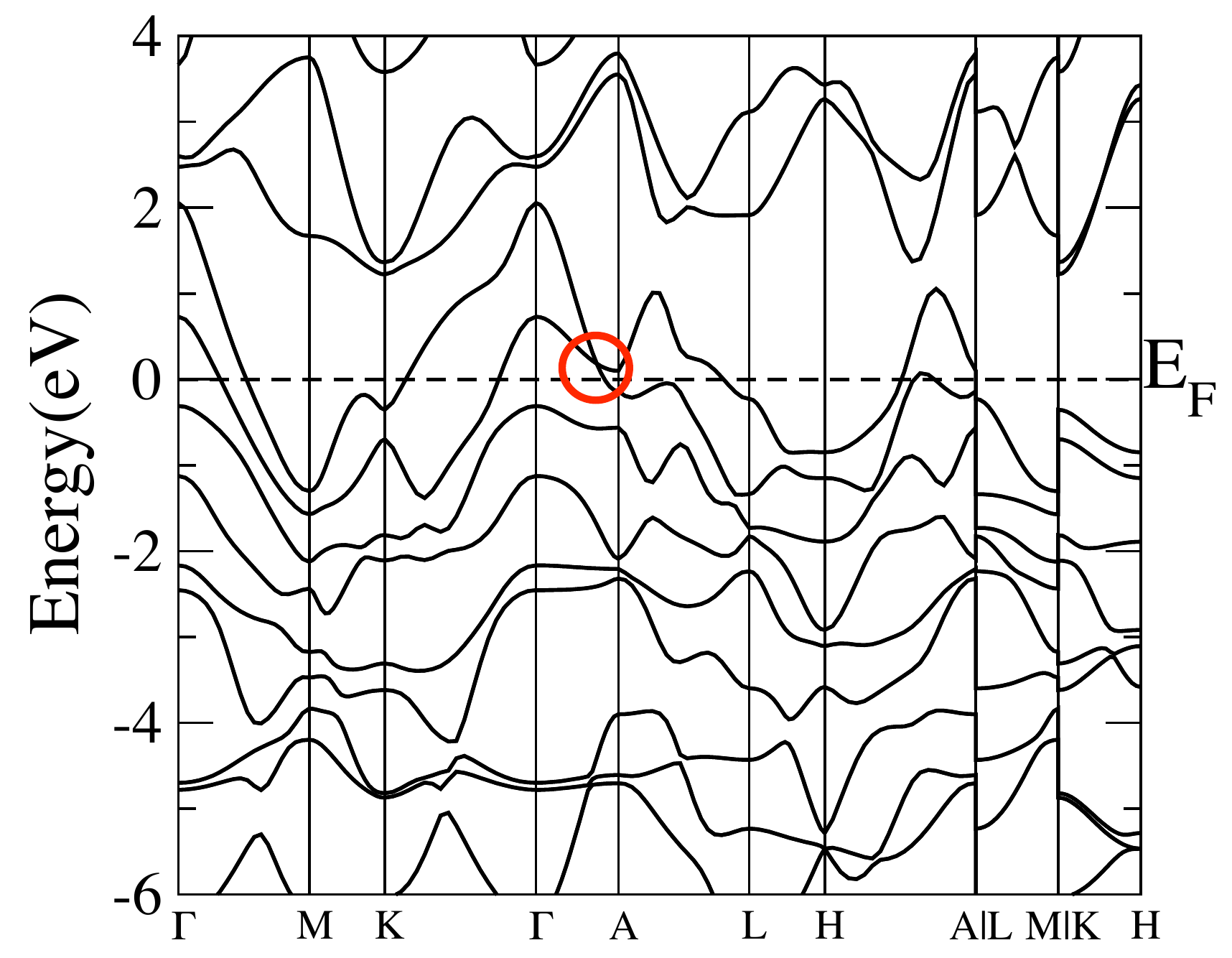}\label{fig:IrTe2}
}\quad
\subfloat[]{
	\includegraphics[height=1.6in]{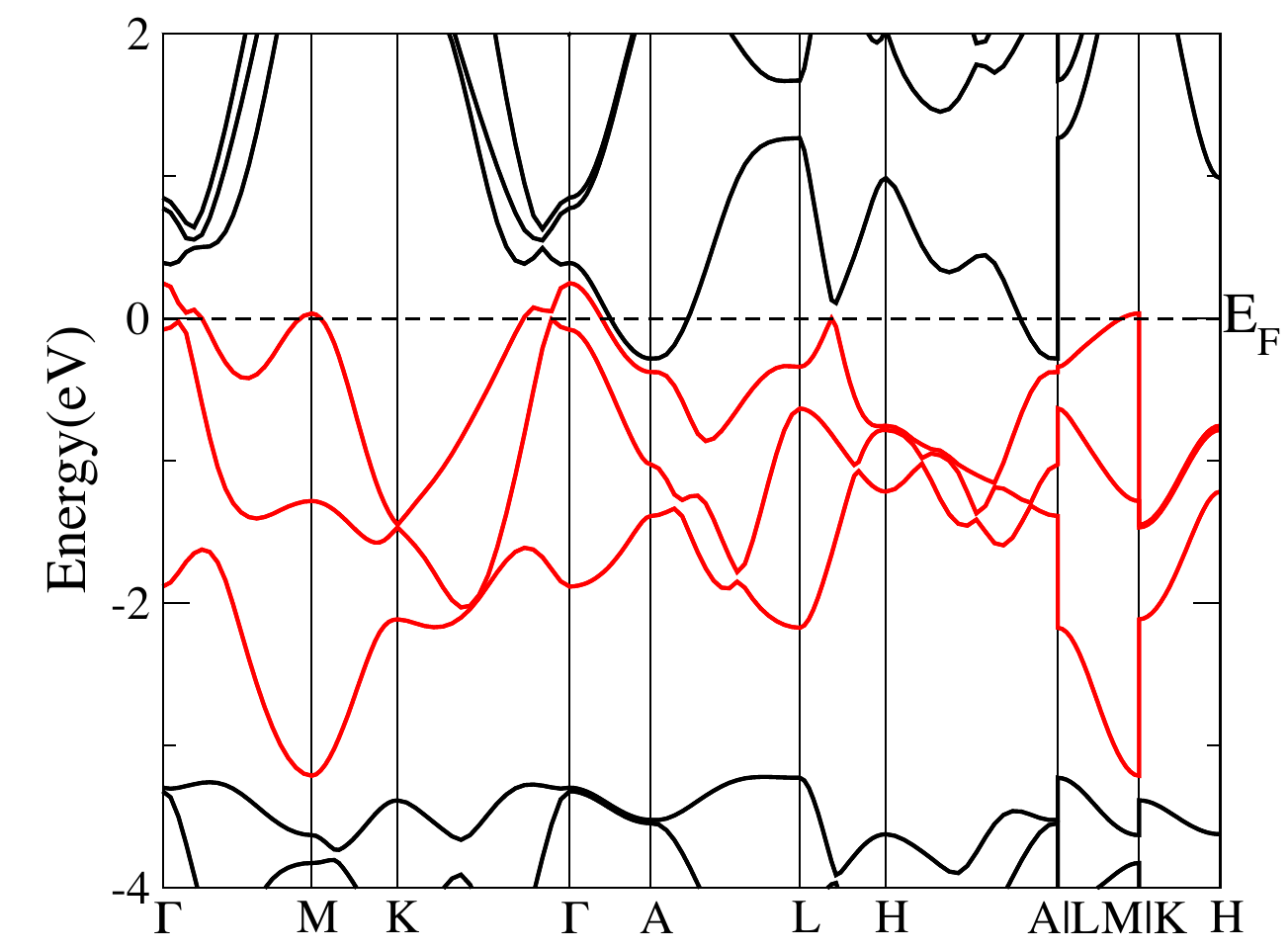}\label{fig:CNb2}
}\quad
\subfloat[]{
	\includegraphics[height=1.6in]{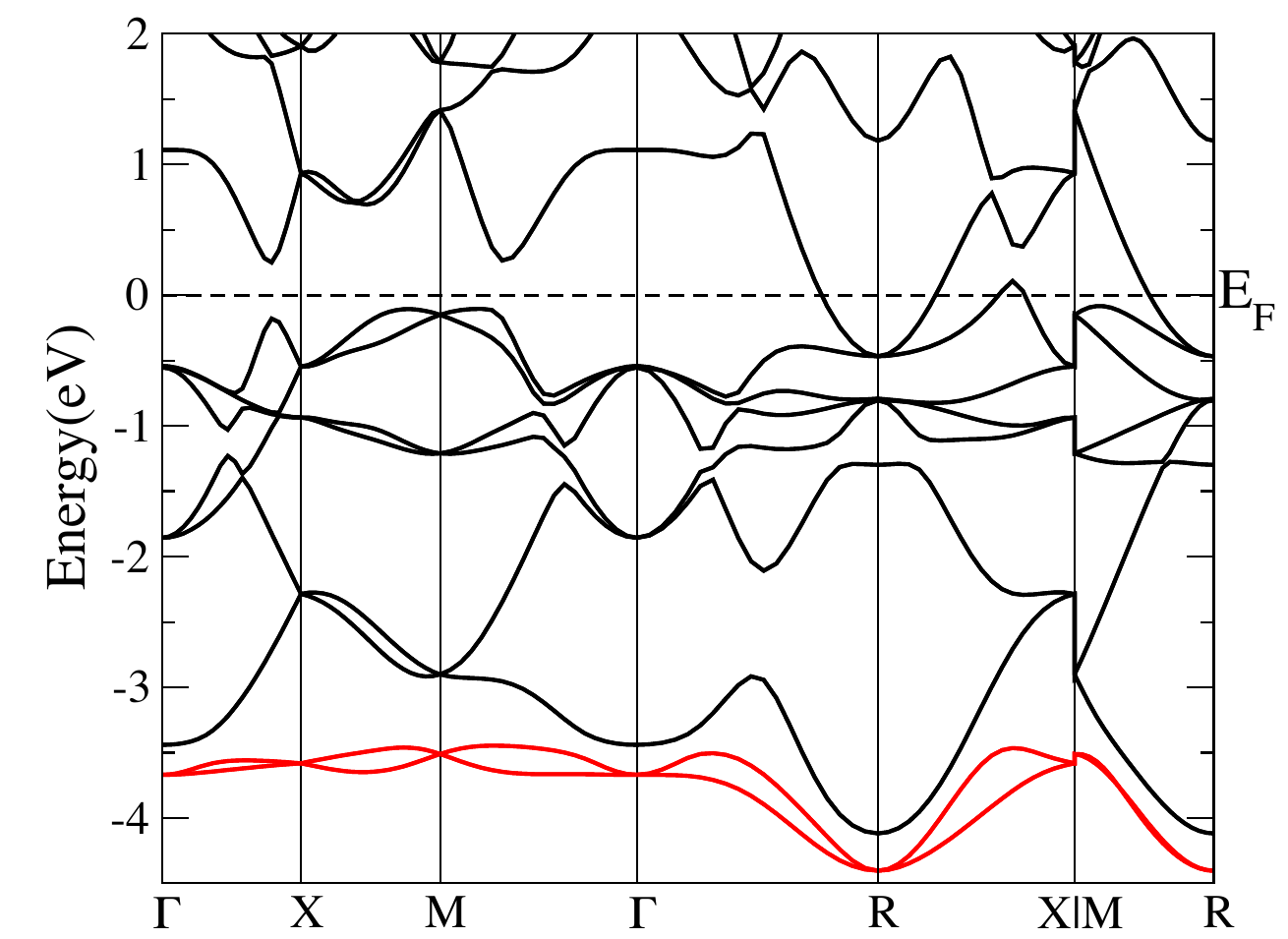}\label{fig:Pb20}
}\quad
\subfloat[]{
	\includegraphics[height=1.6in]{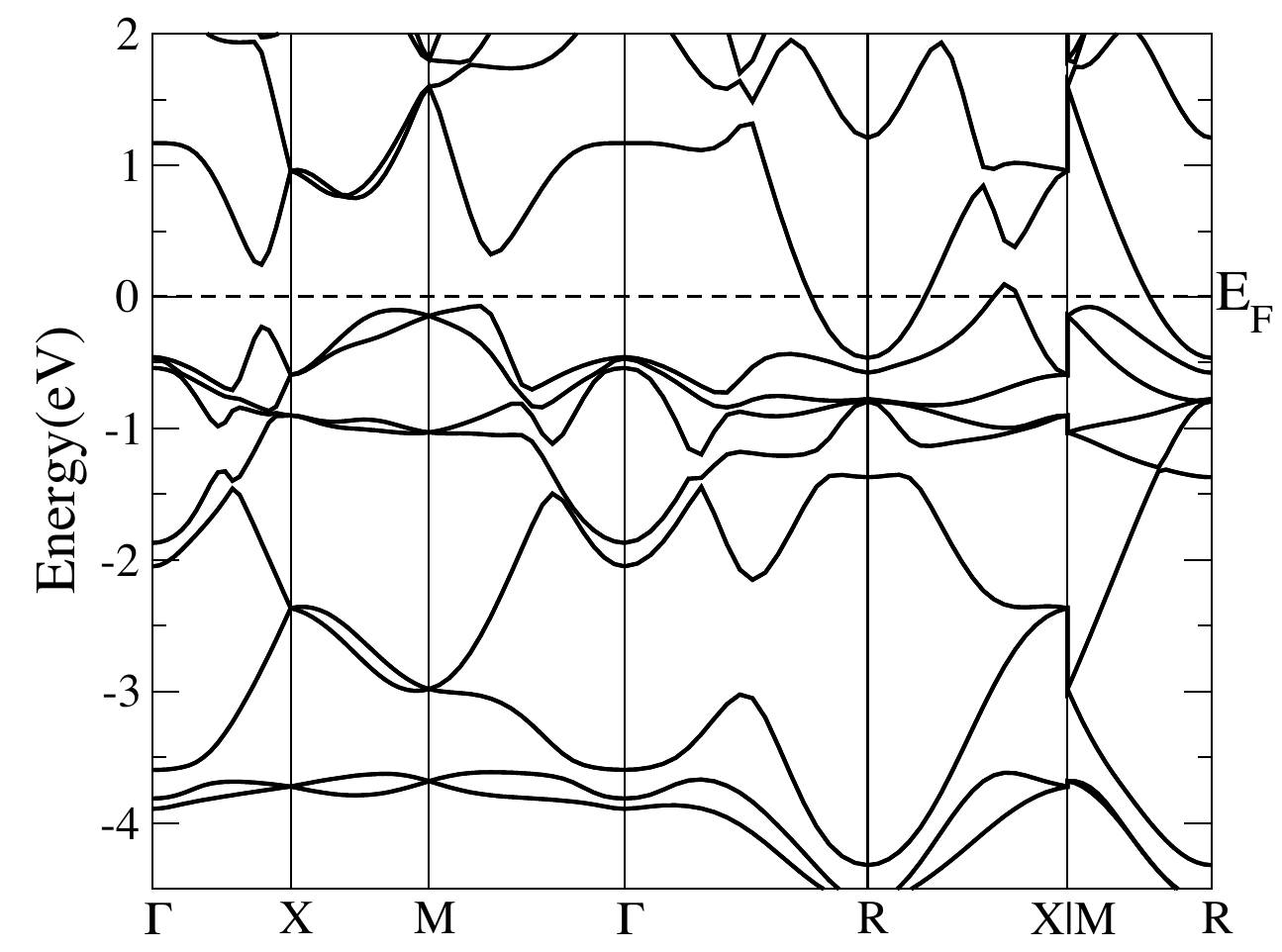}\label{fig:Pb20strain}
}\quad
\subfloat[]{
	\includegraphics[height=1.6in]{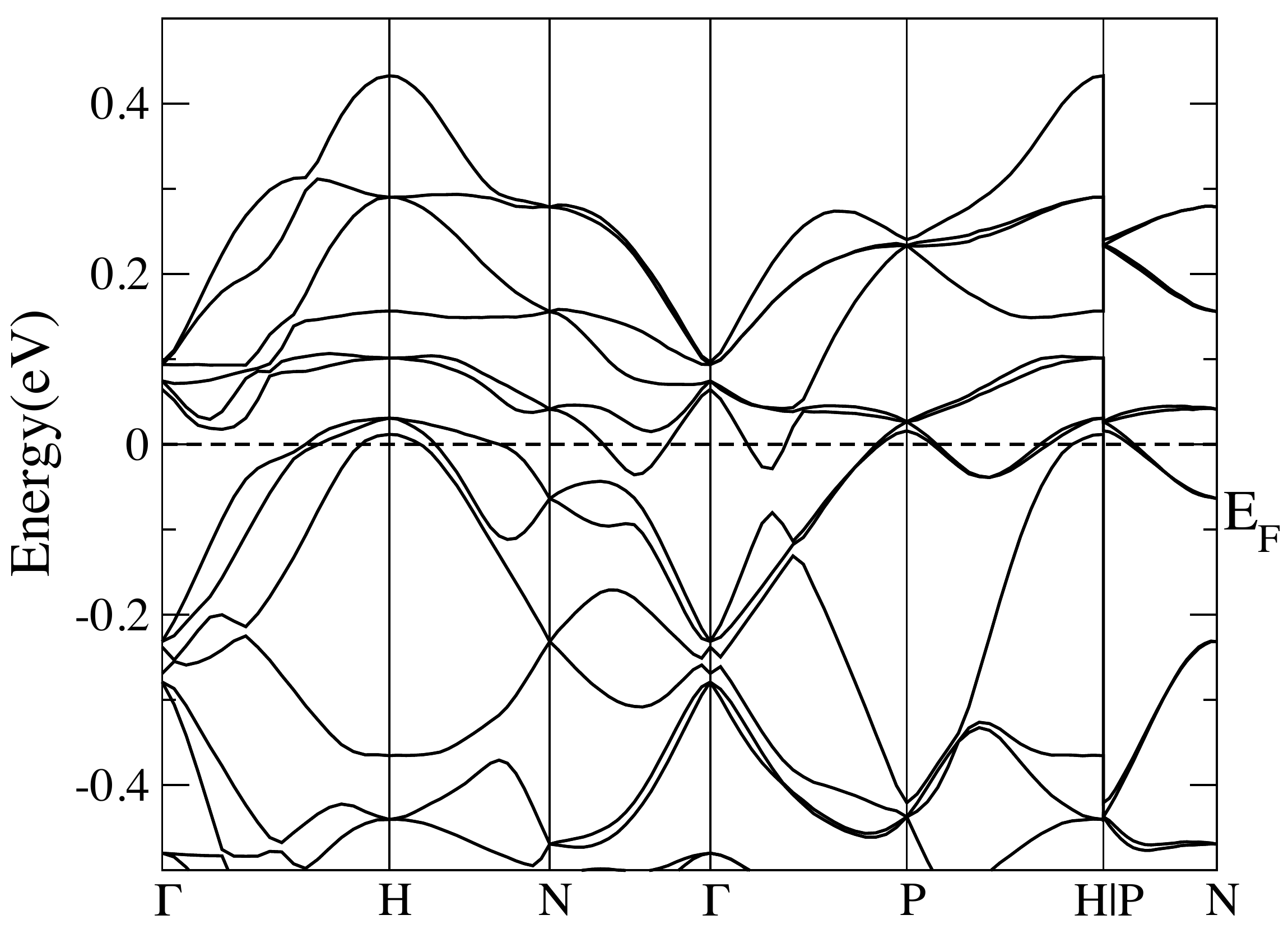}\label{fig:24fold}
}
\caption{Band structures for new topological insulators and semimetals. (a) shows the band structure for IrTe$_2$ in  $P\bar{3}m1$ (164). The red circle highlights the type-II Dirac point near the Fermi level. (b) Shows the band structure for the narrow-gap weak topological insulator CNb$_2$ in the same space group, with the topologically nontrivial valence bands shown in red. (c) gives the band structure for unstrained Pb$_2$O in  $Pn\bar{3}m$ (224). The isolated group of bands near $-3.5$eV shown in red does not form a BR, and hence are topological. (d) gives the band structure of Pb$_2$O under uniaxial strain, which opens a topological gap near the Fermi level. Finally, (e) gives the band structure for Cu$_3$TeO$_6$ in  $Ia\bar{3}$ (206). The twenty-four bands at the Fermi level in this material are half filled, and form the highest-dimensional PEBR allowed for any of the $230$ space groups.
}
\end{figure}

\subsection{Cu$_2$ABX$_4$}\label{subsec:cu2abx4}
The Cu$_2$ABX$_4$ materials all belong to the symmorphic tetragonal space group I$\bar{4}2$m (121). This group is body-centered, and so we take for a basis of lattice vectors
\begin{equation}
\mathbf{e}_1=\half(-a\hat{\mathbf{x}}+a\hat{\mathbf{y}}+c\hat{\mathbf{z}}), \; \mathbf{e}_2=\half(a\hat{\mathbf{x}}-a\hat{\mathbf{y}}+c\hat{\mathbf{z}}),\; \mathbf{e}_3=\half(a\hat{\mathbf{x}}+a\hat{\mathbf{y}}-c\hat{\mathbf{z}}).\label{eq:bctvecs}
\end{equation}
In addition to these translations, the space group is generated by a fourfold roto-inversion $IC_{4z}\equiv S_{4}^-$ about the $z$-axis, and the rotation $C_{2x}$ about the $x$-axis. There are four maximal Wyckoff positions, labelled $2a,2b$, $4c,$ and $4d$ (divided by $2$ for the primitive cell description given in Eq.~(\ref{eq:bctvecs})).  The coordinate triplets of the symmetry equivalent points in the unit cell, with respect to Eq.~(\ref{eq:bctvecs}), are given by:
\begin{align}
\mathbf{q}^{2a}&=(0,0,0),\\
\mathbf{q}^{2b}&=(\half,\half,0),\\
\{\mathbf{q}^{4c}_1,\mathbf{q}^{4c}_2\}&=\{(\half,0,\half),(0,\half,\half)\},\\
\{\mathbf{q}^{4d}_1,\mathbf{q}^{4d}_2\}&=\{(\frac{3}{4},\frac{1}{4},\half),(\frac{1}{4},\frac{3}{4},\half)\}.
\end{align}
with stabilizer groups
\begin{align}
G_{\mathbf{q}^{2a}_1}&\approx G_{\mathbf{q}^{2b}_1}\approx D_{2d} \\
G_{\mathbf{q}^{4c}_1}&\approx D_2 \\
G_{\mathbf{q}^{4d}_1}&\approx S_4.
\end{align}
We note that the $4c$ position with site-symmetry group $D_2$ is exceptional as per Table~\ref{table:dbr}, althoug this will not play a role here. We also will need to consider the non-maximal $8i$ Wyckoff position, with {coordinates}
\begin{equation}
\{\mathbf{q}^{8i}_j\}=\{(x+z,x+z,2x),(z-x,z-x,-2x),(-x-z,x-z,0),(x-z,-x-z,0)\}
\end{equation}
and stabilizer group $C_s${, generated by a single mirror}. As this group is a proper subgroup of the stabilizers $G_{\mathbf{q}^{2a}_1}$ and $G_{\mathbf{q}^{2b}_1}$, composite band representations induced from the $8i$ can be labelled by sums of elementary band representations from either the $2a$ or $2b$ positions. We show the crystal structure for these compounds in Fig.~\ref{fig:cu2abx4}. Furthermore, the character table for the group $D_{2d}$ is shown in Table~\ref{table:D2d}. We will also need the repsresntations of $G_{\mathbf{q}_1^{4d}}\approx S_4$. Since this is an abelian group generated by the single element $IC_{4z}$, all of its representations are one dimensional, and specified by the character $\chi(IC_{4z})$. We list these in Table~\ref{table:s4} below.
\begin{table}[h]
\begin{tabular}{c|c|c|c|c|c|c}
Rep & $E$ & $C_{2z}$ & $IC_{4z}$ & $C_{2x}$ & $m_{110}$ & $\bar{E}$ \\
\hline
$\rho^{2a}_1$ & $1$ &  $1$ &  $1$ &  $1$ &  $1$ & $1$ \Tstrut\\
$\rho^{2a}_2$ & $1$ &  $1$ & $-1$ &  $1$ & $-1$ & $1$ \\
$\rho^{2a}_3$ & $1$ &  $1$ & $-1$ & $-1$ &  $1$ & $1$ \\
$\rho^{2a}_4$ & $1$ &  $1$ &  $1$ & $-1$ & $-1$ & $1$ \\
$\rho^{2a}_5$ & $2$ & $-2$ &  $0$ &  $0$ &  $0$ & $2$ \\
$\bar{\rho}^{2a}_6$ & 2 & 0 & $-\sqrt{2}$ & $0$ & $0$ & $-2$ \\
$\bar{\rho}^{2a}_7$ & 2 & 0 & $\sqrt{2}$ & $0$ & $0$ & $-2$ 
\end{tabular}
\caption{Character table for the point group $D_{2d}$, which is the stabilizer group of both the $2a$ and $2b$ positions in  $I\bar{4}2m$ (121)}\label{table:D2d}
\end{table}
\begin{table}[h]
\begin{tabular}{L|L}
\mathrm{Rep} & IC_{4z} \\
\hline
\rho^{4d}_1 & 1 \Tstrut\\
\rho^{4d}_2 & -1 \\
\rho^{4d}_3 & i \\
\rho^{4d}_4 & -i \\
\bar{\rho}^{4d}_5 & e^{3\pi i /4} \\
\bar{\rho}^{4d}_6 & e^{7\pi i /4} \\
\bar{\rho}^{4d}_7 & e^{5\pi i /4} \\
\bar{\rho}^{4d}_8 & e^{i\pi/4}
\end{tabular}
\caption{Character table for the point group $S_4$, which is the stabilizer group of the $4d$ position in  $I\bar{4}2m$ (121)}\label{table:s4}
\end{table}
\begin{figure}[t]
\centering
\includegraphics[height=2.5in]{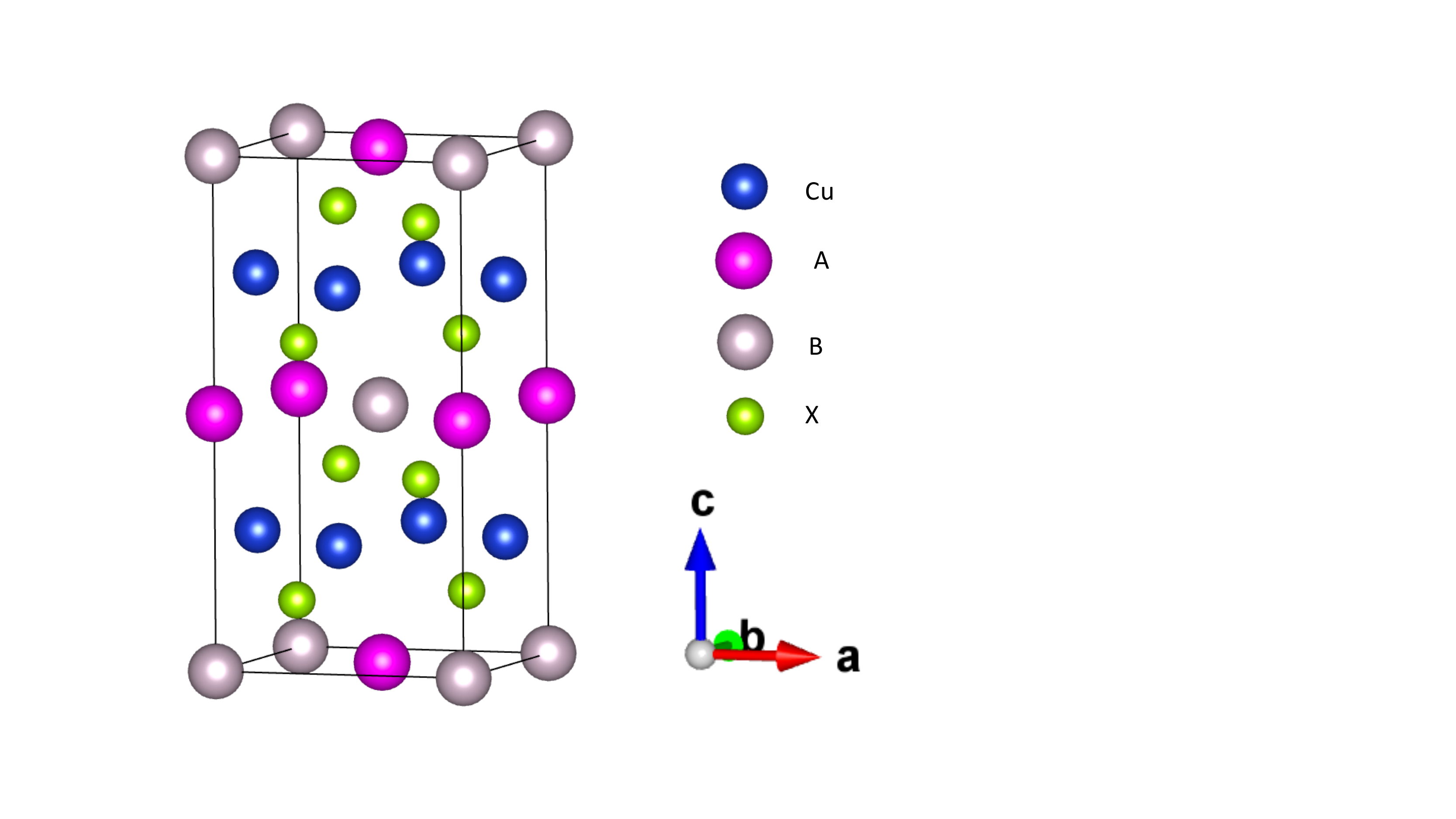}
\caption{Crystal structure of the Cu$_2$ABX$_4$ class of compounds, with A$=$Ge,Sn,Sb, B$=$Zn,Cd,Hg,Cu, and X$=$S,Se,Te. The blue circles represent the Cu atoms at the $4d$ position, which contribute at the Fermi level.}\label{fig:cu2abx4}
\end{figure}

In the particular compounds of interest, the Cu atoms sit at the $4d$ position, the A atoms sit at $2b$, the B atoms at $2a$, and the X atoms at $8i$. By consulting Section~\ref{sec:data}, we see that the elementary band representations induced from the {one-dimensional representations of the stabilizer group of the} $4d$ position respect time-reversal symmetry in momentum space. Because of this, we know from Section~\ref{sec:bandreps} that the physically elementary band representations induced from this site can be disconnected, and hence topological. Furthermore, ab-initio calculations reveal that in this material class, the relevant states near the Fermi level come from $d$ orbitals at the $4d$ position, and $p$ orbitals at the $8i$ position. Our discussion in the main text thus flags this group of materials as prime candidates for topological insulators.

We focus below on three cases {out of this large class of $36$ materials}. First, there is Cu$_2$GeZnS$_4$. Ab initio calculations reveal this to be a large gap (trivial) insulator without spin-orbit coupling, and hence it will remain so for weak SOC. Indeed, we find that the {eighty-four} valence bands nearest the Fermi level transform according to the physical composite band representation $(6\bar{\rho}^{2b}_6\oplus 5\bar{\rho}^{2b}_7\oplus 3\bar{\rho}^{4d}_5\oplus 2\bar{\rho}^{4d}_6\oplus3\bar{\rho}^{4d}_7\oplus 2\bar{\rho}^{4d}_8)\uparrow G$, while the {lowest lying} conduction band transforms according to the physically elementary $\bar{\rho}^{2b}_7\uparrow G$ band representation. In this case, all band representations induced from the $1D$ representations of $G_{\mathbf{q}^{4d}_1}$ are "occupied", and hence the material is topologically trivial. We show the band structure for this material in Fig.~\ref{fig:Cu2GeZnS4}

\begin{figure}[t]
\includegraphics[height=2.5in]{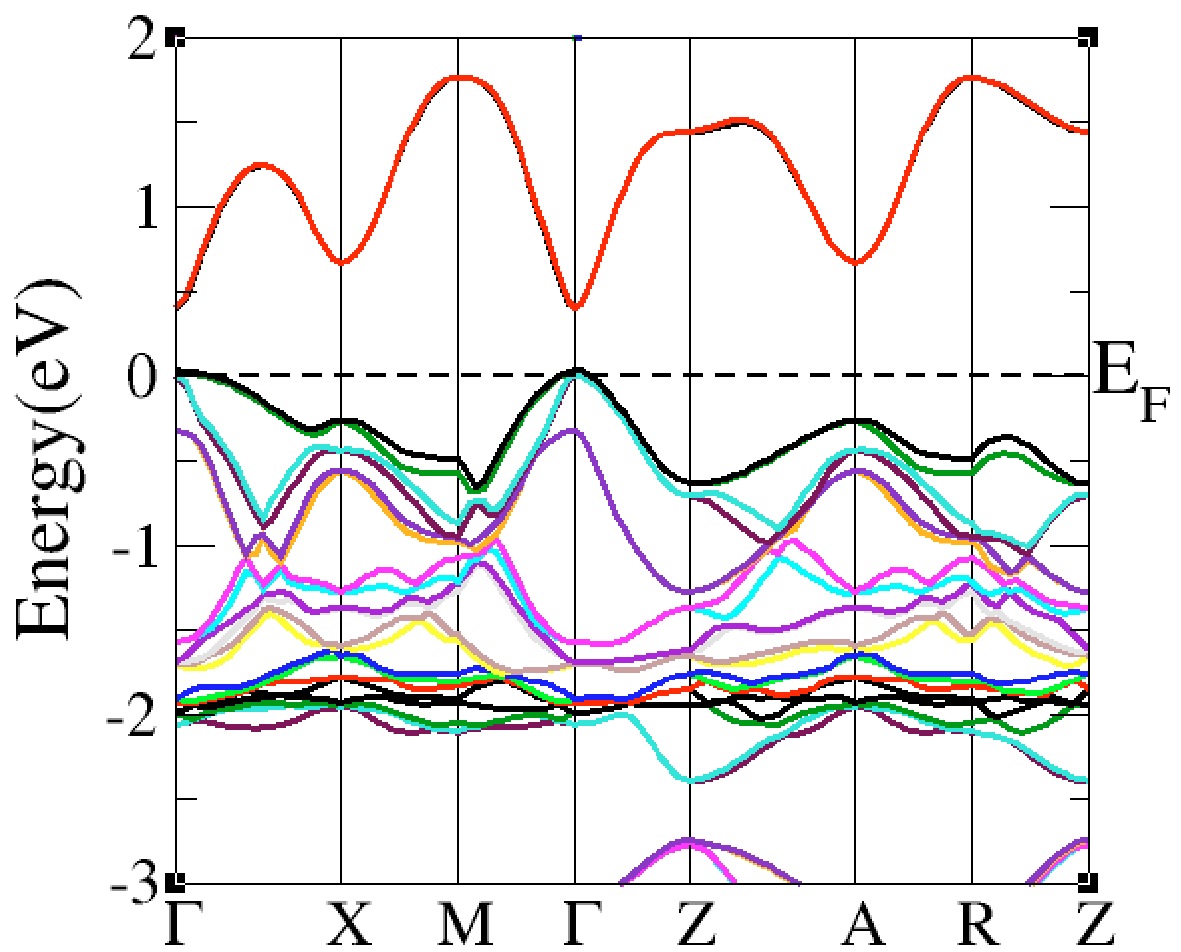}
\caption{Band structure for the topologically trivial insulator Cu$_2$GeZnS$_4$ with spin orbit coupling included. The valence and conduction bands each form separate physical band representations, and so the $0.5eV$ band gap is topologically trivial.}\label{fig:Cu2GeZnS4}
\end{figure}

Instead, let us consider Cu$_2$SbCuS$_4$\cite{cu3sbs4ref}. {Without SOC, this material is a zero-gap semimetal}. Furthermore, when SOC is included, the $\bar{\rho}^{4d}_5\uparrow G$ band representation and the $\bar{\rho}^{2b}_7\uparrow G$ representation are exchanged between the valence and conduction band as compared with Cu$_2$GeZnS$_4$. As such, the conduction band of Cu$_2$SbCuS$_4$ consists of the $\bar{\rho}^{4d}_5\uparrow G$ band representation, induced from the \emph{one-dimensional} $\bar{\rho}^{4d}_5$ site symmetry representation. Thus, this band representation is elementary, but \emph{not} physically elementary. We conclude that this material is a topological insulator. We show the band structure for this material in the left panel of Fig.~\ref{fig:cu2sbcus4}. To confirm {our group-theoretic result}, we have computed the Wilson loop spectrum, shown in the right panel of Fig.~\ref{fig:cu2sbcus4}. The spectrum clearly winds nontrivially throughout the BZ, indicating that Cu$_2$SbCuS$_4$ is a strong topological insulator. Note also that the real-space time-reversal partner $\bar{\rho}^{4d}_7\uparrow G$ band representation is $0.2eV$ below the Fermi level, although the gap in the material is only $0.03eV$. The two time-reversed partner EBRs are shown in red in the inset of Fig.~\ref{fig:cu2sbcus4}. Hence a novel feature of this material is that the ``topological gap'' is much larger than the transport gap.

\begin{figure}[t]
\includegraphics[height=2.5in]{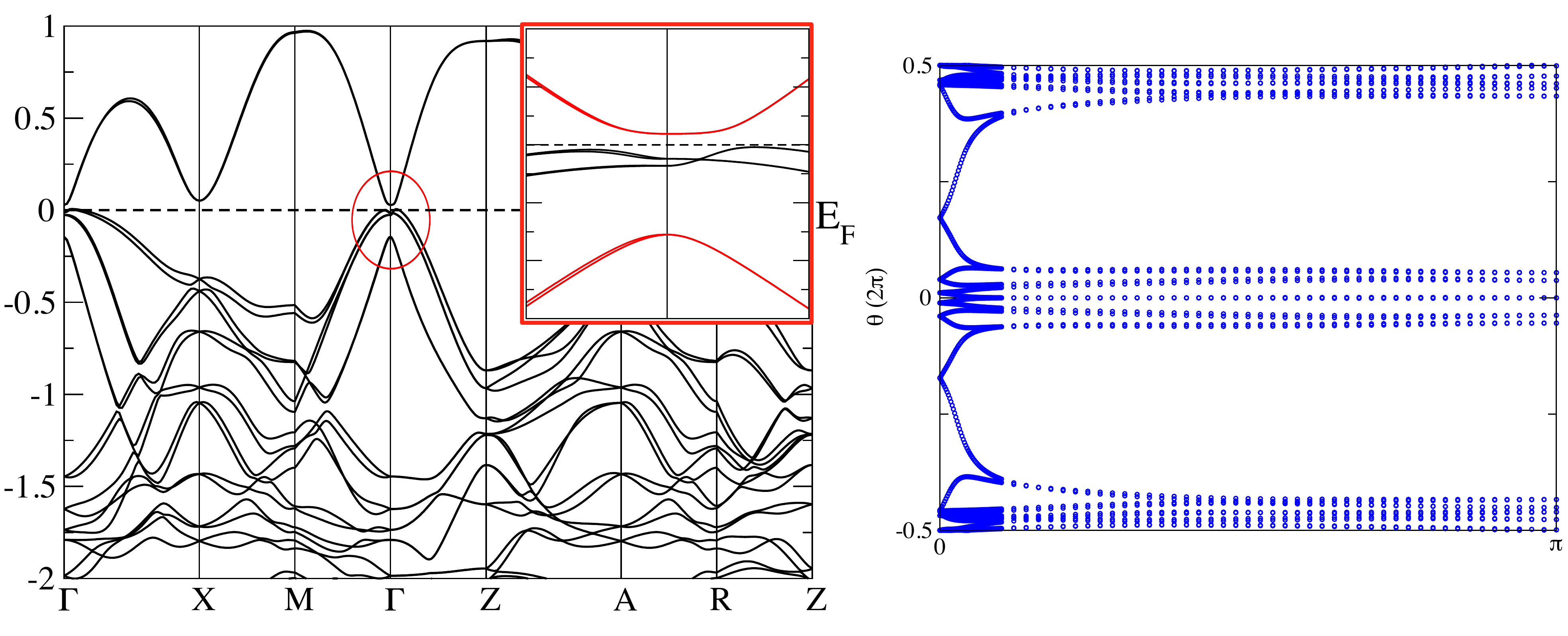}
\caption{Band structure and Wilson loop for the topologically nontrivial compound Cu$_2$SbCuS$_4$. The conduction band here does not form a physically elementary BR induced from a one-dimensional site-symmetry representation. The left panel shows the band structure, with inset showing a zoomed in view of the gap at $\Gamma$. The right panel shows the calculated Wilson loop spectrum. The winding of the Wilson loop shows that this material is a strong topological insulator.}\label{fig:cu2sbcus4}
\end{figure}

Finally, we find Cu$_2$SnHgSe$_4$. We show its band structure and Wilson loop in Fig.~\ref{fig:cu2snhgse4}. It also is a zero-gap semiconductor without SOC which becomes a strong-topological insulator when SOC is turned on. This particular strong TI, however, is not distinguishable from its purely group-theoretic properties: its valence and conduction bands have the same little group representations at every $\mathbf{k}$ point as true physical band representations; however they are topologically nontrivial. To see this, we can calculate their Wilson loop (Berry phase, holonomy), and from it determine any nontrivial topological indices\cite{Freed2013,Shiozaki2017}. {Our discovery of this new TI further highlights the power of our materials search.}
\begin{figure}[t]
\includegraphics[height=2.5in]{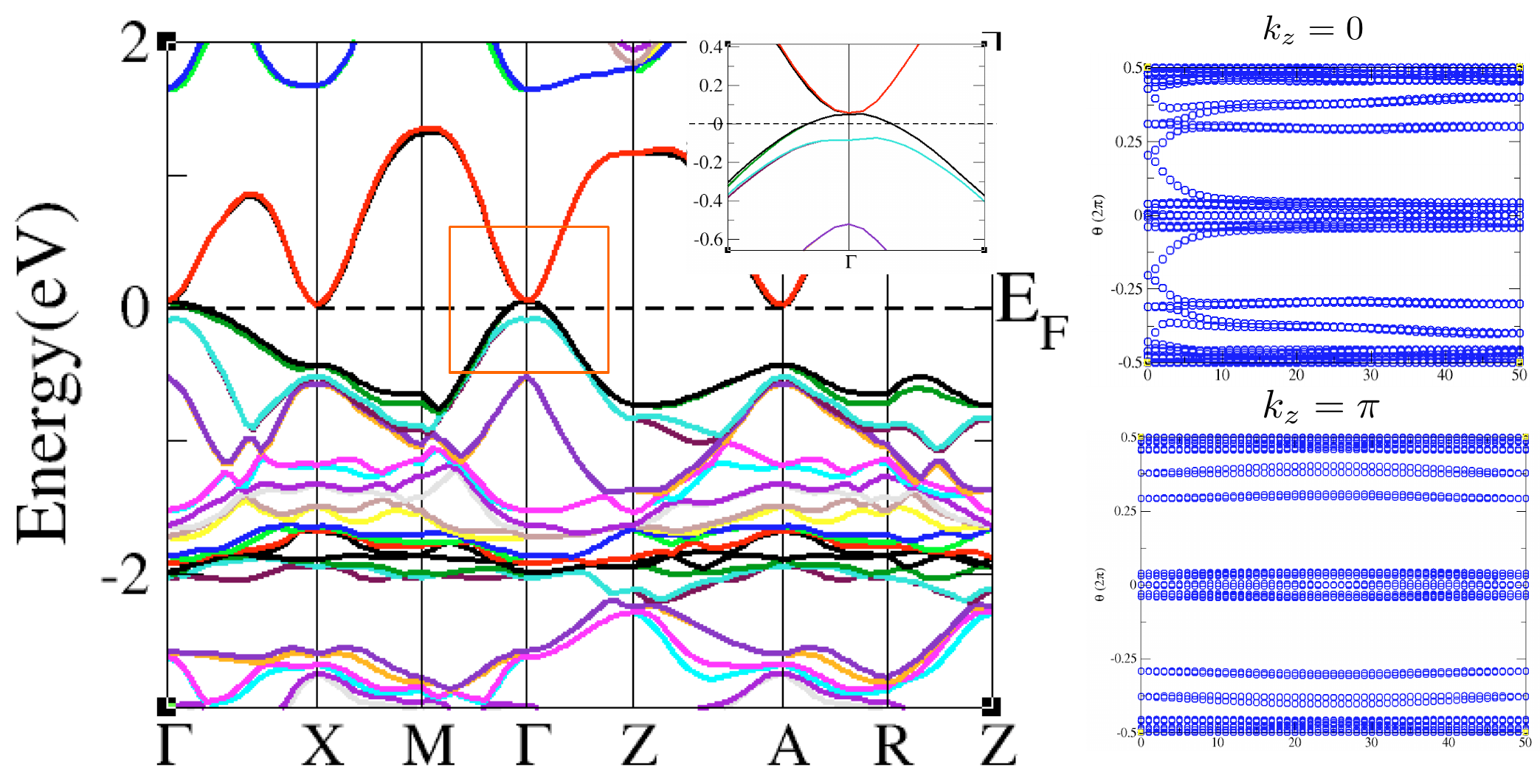}
\caption{Band structure and Wilson loop for the topologically nontrivial compound Cu$_2$SnHgSe$_4$. The left panel shows the band structure, with the inset showing a zoomed in view of the (rather small) gap at the $\Gamma$ point. The right panel shows the Wilson loop calculated and $k_z=0$ and $k_z=\pi$; the winding of the Wilson loops reveals that this compound is a strong topological insulator.}\label{fig:cu2snhgse4}
\end{figure}

\subsection{Square net topological insulators}\label{subsec:bisquare}
Next, we look at topological insulators of the type $(1,2)$ as defined in the main text. These materials are enforced semimetals with a single partially filled elementary band representation without SOC, which then splits into a topologically disconnected composite band representation when spin-orbit coupling is included. We consider square nets of As, Sb, Sn, and Bi which form layered compounds in $P4/nmm$ (129) and $Pnma$  (62) (upon small distortion of the squares). We find approximately $400$ candidate materials of these types, {discovered by targeting our method towards the specific cases of orbitals which can create topological bands}. In each of these classes, the relevant states near the Fermi level come from the $p$-orbitals of the square-net atoms. The maximal positions within the square net layer are still those shown in Fig.~\ref{fig:checker}. Representative crystal structures for these compounds are shown in Figure~\ref{fig:squarenetstructs}.

\begin{figure}
\includegraphics[width=0.2\textwidth]{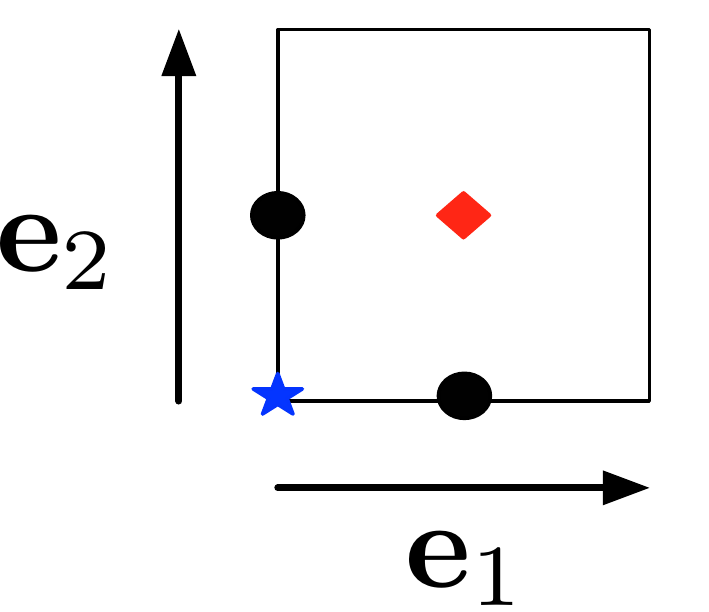}
\caption{Maximal Wyckoff positions in the square net. The blue star indicates the $a$ position at the 2D lattice sites, the red diamond indicates the $b$ position at the center of the square cell, and the black circles denote the $c$ Wyckoff position at the middle of the edges.}\label{fig:checker}
\end{figure}

\begin{figure}[h]
\includegraphics[width=0.5\textwidth]{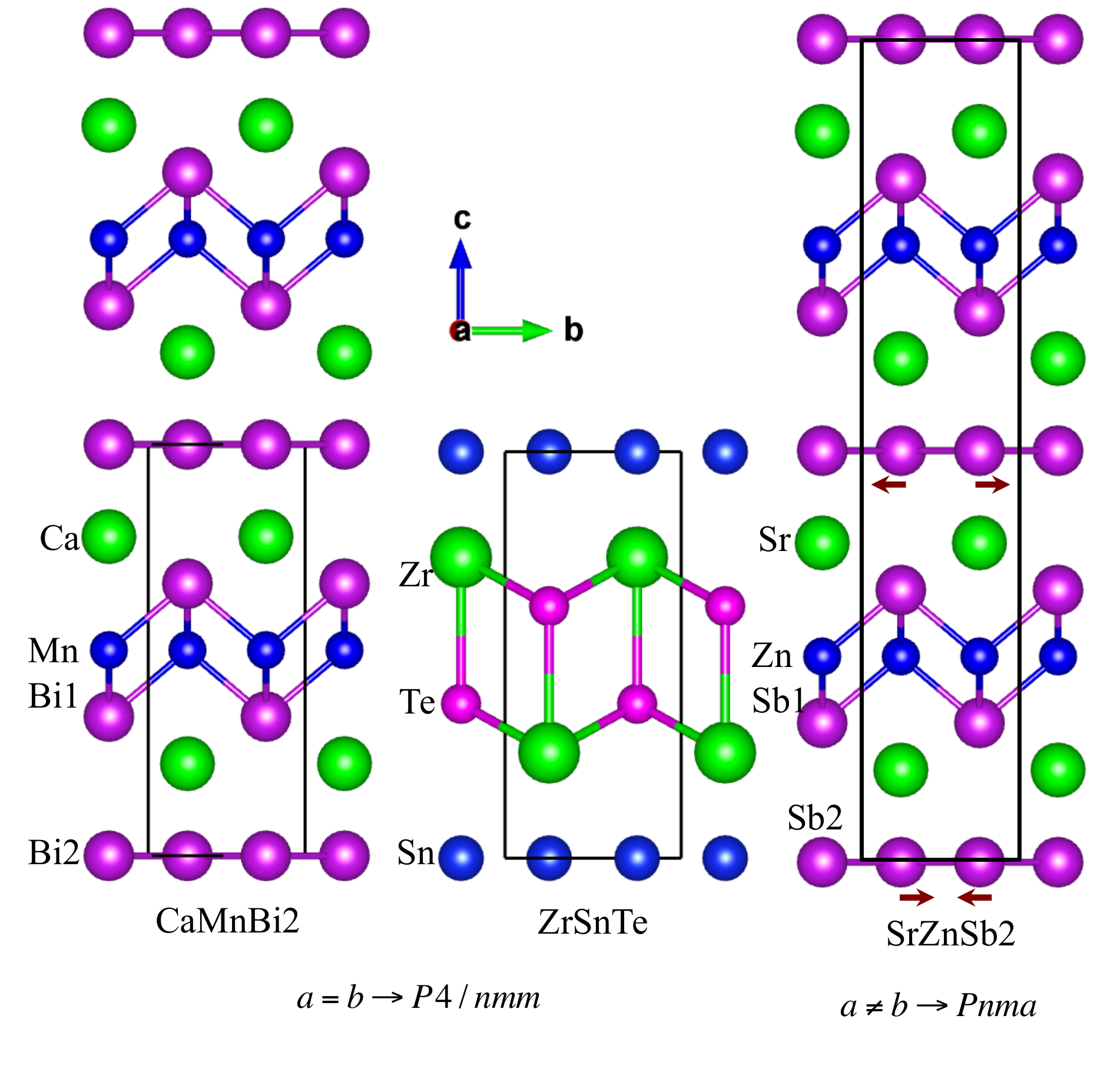}
\caption{Crystal structures for the Bi-square net class of topological insulators. The first and second structures show CaMnBi$_2$ and ZrSnTe in space group $P4/nmm$ (129). In CaMnBi$_2$ the Bi$2$ atoms form the square net, while in ZrSnTe it is the Sn atoms. The third structure shows SrZnSb$_2$ in  $Pnma$ (62). Here it is the atoms labelled Sb$2$ which make up the slightly distorted square net.}\label{fig:squarenetstructs}
\end{figure}

To analyze these materials, we first begin without SOC. Viewing the square net in isolation, we find that the Fermi level sits between the $p_z$ orbital bonding and anti-bonding states, as shown for Bi in Figure~\ref{fig:bielectrons}. However, charge transfer of {two electrons per unit cell} from {the adjacent non-square net layers shown in Fig.~\ref{fig:squarenetstructs} for} each of these materials {fill} the $p_z$ antibonding states, {putting them} below the Fermi level; {at the Fermi level, the $\{p_x,p_y\}$ bonding states are filled, while the antibonding states are empty. However, in these materials, the $\{p_x,p_y\}$ bonding and antibonding states form a single, connected four (per-spin) band PEBR. Thus,} the band structure of each quasi-2D layer has at the Fermi level a single half-filled elementary band representation without SOC, coming from the four $\{p_x,p_y\}$ orbitals per unit cell. This band representation is induced from the two-dimensional representation of the site-symmetry group $D_{2d}$, as indicated in Table~\ref{table:orbtab1}; recall that the character table for this group was given in Table~\ref{table:D2d}. This site-symmetry representation is spanned by $p_x\pm ip_y$ orbitals. The band structure for this band representation in a square net of $Bi^{1-}$ ions is shown in Figure~\ref{fig:binosoc}. Note that at half-filling, there is a linear band crossing along the $\Gamma-M$ line, which is the cross-section of a line-node (line-degeneracy) protected by mirror symmetry.
\begin{figure}[h]
\includegraphics[width=0.5\textwidth]{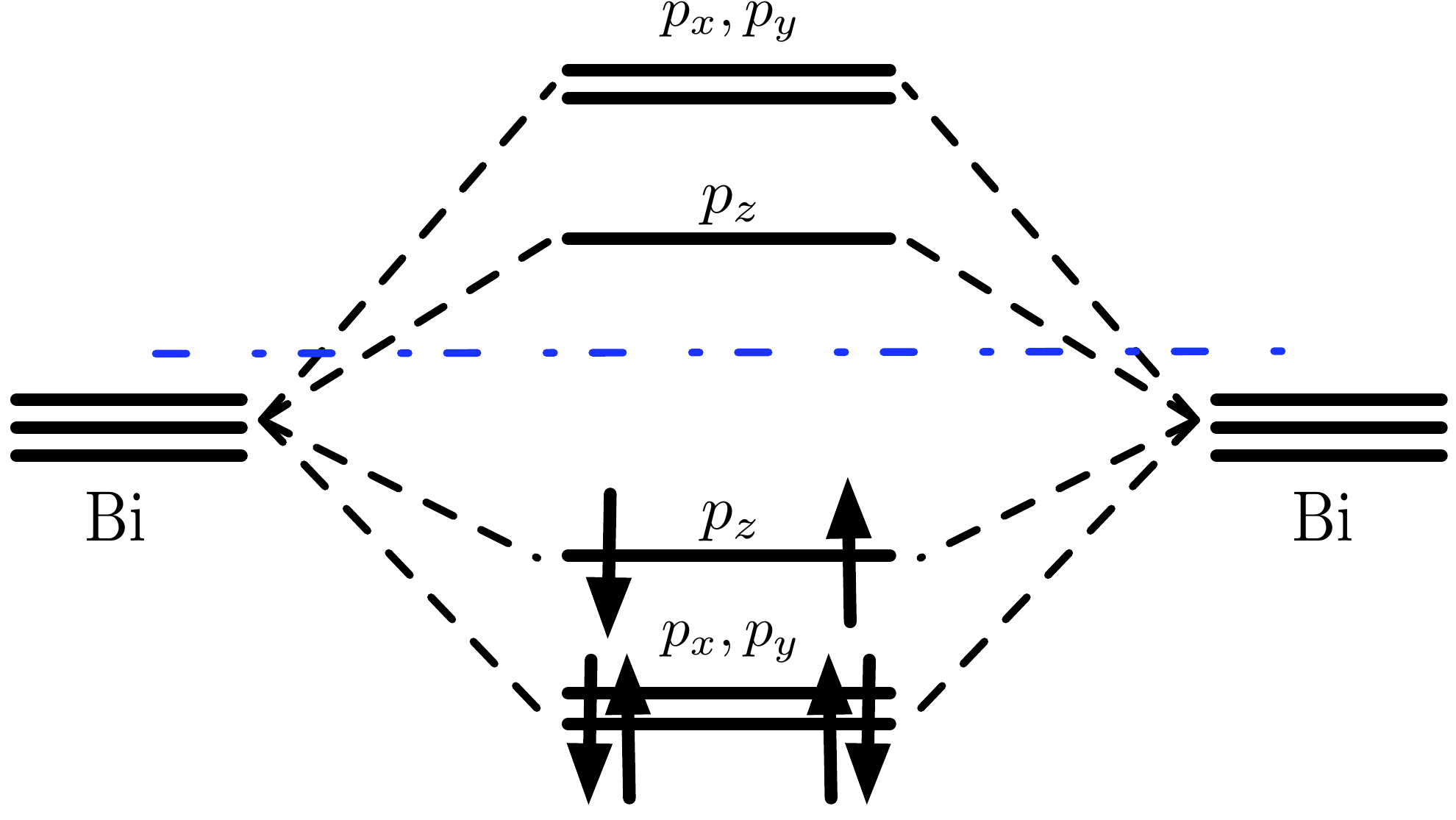}
\caption{Crystal field splitting of levels in the Bi square net. For undoped bismuth with three electrons per atom, the Fermi level sits at the blue dotted-dashed line. Note that the four $\{p_x,p_y\}$ states transform in a single elementary band representation.}\label{fig:bielectrons}
\end{figure}

\begin{figure}[t]
\centering
\subfloat[]{
\includegraphics[width=0.3\textwidth]{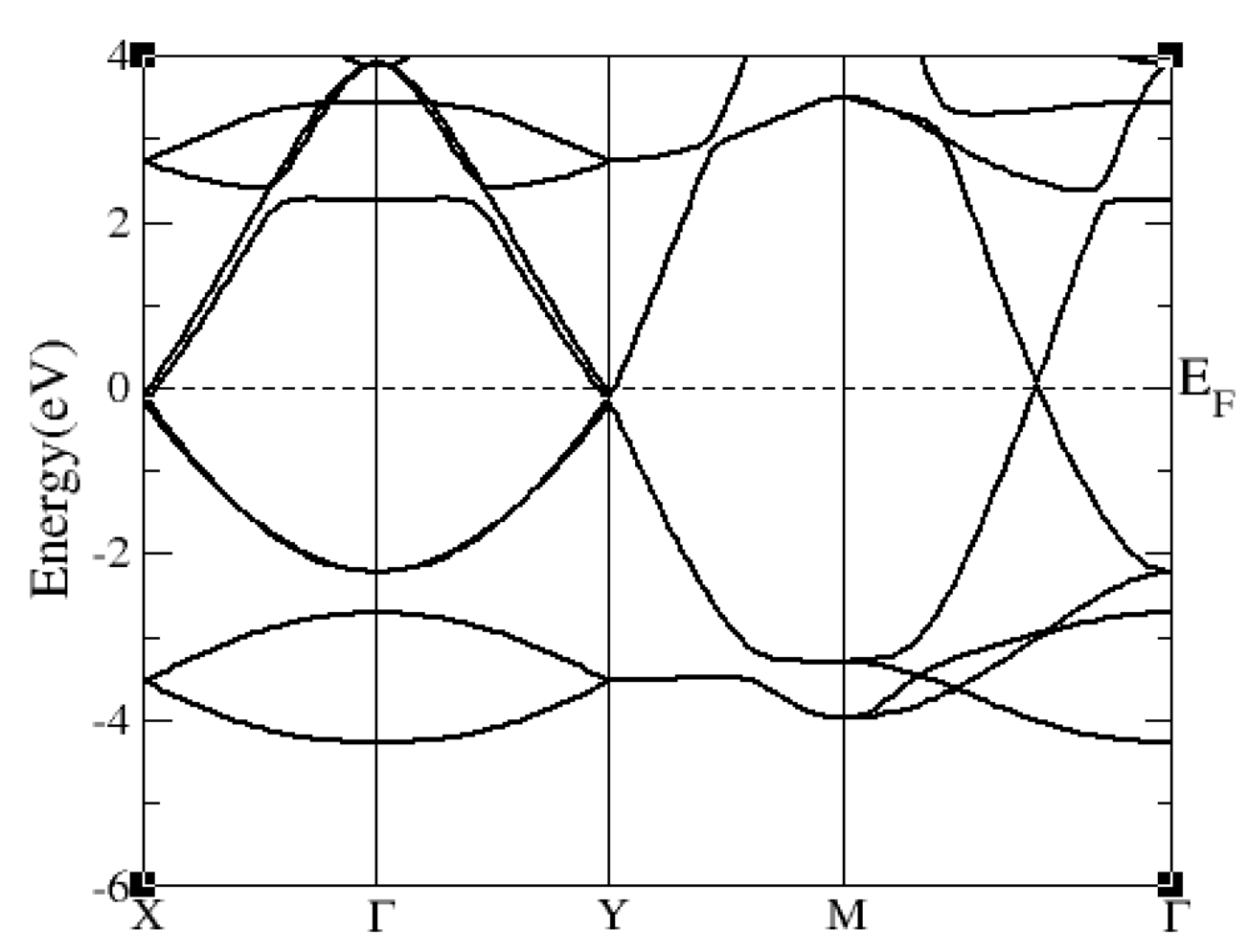}\label{fig:binosoc}
}
\subfloat[]{
\includegraphics[width=0.3\textwidth]{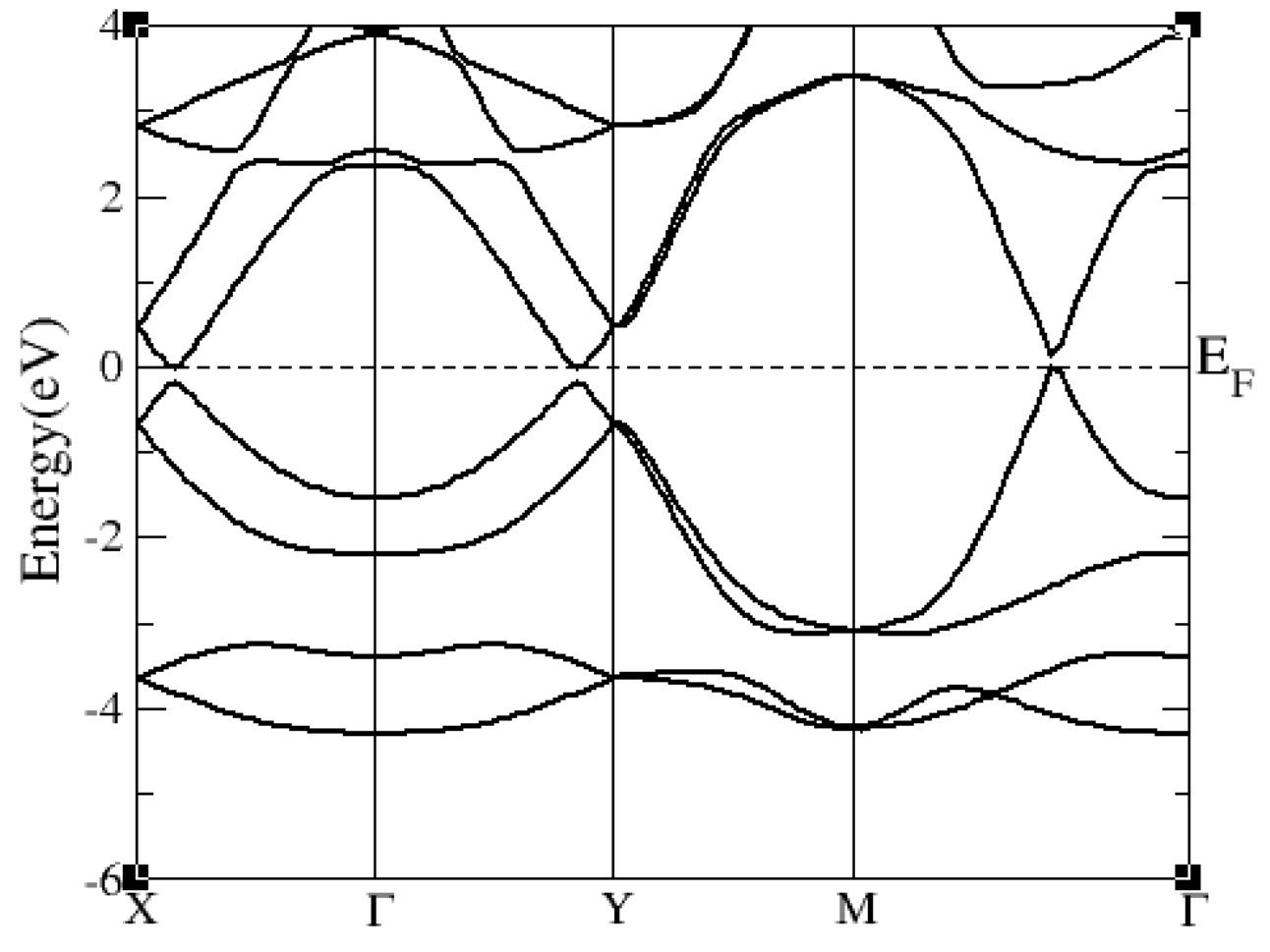}\label{fig:bisoc}
}
\caption{Representative band structure for the bands in the Bi square net induced from $\{p_x,p_y\}$ orbitals. (a) shows the band structure without SOC, showing band crossings at the Fermi level. These gap with infinitesimal SOC into a topologically nontrivial insulator, as shown in (b)}
\end{figure}

This line node is key to the topological nontriviality of these materials when SOC is included. From Table~\ref{table:orbtab2}, we see that with SOC the $\{p_x,p_y\}$ orbitals decompose into the reducible $\bar{\rho}_6\oplus\bar{\rho}_7$ representation of $D_{2d}$, and hence induce a physically composite band representation. Note that the $\bar{\rho}_6$ representation is spanned by $\{|p_x+ip_y,\uparrow\rangle,|p_x-ip_y,\downarrow\rangle\}$ states, while $\bar{\rho}_7$ is spanned by the $\{|p_x+ip_y,\downarrow\rangle,|p_x-ip_y,\uparrow\rangle\}$. Thus, for these two {physically elementary} band representations {($\bar{\rho}_6\uparrow G$ and $\bar{\rho}_7\uparrow G$)} to separate in energy {and give a trivial insulator}, SOC must be large enough to completely separate the initially degenerate spin-up and spin-down states in Fig.~\ref{fig:bisoc}.
\begin{table}[h]
\begin{tabular}{c|c|c|c|c|c|c}
$ D_{2d}$ & $\Gamma$ & $M$ &  $X$ & $\Sigma$ ($\Gamma$-$M$)& $\Delta$ ($\Gamma$-$X$) & d\\
\hline
$\rho_{5 }\uparrow G$ & $\Gamma_{5}^+\oplus \Gamma_{5}^-$ & $M_3\oplus M_4$ & $X_1\oplus X_2$ & $\Sigma_1\oplus \Sigma_2\oplus \Sigma_3\oplus \Sigma_4$ & $\Delta_1\oplus \Delta_2\oplus \Delta_3\oplus \Delta_4$ &4\Tstrut \\
\hline                                                                              
$\bar{\rho}_{6}\uparrow G$ & $\bar{\Gamma}_{6}\oplus \bar{\Gamma}_{9}$ & $\bar{M}_{5}$ & $\bar{X}_{3}\oplus \bar{X}_{4}$ & $2\bar{\Sigma}_{5}$ & $2\bar{\Delta}_{5}$ &4 \Tstrut\\
$\bar{\rho}_{7}\uparrow G$ & $\bar{\Gamma}_{7}\oplus \bar{\Gamma}_{8}$ & $\bar{M}_{5}$ & $\bar{X}_{3}\oplus \bar{X}_{4}$ & $2\bar{\Sigma}_{5}$ & $2\bar{\Delta}_{5}$ &4 \\
\end{tabular}
\caption{Band representations induced by $p$-orbitals in a square net, both with and without spin-orbit coupling. Note that the double-valued band representations are distinguished by the little-group representations they subduce at $\Gamma$. The dimensions of the representations d are shown in the last column of the table}\label{table:squarenetbrs}
\end{table}

However, even arbitrarily small spin orbit coupling will gap the aforementioned line node seen along $\Gamma-M$. In contrast to the trivial gap, this gap is topological -- neither the valence nor the conduction band transform as elementary band representations. To see this concretely, let us examine the case of  $P4/nmm$ (129). In Table~\ref{table:squarenetbrs} we give the little group representations at each high-symmetry point arising from the band representations induced by $\{p_x,p_y\}$ orbitals; $\rho_5$ is the two-dimensional SOC-free representation of $D_{2d}$, while $\bar{\rho}_6$ and $\bar{\rho}_7$ are the two relevant two-dimensional double-valued representations. The key is that without spin orbit coupling, the $\Gamma_5^+$ and $M_3$ little group representations at $\Gamma$ and $M$ respectively "lie" in the valence band, while the $\Gamma_5^-$ and $M_4$ representations "lie" in the conduction band. Next, we note that for arbitrarily small spin-orbit coupling, the spin representations decompose as
\begin{align}
\Gamma_5^+&\rightarrow\bar{\Gamma}_6\oplus\bar{\Gamma}_7\\
\Gamma_5^-&\rightarrow\bar{\Gamma}_8\oplus\bar{\Gamma}_9\\
M_3&\rightarrow\bar{M}_5 \\
M_4&\rightarrow\bar{M}_5.
\end{align}
We thus see that with weak spin-orbit coupling, the valence band contains the $\bar{\Gamma}_6$ and $\bar{\Gamma}_7$ little group representations at $\Gamma$. However, comparing with Table~\ref{table:squarenetbrs}, we see that this is not possible if the valence band is a physically elementary band representation. {While this particular energy ordering was determined from ab-initio calculations, we see that the same analysis holds whenever there is one occupied and one unoccupied little group representation at $\Gamma$ without SOC; this is generically true at half-filling.} We thus deduce that for small spin-orbit coupling, these materials are topological insulators. 

The ubiquity of the square net structure {in nature} allows us to identify hundreds of topological insulators in this class. In space group $P4/nmm$ (129) we find materials in the class of ABX$_2$, with A a rare earth metal, B$=$Cu,Ag and X$=$Bi,As,Sb,P, for a total of 48 candidate materials. Furthermore, the recently discovered topological phase in tetragonal bismuth falls into this class of square-net topological insulators\cite{SSBismuth} [albeit in  $I4/mmm$ (139)]. Additionally, in  $P4/nmm$ (129) we find square-net compounds of the type ABX with A$=$Ti,Zr,Hf, or another rare earth, B$=$Si,Ge,Sn,Pb, and X$=$Os,S,Se,Te. In total, this yields $328$ candidate materials in this space group. 
{
\subsubsection{Distored Square Nets}
Although our analysis has focused primarily on the idealized square net, we can show that topological behavior is insensitive to lattice distortions. We can see this most clearly by examining crystal structures with \emph{distorted} square nets. In particular, we focus on  $Pnma$ (62), which is obtained from the idealized square net in  $P4/nmm$ (129) after an in-plane $C_4$ symmetry-breaking distortion, shown schematically in Fig.~\ref{fig:squarenetstructs}. We} find the 58 new candidate topological insulators LaSbTe, SrZnSb$_2$, and AAgX$_2$, for A a rare-earth metal and X$=$P,As,Sb,Bi. Representative band structures are shown in Fig~\ref{fig:bisquares}, where the topological gap can be clearly seen. We expect all these materials to share a qualitatively similar topological band structure.
\begin{figure}[t]
\centering
\subfloat[]{
	\includegraphics[height=1.8in]{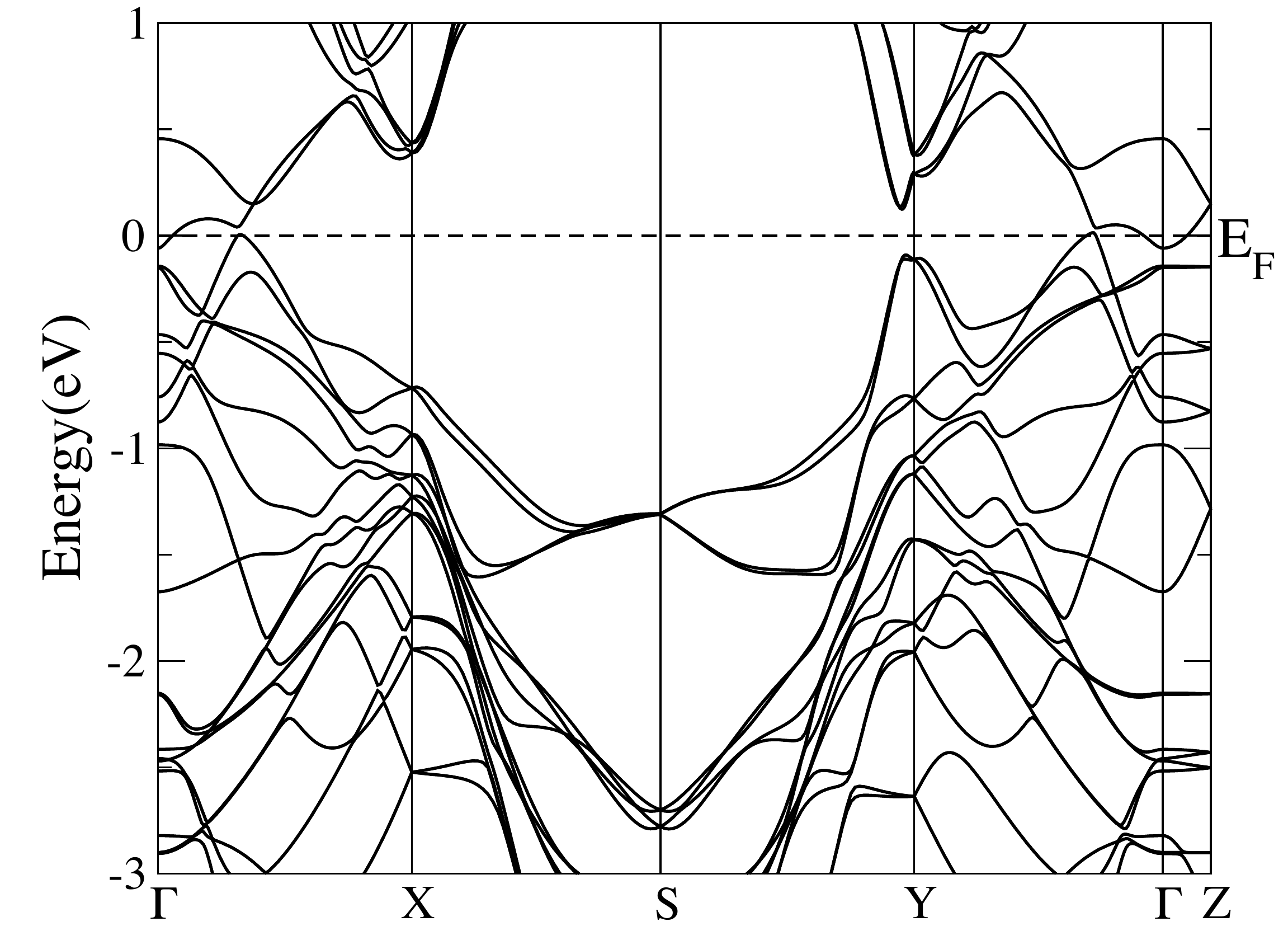}\label{SrZnSb2}
}
\subfloat[]{
	\includegraphics[height=1.8in]{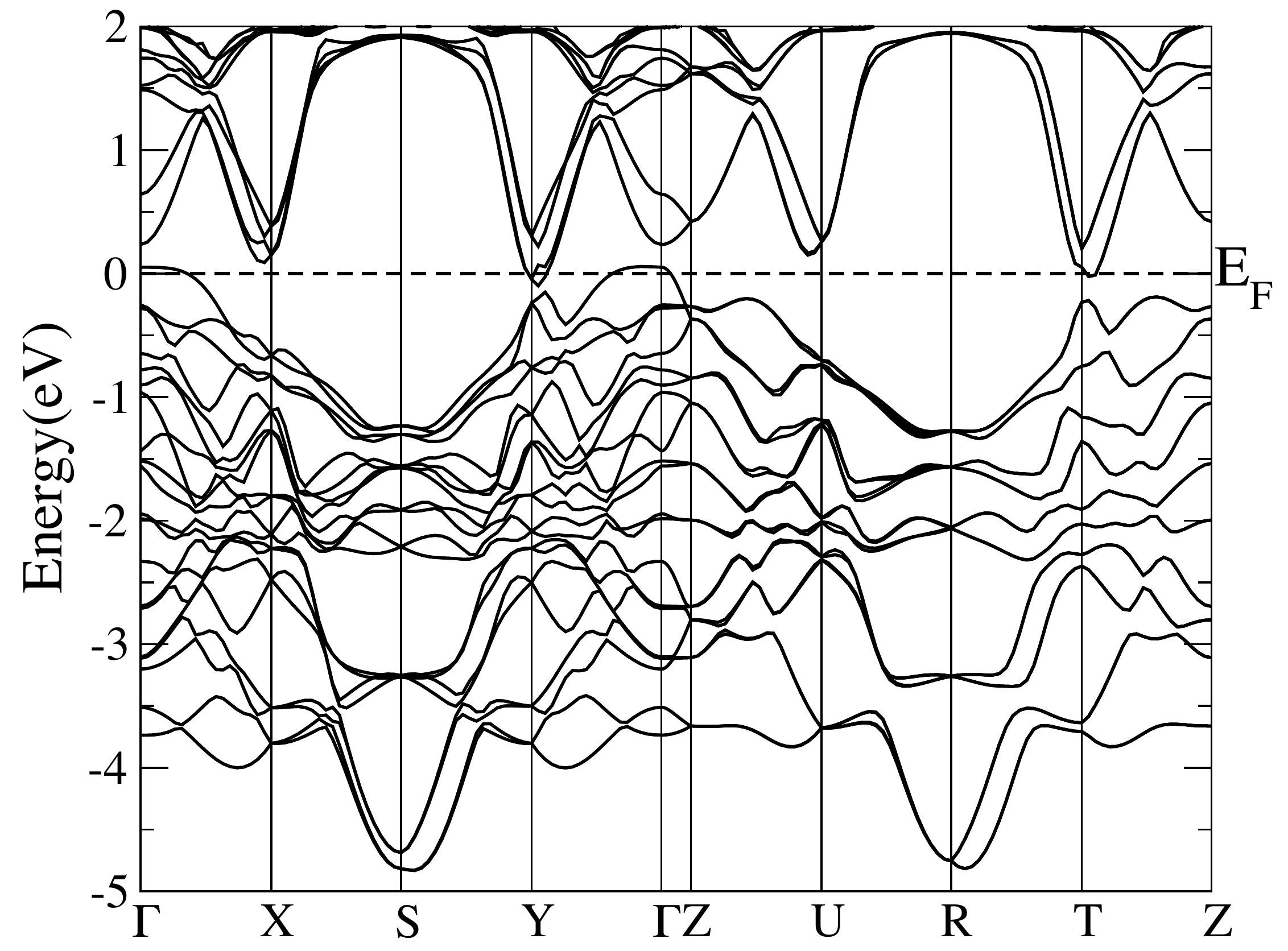}\label{LaSbTe}
}
\caption{Representative band structures for new topologically nontrivial insulators in the distorted Bi- square net structure group. (a) shows the band structure of the $3D$ weak topological insulator SrZnSb$_2$, while (b) shows the band structure of the 3D weak topological insulator LaSbTe.}\label{fig:bisquares}
\end{figure}
We note empirically that the magnitude of this distortion appears to be inversely correlated with the strength of spin-orbit-coupling of the atoms in the square net. We conjecture that this is due to the fact that SOC alone lifts the electronic degeneracy that causes the distortion through the Jahn-Teller effect.}

\subsection{Sixteen-fold connected metals}\label{subsec:metal}
Space group I$\bar{4}3$d  (220) supports a \emph{sixteen-band} physically elementary band representation. In any topologically trivial phase, all sixteen of these bands need to be connected. {We believe this set of high-connectivity bands, far exceeding the minimum connectivity of Refs.~\onlinecite{Watanabe15,Watanabe16}, can lead to robust protected semimetals with large conductivities, strong correlations, Mott physics, and other exotic properties.} We find examples of this band representation, partially filled at the Fermi-level, in the series of compounds A$_{15}$B$_4$, with A$=$Cu,Li,Na and B$=$Si,Ge,Sn,Pb. {It is amusing to note that these materials which we identified with group theory, are also known as promising candidates for the next generation of batteries\cite{batteryref}. Thus, batteries seem to be symmetry-protected (semi-)metals.} The crystal structure for these compounds in a conventional unit cell is shown in Figure~\ref{fig:a15b4struct}. Note that there are two formula units per primitive unit cell.
\begin{figure}[t]
\includegraphics[width=1.5in]{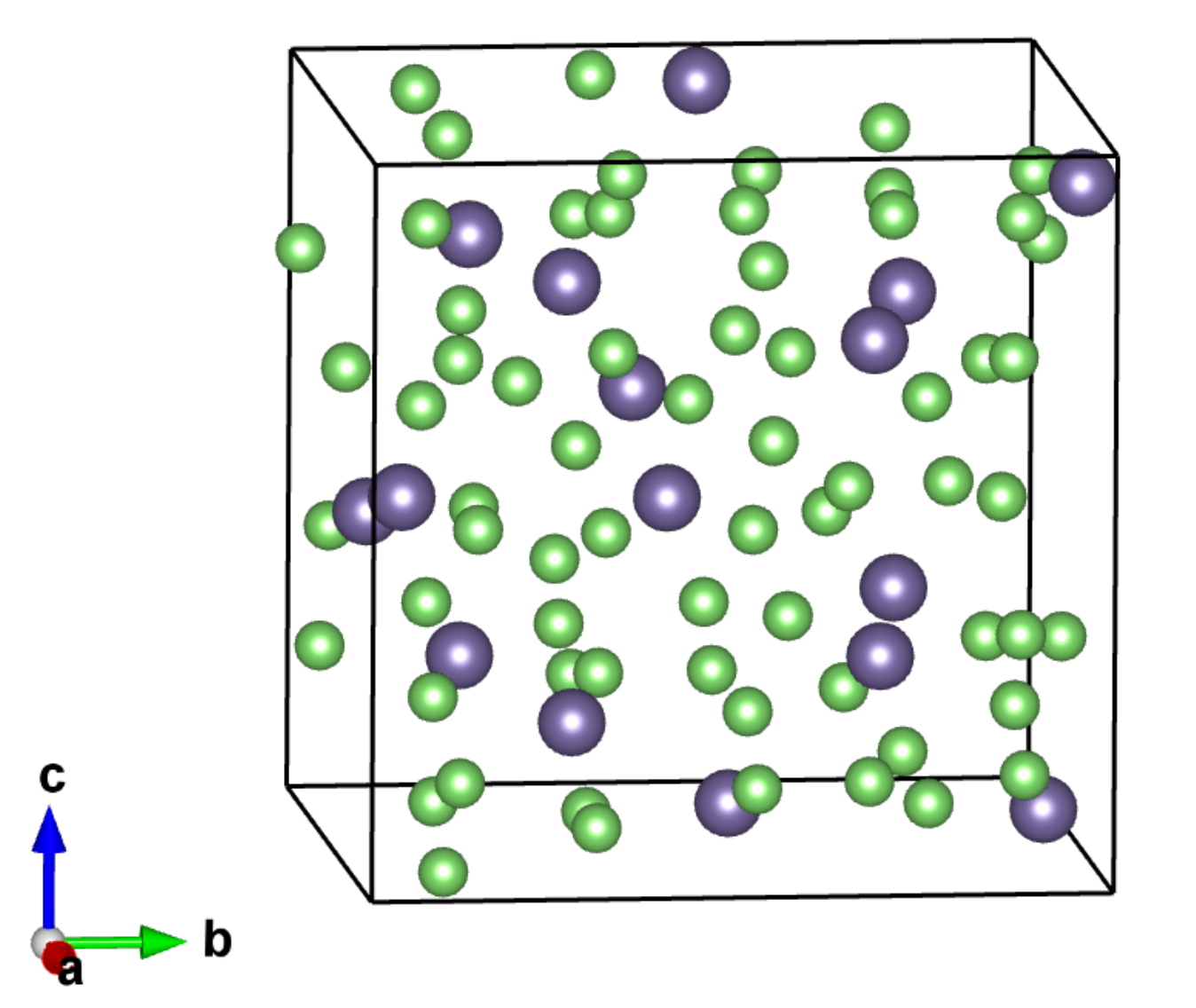}
\caption{Crystal structure of the A$_{15}$B$_4$ class of materials in  $I\bar{4}3d$ (220). One conventional unit cell is shown. The small green circles indicate the location of the $A$ atoms at the $12a$ and $48e$ Wyckoff position. The larger purple circles indicate the B atoms at the $16c$ Wyckoff positon.}\label{fig:a15b4struct}
\end{figure}

To analyze these materials, we first review the basic facts about space group $I\bar{4}3d$ (220). This is a non-symmorphic, body centered cubic space group. We take as primitive basis vectors for the BCC lattice
\begin{equation}
\mathbf{e}_1=\frac{a}{2}(-\hat{\mathbf{x}}+\hat{\mathbf{y}}+\hat{\mathbf{z}}),\; 
\mathbf{e}_2=\frac{a}{2}(\hat{\mathbf{x}}-\hat{\mathbf{y}}+\hat{\mathbf{z}}),\; 
\mathbf{e}_3=\frac{a}{2}(\hat{\mathbf{x}}+\hat{\mathbf{y}}-\hat{\mathbf{z}}), \label{eq:bccvecs}
\end{equation}
which we recognize as Eq.~(\ref{eq:bctvecs}) with the lattice constants $a=c$. In addition to the translations, space group $I\bar{4}3d$ (220) is generated by the cubic threefold rotation $\{C_{3,111}|000\}$ about the $[111]$ axis, the four-fold roto-inversion $\{IC_{4,011}|\half 00\}$ about the $\hat{\mathbf{x}}=\mathbf{e}_2+\mathbf{e}_3$ axis, and the mirror $\{m_{1\bar{1}0}|\half\half\half\}$ that sends $\mathbf{\hat{x}}\leftrightarrow\mathbf{\hat{y}}$.

There are three maximal Wyckoff positions in this space group, denoted $12a,12b$ and $16c$, with multiplicity $6,6$ and $8$ respectively in the primitive unit cell description. Also note that there is the non-maximal $48e$ Wyckoff position, with multiplicity $24$. The A atoms sit at the $12a$ and $48e$ positions, while the B atoms sit at the $16c$ position. Since the electrons near the Fermi energy come from the B atoms, we will here be interested only in the $16c$ position. It has representative coordinate $\mathbf{q}_1^{16c}=(x,x,x)$ in terms of the lattice vectors Eq.~(\ref{eq:bccvecs}); it is clear that the stabilizer group is $G_{\mathbf{q}^{16c}_1}\approx C_3$, generated by the threefold rotation $\{C_{3,111}|000\}$. The coordinate triplets of its symmetry equivalent points in the primitive unit cell  are obtained by the repeated action of $\{IC_{4,011}|\half 00\}$ and $\{m_{1\bar{1}0}|\half\half\half\}$. Because the stabilizer group $C_3$ is abelian, its double-valued representations are all one-dimensional, and specified by the character $\chi(\{C_{3,111}|000\})$. The three possible double-valued representations are
\begin{align}
\bar{\rho}^{16c}_4(\{C_{3,111}|000\})&=-1, \\
\bar{\rho}^{16c}_5(\{C_{3,111}|000\})&=e^{-i\pi/3}, \\
\bar{\rho}^{16c}_6(\{C_{3,111}|000\})&=e^{i\pi/3}.
\end{align}  
Consulting Section~\ref{sec:data}, we see that in the physically elementary band representation $(\bar{\rho}^{16c}_5\oplus\bar{\rho}^{16c}_6)\uparrow G$, Kramers's theorem forces connection between bands coming from the $\bar{\rho}^{16c}_5\uparrow G$ and $\bar{\rho}^{16c}_6\uparrow G$ (non-physically) elementary band representations. As such, in any trivial phase, this band representation is sixteen-fold connected.

In the A$_{15}$B$_4$ class of materials, the B atoms sit at the $16c$ Wyckoff position. In the particular examples of Cu$_{15}$Si$_4$, Li$_{15}$Ge$_4$, Li$_{15}$Si$_4$, Na$_{15}$Sn$_4$, and Na$_{15}$Pb$_4$,the relevant states at the Fermi level are  precisely the B atom $p$-states, of which there are $48$ per unit cell. Due to charge transfer with the A atoms, there are $46$ electrons filling these states. $32$ out of those $46$ electrons go into filled valence bands, leaving $14$ electrons to fill a band of connectivity $16$. From Table~\ref{table:orbtab2}, we see that these yield bands transforming in the $(2\bar{\rho}^{16c}_4\oplus2\bar{\rho}^{16c}_5\oplus2\bar{\rho}^{16c}_6)\uparrow G$ composite band representation. In the materials listed above, ab-initio calculations reveal that the band representation closest to the Fermi-level is precisely the sixteen-branched $(\bar{\rho}^{16c}_5\oplus\bar{\rho}^{16c}_6)\uparrow G$ band representation, which by electron counting is $14/16=7/8$ filled. These materials are thus truly {protected metals}.

\begin{figure}[t]
\centering
\subfloat[]{
	\includegraphics[width=0.3\textwidth]{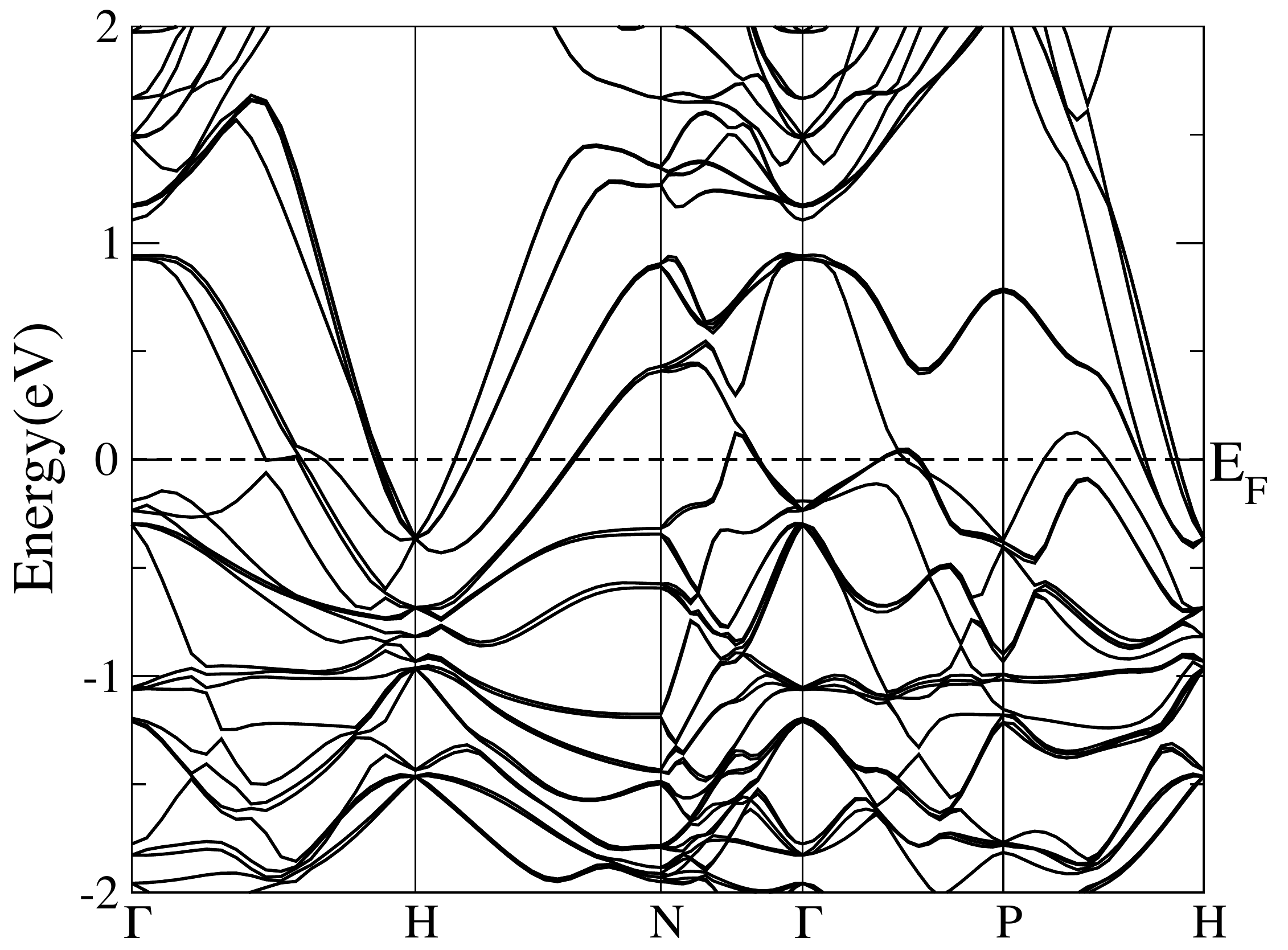}\label{fig:cu15si4}
}
\subfloat[]{
	\includegraphics[width=0.3\textwidth]{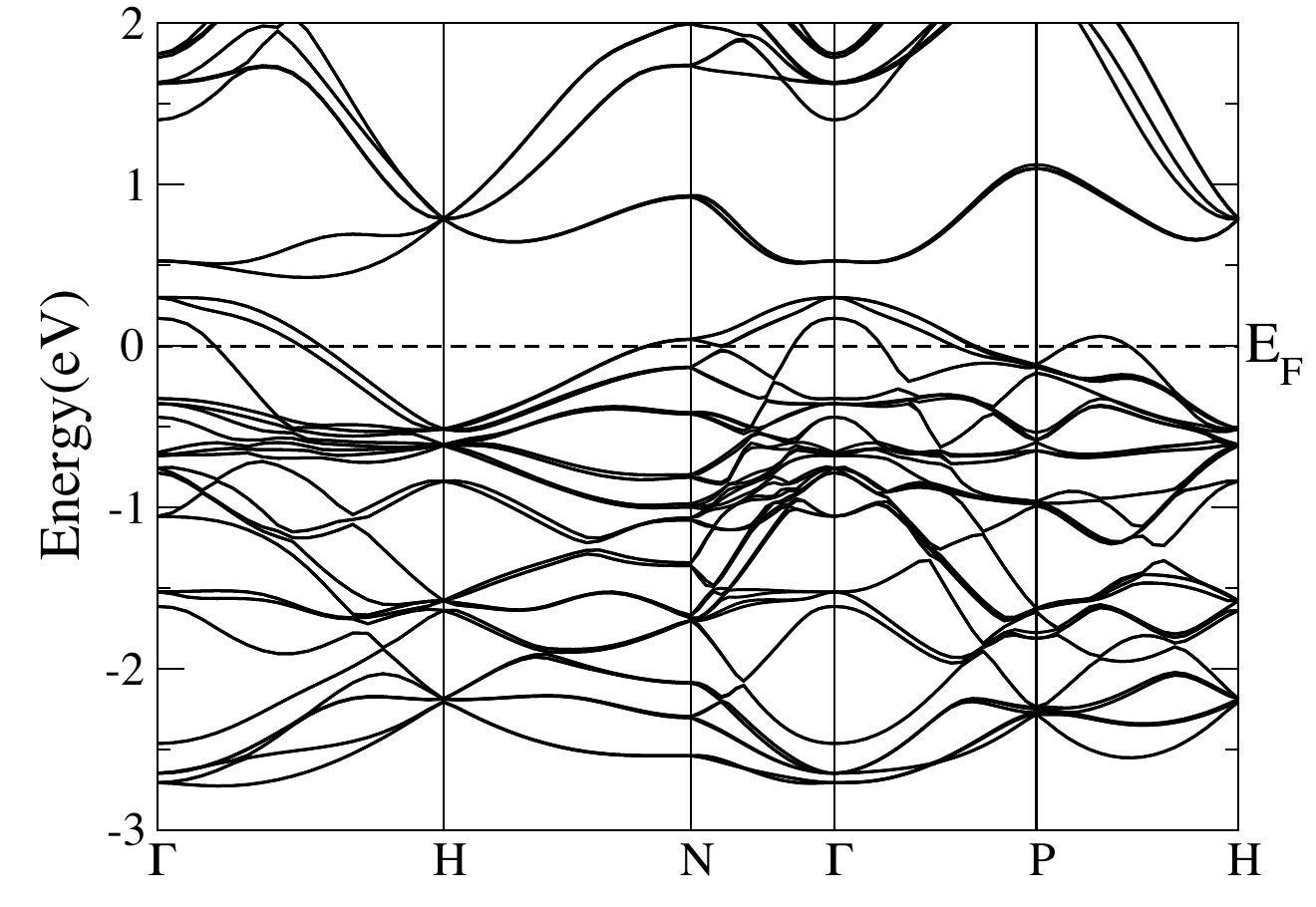}\label{fig:li15ge4}
}
\subfloat[]{
	\includegraphics[width=0.3\textwidth]{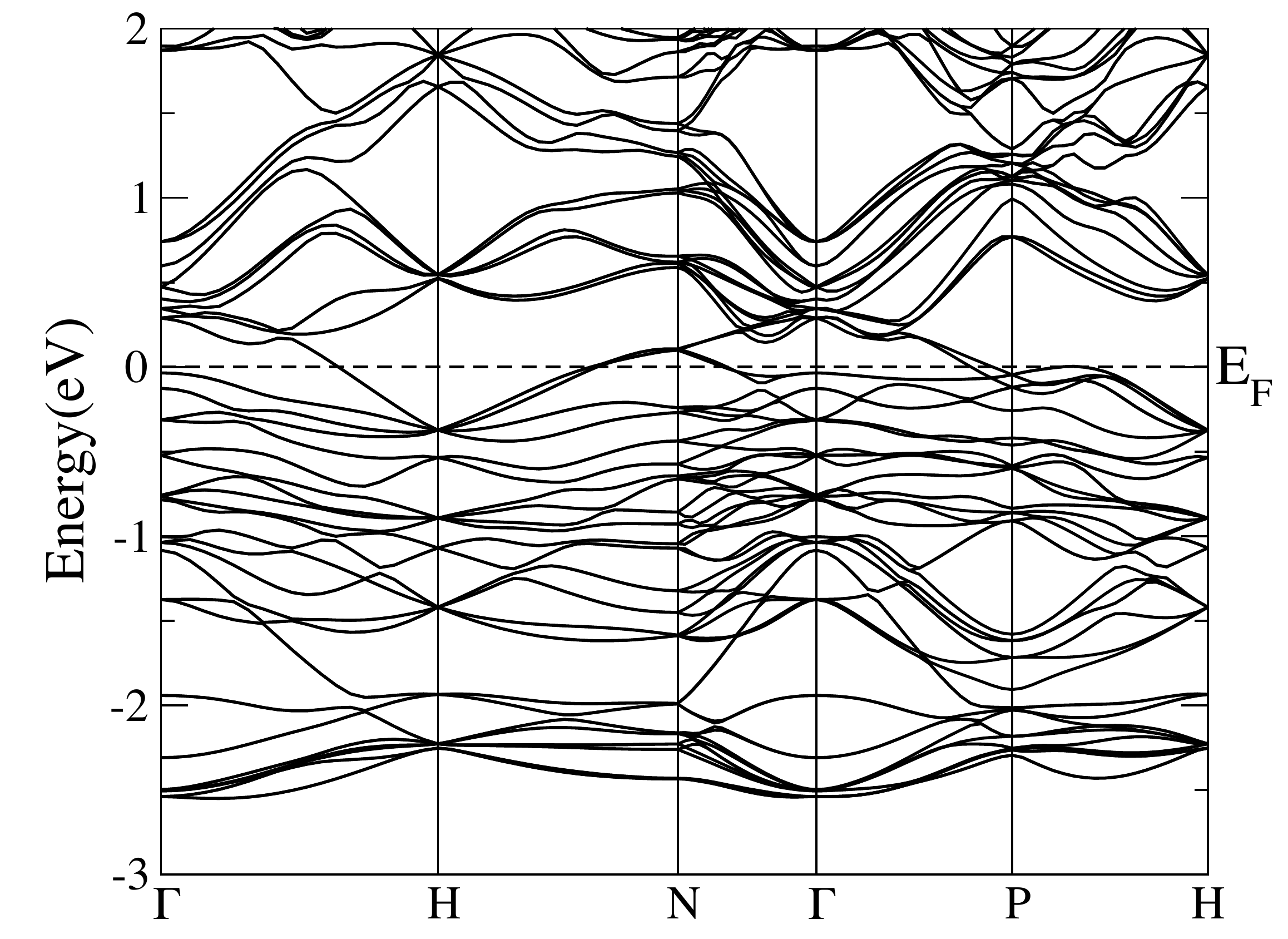}\label{fig:na15pb4}
}
\caption{Band structures with spin-orbit coupling for the sixteen-fold connected metals in the A$_{15}$B$_4$ structure group. Spin-orbit coupling has been included in all calculations. (a) shows the band structure for Cu$_{15}$Si$_4$; the Cu d-orbitals can be seen far below the Fermi level. (b) shows the band structure for Li$_{15}$Ge$_4$. Finally, (c) shows the band structure for Na$_{15}$Pb$_4$.}\label{fig:a15b4}
\end{figure}

We show band structures for these highly-connected metals in Fig.~\ref{fig:a15b4}. These band structures show clearly that the interconnections between these bands are mediated by the exotic degeneracies of Ref.~\onlinecite{Bradlyn2016}. In particular, we see that at the $P$ point these materials host a threefold degenerate fermion, while at the $H$ point they have eightfold degenerate excitations. In fact, {our} site-symmetry tables of Ref.~\onlinecite{GroupTheoryPaper} reveal that the Kramers-enforced connection between the $\bar{\rho}^{16c}_5\uparrow G$ and $\bar{\rho}^{16c}_6\uparrow G$ band representations occurs precisely at an eightfold degeneracy point at $H$.
}  
\vspace{-4ex}
\subsection{Twenty-fourfold connected metals}
\vspace{-3ex}
An exhaustive search of the dimensions of all 10,403 EBRs and PEBRs shows that the greatest number of bands that are forced to be connected by symmetry in a topologically trivial phase is $24$. An example of this occurs in  $Ia\bar{3}$ (206). In this group the $24d$ maximal Wyckoff position has multiplicity two, and site-symmetry group isomorphic to $C_2$. In spin-orbit coupled systems with TR symmetry, the two-dimensional physically irreducible $\bg_3\oplus\bg_4$ representation of this site-symmetry group thus induces a twenty-four band PEBR. In any topologically trivial phase, all twenty-four of these bands \emph{must} be interconnected. In Fig.~\ref{fig:24fold} we show the band structure of Cu$_3$TeO$_6$, which we calculate, has this PEBR half-filled at the Fermi level. Although interaction effects cause this material to be a Mott insulator\cite{24foldmott}, expect other materials with this PEBR near the Fermi level may exhibit exotic fillings due to charge-transfer effects. 
\vspace{-2ex}
\section{Supplementary data}\label{sec:data}
\vspace{-4.5ex}
\begin{table}[H]
\centering
\begin{tabular}{ccccc}
Site symmetry group & Reducing group&  intersection group & Rep dimension & Space Group Number \\
{ ($G_\mathbf{q}$)} & { ($G_{\mathbf{q}'}$)}& {($G_0$)}  &  &  \\
\hline
$D_3$ & $C_{3i}$& $C_3$& $2$ & $163, 165, 167, 228, 230$ \Tstrut\\
 & $T_h$&$C_3$ &$2$ & $223$ \\
 & $O$ &$C_3$ &$2$ &$211$ \\
 & $T$ &$C_3$ &$2$ & $208, 210, 228$ \\
 & $C_{3h}$ &$C_3$ & $2$ & $188, 190, 192, 193$\\ 
 \hline
$D_6$ &$C_{6h}$ &$C_{6}$ &$2$ & $192$\Tstrut \\
\hline
 $D_4$ &$O$ & $C_{4}$& $2$ & $207,211,222$ \Tstrut\\
  &$C_{4h}$ & $C_{4}$&$2$ & $124,140$ \\
  \hline
  \hline
$D_{2d}$ &$D_{4h}$ &$C_{2v}$ &$2$ & $229$ \Tstrut\\
 &$T_h$ &$C_{2v}$ &$2$ & $226$ \\
 & $T_d$ &$C_{2v}$ &$2$ & $215,217,224$ \\
 & $D_{2h}$ &$C_{2v}$ &$2$ &$131, 132, 139, 140, 223$
\end{tabular}
\caption{Maximal Wyckoff positions which yield \emph{composite} band representations for the single groups and thus do not need to be considered in a search for elementary band reps; computed by Bacry, Michel, and Zak\cite{Bacry1988}. {Point group symbols are given in Schoenflies notation.\cite{Cracknell}} The first column gives the maximal site-symmetry group, $G_\mathbf{q}$, which induces the composite representation. The second column gives the site-symmetry group, $G_{\mathbf{q}'}$, into whose band representations this composite representation can be reduced. The third column gives the intersection group, $G_0=G_{\mathbf{q}}\cap G_{\mathbf{q}'}$. The fourth column gives the dimension of the irrep which induces the composite band rep. The fifth column indicates the space groups for which this occurs.
{With (spinless) time-reversal, only the groups below the double line yield composite physical band representations (and do not need to be considered in a search for physically elementary band reps).}
}\label{table:sbr}
\end{table}

\begin{table}[H]
\centering
\begin{tabular}{ccccc}
Site symmetry group & Reducing group&  Intersection group & Rep dimension & Space Group Number \\
{($G_\mathbf{q}$)} & {($G_{\mathbf{q}'}$)}& {($G_0$)}  &  &  \\
\hline
$T_d$ & $D_{3d}$ &$C_{3v}$ & $4$ & $224, 227^*$ \Tstrut\\
 & $O_h$ & $C_{3v}$ & $4$ & $225$\\
\hline
$D_3$ & $T_h$& $C_3$ & $2$ & $223$\Tstrut\\
 &$O$ &$C_3$ & $2$ & $211$ \\
 &$T$ &$C_3$ & $2$ &$208,210,228$ \\
 & $C_{3h}$ &$C_3$ &$2$ &$188^*, 190^*, 192^*, 193^*$\\
 & $C_{3i}$ &$C_3$ &$2$& $163^*, 165^*, 167^*, 228^*, 230^*$\\
 \hline
$D_{3h}$ & $D_{3d}$ &$C_{3v}$ & $2$ & $193^*,194^*$ \Tstrut\\
\hline
$D_6$ &$C_{6h}$ &$C_6$ &$2$ & $192^*$\Tstrut\\
\hline
$D_4$ & $O$ &$C_4$ &$2$ & $207,211, 222$\\
 & $C_{4h}$ &$C_4$ &$2$ &$124^*,140^*$\\
 \hline
$C_{2v}$ &$C_{6v}$ &$C_s$  &$2$ & $183$\Tstrut\\
&$C_{3v}$ &$C_s$ & $2$ & $183$\\
 &$C_{2h}$ &$C_s$ & $2$ & $51^*,  63^*,  67^*,  74^*, 138^*$\\
 & $C_{4v}$ &$C_s$ &$2$&$99, 107$ \\
 & $D_{2d}$ & $C_s$ & $2$ & $115,137$ \\
 \hline
$D_{2}$ &$T$& $C_2$&$2$ & $195,197, 201, 208, 209, 218$\Tstrut\\
 & $D_6$ &$C_2$ &$2$&$177,192$\\
 & $D_3$ &$C_2$ & $2$ & $177, 192, 208, 211, 214, 230$\\
 & $S_{4}$ &$C_2$ &$2$&$112^*, 116^*, 120^*, 121, 126^*, 130^*, 133^*, 138^*, 142^*, 218^*, 230^*$ \\
 & $D_{2d}$ &$C_2$ &$2$& $111, 121, 132, 134, 224$ \\
 &$C_{2h}$ &$C_2$ &$2$ & $49^*,  66^*,  67^*,  69^*,  72^*, 124^*, 128^*, 132, 134, 135^*, 138^*, 192^*$ \\
 & $D_4$ &$C_2$ & $2$ & $89,  97, 124, 126, 211$ \\
 & $D_{3d}$ & $C_2$ & $2$ & $224$ \\
 & $O$ & $C_2$ & $2$ & $209$
\end{tabular}
\caption{Maximal Wyckoff positions which yield band representations that {we find to be} equivalent to \emph{composite} band representations for the double groups. 
The first column gives the maximal site-symmetry group,  $G_\mathbf{q}$, which induces the composite representation. {Point groups symbols are given using Schoenflies notation\cite{Cracknell} (e.g $C_s$ is the point group generated by reflection).} The second column gives the site-symmetry group, {$G_{\mathbf{q}'}$}, into whose band representations this composite representation can be reduced. The third column gives the intersection group, $G_0=G_{\mathbf{q}}\cap G_{\mathbf{q}'}$. The fourth column gives the dimension of the irrep which induces the composite band rep. The fifth column indicates the space groups for which this occurs.
An asterisk ($*$) indicates that while the band rep {is disconnected in momentum space when time-reversal symmetry is ignored, there are extra connectivity constraints imposed by Kramers's theorem when TR is present.} This can occur when the {representation $\sigma$ of $G_0$ induces two one-dimensional representations of $G_\mathbf{q}'$} that are not momentum-space time reversal invariant in isolation.}
\label{table:dbr}
\end{table}

\begin{table}[H]
\centering
\begin{tabular}{ccccc}
Site symmetry group ($G_\mathbf{q}$) & Reducing group ($G_\mathbf{q}'$) & Space Group Number \\
\hline
$S_4$ & $C_{2h}$ & $84,87,135,136$ \Tstrut\\
& $D_2$ & $112,116,120,121,126,130,133,138,142,218,230$ \\
& $D_4$ & $222$ \\
& $D_{2d}$ & $217$ \\
& $T$ & $219,228$  \\
\end{tabular}
\caption{Additional exceptional band representations with time-reversal. In all cases, the exceptional representation is the physically irreducible two-dimensional representation of $S_4$. For the space groups listed in this table, this band representation decomposes through $G_0=C_2$ into a composite band representation induced from the reducing group $G_\mathbf{q}'$. The first column gives the reducing group, while the second column gives the associated space groups for which the exception occurs.}\label{table:zakwaswrong}
\end{table}

\begin{table}[H]
\centering
{
\begin{tabular}{L|L|L|L|L}
\mathrm{PG} & \mathrm{PG Symbol}& s & p & d \\
\hline
C_1 & 1&\g_1 & 3\g_1 & 5\g_1 \Tstrut\\
C_i & \bar{1}& \g_1^+ & 3\g_1^- & 5\g_1^+\\
\hline
C_2 &2& \g_1 & \g_1\oplus2\g_2 & 3\g_1\oplus2\g_2 \Tstrut\\
C_s &m& \g_1 & 2\g_1 \oplus\g_2 & 3\g_1 \oplus2\g_2 \\
C_{2h} &2/m& \g_1^+ & \g_1^-\oplus2\g_2^- & 3\g_1^+\oplus2\g_2^+\\
\hline
D_2 &222& \g_1 & \g_2\oplus\g_3\oplus\g_4 & 2\g_1\oplus\g_2\oplus\g_3\oplus\g_4 \Tstrut\\
C_{2v} &mm2& \g_1 & \g_1\oplus\g_3\oplus\g_4 & 2\g_1\oplus\g_2\oplus\g_3\oplus\g_4\\
D_{2h} &mmm& \g_1^+ & \g_2^-\oplus\g_3^-\oplus\g_4^- & 2\g_1^+\oplus\g_2^+\oplus\g_3^+\oplus\g_4^+ \\
\hline
C_4 &4& \g_1 & \g_1\oplus\g_3\oplus\g_4 & \g_1\oplus2\g_2\oplus\g_3\oplus\g_4 \Tstrut\\
S_4 &\bar{4}& \g_1 & \g_2\oplus\g_3\oplus\g_4 & \g_1\oplus2\g_2\oplus\g_3\oplus\g_4 \\
C_{4h} &4/m& \g_1^+ & \g_1^-\oplus\g_3^-\oplus\g_4^- & \g_1^+\oplus2\g_2^+\oplus\g_3^+\oplus\g_4^+ \\
\hline
D_4 &422& \g_1 & \g_3\oplus\g_5 & \g_1\oplus\g_2\oplus\g_4\oplus\g_5\Tstrut \\
C_{4v} &4mm& \g_1 & \g_1\oplus\g_5 & \g_1\oplus\g_2\oplus\g_3\oplus\g_5 \\
D_{2d} &\bar{4}2m& \g_1 & \g_3\oplus\g_5 & \g_1\oplus\g_2\oplus\g_3\oplus\g_5 \\
D_{4h} &4/mmm & \g_1^+ & \g_3^-\oplus\g_5^- & \g_1^+\oplus\g_2^+\oplus\g_4^+\oplus\g_5^+ \\
\hline
C_3 &3& \g1 & \g_1\oplus\g_2\oplus\g_3 & \g_1\oplus2\g_2\oplus2\g_3 \Tstrut\\
C_{3i} &\bar{3}& \g1^+ & \g_1^-\oplus\g_2^-\oplus\g_3^- & \g_1^+\oplus2\g_2^+\oplus2\g_3^+ \\
\hline
D_3 &32& \g_1 & \g_2\oplus\g_3 & \g_1\oplus2\g_3 \Tstrut\\
C_{3v} &3m& \g_1 & \g_1 \oplus\g_3 & \g_1\oplus2\g_3 \\
D_{3d} &\bar{3}m& \g_1^+ & \g_2^-\oplus\g_3^- & \g_1^+\oplus2\g_3^+ \\
\hline
C_6 &6& \g_1 & \g_1\oplus\g_3\oplus\g_5 & \g_1\oplus\g_3\oplus\g_4\oplus\g_5\oplus\g_6 \Tstrut\\
C_{3h} &\bar{6}& \g_1 & \g_2\oplus\g_3\oplus\g_5 & \g_1\oplus\g_3\oplus\g_4\oplus\g_5\oplus\g_6 \\
C_{6h} &6/m& \g_1^+ & \g_1^-\oplus\g_3^-\oplus\g_5^- & \g_1^+\oplus\g_3^+\oplus\g_4^+\oplus\g_5^+\oplus\g_6^+ \\
\hline
D_6 &622& \g_1 & \g_2\oplus\g_6 & \g_1\oplus\g_5\oplus\g_6 \Tstrut\\
C_{6v} &6mm& \g_1 & \g_2\oplus\g_1 & \g_1\oplus\g_5\oplus\g_6 \\
D_{3h} &\bar{6}2m& \g_1 & \g_3\oplus\g_5 & \g_1\oplus\g_5\oplus\g_6 \\
D_{6h} &6/mmm& \g_1^+ & \g_2^-\oplus\g_6^- & \g_1^+\oplus\g_5^+\oplus\g_6^+ \\
\hline
T &23& \g_1 & \g_4 & \g_2\oplus\g_3\oplus\g_4 \Tstrut\\
T_h &m\bar{3}& \g_1^+ & \g_4^- & \g_2^+\oplus\g_3^+\oplus\g_4^+\\
\hline
O &432& \g_1 & \g_4 & \g_3\oplus\g_5 \Tstrut\\
T_d &\bar{4}3m& \g_1 & \g_4 & \g_3\oplus\g_4 \\
O_h &m\bar{3}m& \g_1^+ & \g_4^- & \g_3^+\oplus\g_5^- \\
\end{tabular}}
\caption{Decompositions of the representations spanned by spinless $s,p$ and $d$ orbitals into point group representations\cite{PointGroupTables}. The first column gives the point group symbol in Schoenflies notation, listed in the conventional order, and the second column gives point group symbol in Hermann-Mauguin notation. $s$ orbitals transform in the point group representation listed in the third column. $p$ orbitals transform in the representation listed in the fourth column, and $d$ orbitals transform in the representation listed in the last column.  The representation labels correspond to the labelling of little group representations at the $\Gamma$ point; the notation matches the Bilbao Crystallographic Server \cite{ssc,cdml,Bilbao1}}\label{table:orbtab1}.
\end{table}
\begin{table}[H]
\centering
\begin{tabular}{L|L|L|L|L}
\mathrm{PG}&\mathrm{PG Symbol} & s & p & d \\
\hline
C_1 &1& 2\bg_2 & 6\bg_2 & 10\bg_2\Tstrut \\
C_i &\bar{1}& 2\bg_3 & 6\bg_2 & 10\bg_3 \\
\hline
C_2 &2& \bg_3\oplus\bg_4 & 3\bg_3\oplus3\bg_4 & 5\bg_3 \oplus 5\bg_4 \Tstrut\\
C_s &m& \bg_3\oplus\bg_4 & 3\bg_3\oplus3\bg_4 & 5\bg_3 \oplus 5\bg_4 \\
C_{2h} &2/m& \bg_3\oplus\bg_4 & 3\bg_5\oplus3\bg_6 & 5\bg_3 \oplus 5\bg_4 \\
\hline
D_2 &222& \bg_{5} & 3\bg_5 & 5\bg_5\Tstrut \\
C_{2v} &mm2& \bg_{5} & 3\bg_5 & 5\bg_5 \\
D_{2h} &mmm&  \bg_{5} & 3\bg_6 & 5\bg_5 \\ 
\hline
C_4 &4& \bg_6 \oplus \bg_8 & 2\bg_6\oplus2\bg_8\oplus\bg_5\oplus\bg_7 & 2\bg_6\oplus2\bg_8\oplus3\bg_5\oplus3\bg_7\Tstrut\\
S_4 &\bar{4}& \bg_6 \oplus \bg_8 & \bg_6\oplus\bg_8\oplus2\bg_5\oplus2\bg_7 & 2\bg_6\oplus2\bg_8\oplus3\bg_5\oplus3\bg_7\\
C_{4h} &4/m& \bg_6 \oplus \bg_8 & 2\bg_{10}\oplus2\bg_{12}\oplus\bg_{9}\oplus\bg_{11} & 2\bg_6\oplus2\bg_8\oplus3\bg_5\oplus3\bg_7\\
\hline
D_4 &422& \bg_7 & 2\bg_7\oplus\bg_6 & 2\bg_7\oplus3\bg_6 \Tstrut\\
C_{4v} &4mm& \bg_7 & 2\bg_7\oplus\bg_6 & 2\bg_7\oplus3\bg_6 \\
D_{2d} &\bar{4}2m& \bg_7 & 2\bg_6\oplus\bg_7 & 2\bg_7\oplus3\bg_6 \\
D_{4h} &4/mmm& \bg_7 & 2\bg_8\oplus\bg_9 & 2\bg_7\oplus3\bg_6 \\
\hline
C_3 &3& \bg_5 \oplus \bg_6 & 2\bg_5 \oplus 2\bg_6 \oplus2\bg_4 & 3\bg_5 \oplus 3\bg_6 \oplus4\bg_4\Tstrut\\
C_{3i} &\bar{3}& \bg_5 \oplus \bg_6 & 2\bg_7 \oplus 2\bg_8 \oplus2\bg_9 & 3\bg_5 \oplus 3\bg_6 \oplus4\bg_4\\
\hline
D_3 &32& \bg_6 & 2\bg_6\oplus\bg_4\oplus\bg_5 & 3\bg_6 \oplus 2\bg_4 \oplus 2\bg_5\Tstrut \\
C_{3v} &3m& \bg_6 & 2\bg_6\oplus\bg_4\oplus\bg_5 & 3\bg_6 \oplus 2\bg_4 \oplus 2\bg_5 \\
D_{3d} &\bar{3}m &\bg_8 & 2\bg_9 \oplus \bg_6 \oplus \bg_7 & 3\bg_8 \oplus 2\bg_4 \oplus 2\bg_5 \\
\hline
C_6 &6& \bg_{10}\oplus\bg_{11} & 2\bg_{10}\oplus2\bg_{11}\oplus\bg_7\oplus\bg_8 & 2\bg_{10}\oplus2\bg_{11}\oplus2\bg_7\oplus2\bg_8\oplus\bg_9\oplus\bg_{12}\Tstrut \\
C_{3h} &\bar{6}& \bg_{10}\oplus\bg_{11} & 2\bg_{10}\oplus2\bg_{11}\oplus\bg_7\oplus\bg_8 & 2\bg_{10}\oplus2\bg_{11}\oplus2\bg_7\oplus2\bg_8\oplus\bg_9\oplus\bg_{12} \\
C_{6h} &6/m& \bg_{10}\oplus\bg_{11} & 2\bg_{16}\oplus2\bg_{17}\oplus\bg_{13}\oplus\bg_{14} & 2\bg_{10}\oplus2\bg_{11}\oplus2\bg_7\oplus2\bg_8\oplus\bg_9\oplus\bg_{12} \\
\hline
D_6 &622& \bg_9 & 2\bg_9 \oplus \bg_7 & 2\bg_9\oplus2\bg_7\oplus\bg_8 \Tstrut\\
C_{6v} &6mm& \bg_9 & 2\bg_9 \oplus \bg_7 & 2\bg_9\oplus2\bg_7\oplus\bg_8 \\
D_{3h} &\bar{6}2m& \bg_9 & 2\bg_9 \oplus \bg_7 & 2\bg_9\oplus2\bg_7\oplus\bg_8 \\
D_{6h} &6/mmm& \bg_9 & 2\bg_{12} \oplus \bg_{10} & 2\bg_9\oplus2\bg_7\oplus\bg_8 \\
\hline
T &23& \bg_5 & \bg_5\oplus\bg_6\oplus\bg_7 & \bg_5\oplus2\bg_6\oplus2\bg_7 \Tstrut\\
T_h &m\bar{3}& \bg_5 & \bg_8\oplus\bg_9\oplus\bg_{10} & \bg_5\oplus2\bg_6\oplus2\bg_7 \\
\hline
O &432& \bg_6 & \bg_8\oplus\bg_6 & 2\bg_8\oplus\bg_7 \Tstrut\\
T_d &\bar{4}3m& \bg_6 & \bg_8\oplus\bg_6 & 2\bg_8\oplus\bg_7 \\
O_h &m\bar{3}m& \bg_6 & \bg_8\oplus\bg_{11} & 2\bg_{10}\oplus\bg_7 \\
\end{tabular}
\caption{Decompositions of the representations spanned by  spinful $s,p$ and $d$ orbitals (assuming spin-1/2 electrons) into point group representations\cite{PointGroupTables}. The first column gives the point group symbol in Schoenflies notation, listed in the conventional order, and the second column gives point group symbol in Hermann-Mauguin notation. $s$ orbitals transform according to  the point group representation listed in the third column. $p$ orbitals transform  according to  the representation listed in the fourth column, and $d$ orbitals transform  according to  the representation listed in the last column. The representation labels correspond to the labelling of little group representations at the $\Gamma$ point; the notation matches the Bilbao Crystallographic Server \cite{ssc,cdml,Bilbao1}.}\label{table:orbtab2}
\end{table}

\begin{table}[H]
\small
\centering
\label{my-label}
\begin{tabular}{lclclclclc}\toprule
{SG} &{Mat.} &{SG} &{Mat.} &{SG} &{Mat.} &{SG} &{Mat.} &{SG} &{Mat.}   \\
\hline\hline
\\
2  $P\bar{1}     $  & IrTe$_2$            & 92  $P4_{1}2_{1}2   $    & La$_5$Si$_4$         & 146 $R3            $ & SnAu$_5$              & 178  $P6_{1}22       $  & Ir$_3$Zr$_5$          & 221 $Pm{\bar 3}m$  & LaIn$_3$   \\
4  $P2_{1}  $  & Ge$_2$LaPt$_2$      & 100 $P4bm           $    & La$_5$S$_7$          & 147 $P{\bar 3}     $ & NW$_2$                & 180  $P6_{2}22       $  & Ge$_2$Ta              & 223 $Pm{\bar 3}n$  & IrTi$_3$\\
13 $P2/c    $  & AuCrTe$_4$	     & 103 $P4cc           $    & TaTe$_4$             & 148 $R{\bar 3}     $ & Ir$_3$Te$_8$          & 182  $P6_{3}22       $  & Ni$_3$N               & 224 $Pn{\bar 3}n$  & AgO$_2$\\
14 $P2_{1}/c$  & AgF$_4$Na$_2$       & 109 $I41md          $    & LaPtSi               & 149 $P312          $ & TiO$_3$               & 185  $P6_{3}cm       $  & IrMg$_3$              & 225 $Fm{\bar 3}m$  & BiLa\\
26 $Pmc21   $  & In$_4$LaPd$_2$      & 113 $P{\bar 4}2_{1}m$    & Na$_5$Sn             & 150 $P321          $ & Li$_7$Pb$_2$          & 186  $P6_{3}mc       $  & Au$_3$Sr$_7$          & 226 $Fm{\bar 3}c$  & NaZn$_13$\\
34 $Pnn2    $  & CoTe$_2$            & 120 $I{\bar4}c2     $    & K(SnAu$_2$)$_2$      & 152 $P3_{1}21      $ & Ga$_3$Ni$_13$Ge$_6$   & 187  $P{\bar 6}m2    $  & LiZnGe                & 227 $Fd{\bar 3}m$  & RbBi$_2$\\
36 $Cmc2_{1}$  & AsNi	             & 122 $I{\bar4}d      $    & FeAgS$_2$            & 155 $R32           $ & Ni$_3$S$_2$           & 188  $P{\bar 6}c2    $  & LiScI$_3$             & 230 $Ia{\bar 3}d$  & Ga$_4$Ni$_3$\\
39 $Aem2    $  & LaS                 & 123 $P4/mmm         $    & InSePd$_5$           & 157 $P31m          $ & AuCd                  & 189  $P{\bar 6}2m    $  & GaAg$_2$              &                     & \\
43 $Fdd2    $  & Ge$_5$Y$_3$         & 128 $P4/mnc         $    & CSc$_3$              & 159 $P31c          $ & IrLi$_2$Si$_3$        & 190  $P{\bar 6}2c    $  & HfSnRh                &                     & \\
52 $Pnna    $  & Bi$_3$Sr$_2$        & 129 $P4/nmm         $    & LaTe$_2$             & 160 $R3m           $ & As$_3$Sn$_4$          & 191  $P6/mmm         $  & Ga$_2$La              &                     & \\
55 $Pbam    $  & Al$_3$Pt$_5$        & 130 $P4/ncc         $    & Ge$_3$La$_5$         & 161 $R3c           $ & Li$_2$ReO$_3$         & 193  $P6{\bar 3}/mcm $  & Sr$_5$Sb$_3$          &                     & \\
58 $Pnnm    $  & AlAu$_2$            & 131 $P4_{2}/mmc     $    & La(BC)$_2$           & 162 $P{\bar 3}1m   $ & Ag$_5$(PbO$_3$)$_2$   & 194  $P6{\bar 3}/mmc $  & Ge$_3$Li$_2$Zn        &                     & \\
59 $Pnmm    $  & Ag$_3$Sn            & 136 $P4_{2}/mnm     $    & ReO$_2$              & 164 $P{\bar 3}m1   $ & Ag$_2$F               & 198  $P2_{1}3        $  & NiAsS                 &                     & \\
61 $Pbca    $  & AgF$_2$             & 138 $P4_{2}/mcm     $    & Ge$_7$La$_11$Mg$_2$  & 165 $P{\bar 3}c1   $ & Ca$_5$CuPb$_3$        & 200  $Pm{\bar 3}     $  & Au$_6$In$_5$Na$_2$    &                     & \\
62 $Pnma    $  & AgSr                & 139 $I4/mmm         $    & LiTlPd$_2$           & 166 $R{\bar 3}m    $ & Zr$_2$Te$_2$P         & 205  $Pa{\bar 3}     $  & PdN$_2$               &                     & \\
63 $Cmcm    $  & BiZr                & 140 $I4/mcm         $    & Te$_3$Tl$_5$         & 167 $R{\bar 3}c    $ & Ir$_3$Mg$_13$         & 206  $Ia{\bar 3}     $  & Mg$_3$Bi$_2$          &                     & \\
64 $Cmce    $  & Al$_3$Ge$_4$La$_2$  & 141 $I4_{1}/amd     $    & NiTi$_2$             & 173 $P6_{3}        $ & AlCaSi                & 212  $P4_{3}32       $  & BaSi$_2$              &                     & \\
65 $Cmce    $  & Al$_3$Ge$_4$La$_2$  & 142 $I4_{1}/acd     $    & IrSn$_4$             & 174 $P{\bar 6}     $ & Li$_2$Ni$_12$P$_7$    & 213  $P4_{2}32       $  & Ni$_2$W$_3$N          &                     & \\
74 $Imma    $  & La$_3$Pd$_4$Si      & 143 $P3             $    & TiNi                 & 175 $P6/m          $ & Rb$_4$SnTe$_4$        & 214  $P4_{2}32       $  & La$_3$SbI$_3$         &                     & \\
84 $P4_{2}/m$  & AlNi$_4$Zr$_5$      & 144 $P3_{1}         $    & IrGe$_4$              & 176 $P6_{3}/m      $ & V$_3$S$_4$            & 215  $P{\bar 4}3m    $  & Li$_8$Al$_3$Si$_5$    &                     & \\
\hline
\end{tabular}
\caption{Excerpt of semimetal candidates, with electron filling \emph{smaller} than the number of bands in the smallest PEBR. This criteria ensures that all materials shown are partially filled (semi-)metals with SOC. A complete list will be presented in a future work.}\label{table:semimetals}                           
\end{table}

\section{Table of EBRs and PEBrs}

Here we give the table of elementary and physically elementary band representations induced from the maximal Wykoff positions in all 230 space groups in a condensed form. The column labeled ``SG'' gives the space group number. ``MWP'' gives the standard name of the maximal Wyckoff position, and ``WM'' gives its multiplicity in the primitive cell. ``PG'' is the point group number of for the site symmetry group, and ``Irrep'' gives the name of the site-symmetry group representation from which each band representation is induced. The reperesentations are labelled using the notation of Stokes, Cordes, and Campbell\cite{ssc}. The column ``Dim'' denotes the dimension of the point stabilizer group irrep. The column ``KR'' denotes whether the band representation is also a physical band representation. Those with a ``$1$'' in this column are PEBRs as is, Those with a ``$2$'' join with copies of themselves when TR symmetry is included. Finally, EBRs labelled by ``$f$'' (for first) pair with their conjugate BR labelled by ``$s$'' (and listed directly below) when TR symmetry is added. 
The column labelled ``Bands'' gives the total number of bands in the physical band representation (to obtain the number of bands in the EBR without TR, divide this number by $1$ if the entry in KR is $1$, and $2$ otherwise). The column ``Re'' indicates whether the given band representation can be made time-reversal invariant in momentum space: a $1$ in this column indicates that TR symmetry is satisfied at each $\mathbf{k}$ point, while a $2$ indicates that the given band representation must be connected in momentum space with its TR conjugate. In particular, those band representations induced from $1d$ site-symmetry representations and with a $1$ in the ``Re'' column are prime candidates for topological insulators, as discussed in Section~IV.~A of the main text. Finally, the columns ``E'' and ``PE'' indicate whether the given band representation is an exception (in the language of Sec.~\ref{sec:bandreps} and Tables~\ref{table:sbr}, \ref{table:dbr}, and \ref{table:zakwaswrong}), with and without TR symmetry respectively. An ``e'' in either of these columns indicates elementary, while a ``c'' indicates composite. This full set of data can be accessed in uncondensed form through the BANDREP program on the Bilbao Crystallographic Server\cite{progbandrep}.
\setlength{\LTcapwidth}{\linewidth}

{\tiny
\begin{longtable}{LLLLLLLLLLL||LLLLLLLLLLL}
\mathrm{SG} &\mathrm{MWP} &\mathrm{WM} &\mathrm{PG} &\mathrm{Irrep}&\mathrm{Dim} &\mathrm{KR} &\mathrm{Bands} & \mathrm{Re}&\mathrm{E}&\mathrm{PE} &\mathrm{SG} &\mathrm{MWP} &\mathrm{WM} &\mathrm{PG} &\mathrm{Irrep}&\mathrm{Dim} &\mathrm{KR} &\mathrm{Bands} & \mathrm{Re}&\mathrm{E}&\mathrm{PE}\\
\hline\hline
\\
1 &  1a & 1&  1&  \g_{1}&  1&  1&  1&  1&  e&  e&  	131 &  2d & 2&  8&  \g_{2}^-&  1&  1&  2&  1&  e&  e \\ 
1 &  1a & 1&  1&  \bg_{2}&  1&  2&  2&  2&  e&  e&  	131 &  2d & 2&  8&  \g_{4}^+&  1&  1&  2&  1&  e&  e \\ 
2 &  1a & 1&  2&  \g_{1}^+&  1&  1&  1&  1&  e&  e&  	131 &  2d & 2&  8&  \g_{4}^-&  1&  1&  2&  1&  e&  e \\ 
2 &  1a & 1&  2&  \g_{1}^-&  1&  1&  1&  1&  e&  e&  	131 &  2d & 2&  8&  \g_{3}^+&  1&  1&  2&  1&  e&  e \\ 
2 &  1a & 1&  2&  \bg_{3}&  1&  2&  2&  2&  e&  e&  	131 &  2d & 2&  8&  \g_{3}^-&  1&  1&  2&  1&  e&  e \\ 
2 &  1a & 1&  2&  \bg_{2}&  1&  2&  2&  2&  e&  e&  	131 &  2d & 2&  8&  \bg_{5}&  2&  1&  4&  1&  e&  e \\
\hline		
\end{longtable}		
}		

\bibliography{connectivity-supp}